\documentclass[%
 reprint,
superscriptaddress,
 amsmath,amssymb,
 aps, spanish,
prb,
]{revtex4-2}

\usepackage{hyperref}
\usepackage[utf8]{inputenc}
\hypersetup{
    colorlinks=true,
    citecolor=blue,
    linkcolor=blue,
    urlcolor=blue
    }

\usepackage{soul}
\usepackage{booktabs} 
\usepackage{graphicx}
\usepackage{epstopdf}
\usepackage{dcolumn}
\usepackage{bm}
\usepackage[dvipsnames]{xcolor}
\usepackage{float} 

\usepackage[normalem]{ulem}

\begin{document}

\preprint{APS/123-QED}

\title{
Domain wall dynamics driven by a transient laser-induced magnetisation
}

\author{Paul-Iulian Gavriloaea}
\email{paul.gavriloaea@csic.es}
\affiliation{Instituto de Ciencia de Materiales de Madrid, CSIC, Cantoblanco, 28049 Madrid, Spain}
	\author{Elías Saugar}
\affiliation{Instituto de Ciencia de Materiales de Madrid, CSIC, Cantoblanco, 28049 Madrid, Spain}
	\author{Rubén Otxoa}
\affiliation{Hitachi Cambridge Laboratory, CB3 OHE Cambridge, United Kingdom}
\affiliation{Donostia International Physics Center, 20018 Donostia San Sebastian, Spain}
	\author{Oksana Chubykalo-Fesenko}
\affiliation{Instituto de Ciencia de Materiales de Madrid, CSIC, Cantoblanco, 28049 Madrid, Spain}

\begin{abstract}
One of  the fundamental effects of the laser-matter interaction is the appearance of  an induced \textit{transient magnetisation}. 
While the underlying phenomena differ in their microscopic origin and cover a diverse array of materials, here we address a fundamental question 
about the possibility to drive domain-wall  dynamics on the femtosecond timescale of the exchange interactions solely by longitudinal changes of the magnetic moments.
 We verify the viability of this hypothesis in the case of a generic ferromagnetic system described in the framework of the high-temperature micromagnetic model based on the Landau-Lifshitz-Bloch  equation. 
The effect is  investigated in a  1D model at constant temperature as well as in a full micromagnetic framework considering realistic laser-induced heating. 
Our results demonstrate that domain-wall deformation in a femtosecond timeframe leads to the displacement of the wall on a larger timescale up to nanoseconds accompanied by a release of excess energy in the form of spin waves. The domain wall deformation leads to the appearance of a magnetisation gradient across the wall which promotes the motion towards the region consisting of spins with decreased magnetisation length. The total displacement is enhanced at larger temperatures and smaller damping due to an increase of the longitudinal relaxation time which ensures the longer presence of the induced magnetisation gradient. 
We also demonstrate an enhanced domain wall motion in the presence of the Dzyaloshinskii-Moriya interaction attributed to augmented magnonic torques. Our results are important towards the understanding of ultrafast magnetism phenomena on the sub-picosecond timescale.

\end{abstract}

\maketitle


\section{\label{sec:level1}Introduction}

        The light-matter interaction holds the key to the fastest and least dissipative magnetisation dynamics observed so far. A cornerstone experimental result, the sub-picosecond demagnetisation of  Ni thin films obtained by Beaurepaire \textit{et al.} in 1996 \cite{Beaurepaire1996UltrafastNickel}, sparked the ever-growing interest for laser induced, ultra-fast magnetisation dynamics, a field of investigation already in its third decade of existence. The optical manipulation of the magnetic order parameter has been explored in a wide variety of materials including metals \cite{Beaurepaire1996UltrafastNickel,Koopmans2000UltrafastOptics,Hohlfeld2001FastPulses,Stanciu2006UltrafastCompensation,Stanciu2007All-opticalLight,Radu2011TransientSpins,Eschenlohr2013UltrafastDemagnetization,John2017MagnetisationPulses}, dielectrics \cite{Kimel2002UltrafastCoupling,Hansteen2005FemtosecondFilms,Atoneche2010LargeGarnet,Stupakiewicz2017UltrafastMedium,Stupakiewicz2019SelectionGarnet} and semiconductors \cite{Kojima2003ObservationMeasurementsb,Wang2007Light-inducedModes,Nemec2012ExperimentalTorque,Ramsay2015OpticalSemiconductorb,Ishii2018UltrafastFilm}. However,  a complete understanding of strongly non-equilibrium magnetism is yet to be achieved, leading often to divergent opinions on the microscopic origin of the observed phenomena. 
        \newline\indent
        While the exhaustive analysis of light-induced effects sets itself as a laborious task, the overall complexity may be reduced by isolating and classifying several mechanisms based on their similar physical nature. 
One such distinction was discussed in the work of Kirilyuk \textit{et al.} \cite{Kirilyuk2010UltrafastOrder}, where an ultra-fast laser pulse excitation is seen to give rise firstly to a class of thermally induced effects, in which the energy carried by the photon system is transferred to the electron and phonon baths, leading to an ultrarapid heating of the sample. Irrespective of the laser pulse polarisation, the heating alone can trigger demagnetisation \cite{Beaurepaire1996UltrafastNickel,Kojima2003ObservationMeasurementsb,Kimel2002UltrafastCoupling}, precession \cite{Koopmans2000UltrafastOptics,Stanciu2006UltrafastCompensation,Wang2007Light-inducedModes} or complete switching of the magnetic order parameter \cite{Hohlfeld2001FastPulses,Radu2011TransientSpins,Ostler2012UltrafastFerrimagnet}. Secondly, they identify the broader class of non-thermal effects which are intrinsically dependent on the polarisation of the laser pulse.
        In the grand picture, non-thermal effects can induce changes in the magnetocrystalline anisotropy of garnet systems by modifying the charge distribution of magnetic ions \cite{Stupakiewicz2017UltrafastMedium,Stupakiewicz2019SelectionGarnet}, as well as they allow the dynamic control of the magnetisation in ferromagnetic semiconductors for excitations above the band gap level \cite{Nemec2012ExperimentalTorque,Ramsay2015OpticalSemiconductorb}. Furthermore, they can lead to switching via an inverse optomagnetic route in metals and dielectrics \cite{Hansteen2005FemtosecondFilms,John2017MagnetisationPulses} or give rise to spin polarisation on the basis of light-induced phonon excitation in antiferromagnetic insulators \cite{Disa2020PolarizingField,Afanasiev2021UltrafastPhonons,Formisano2022Laser-inducedCoF2}. Although in practical situations the heating of the laser-excited sample is always present to some extent, we shall further employ the \textit{non-thermal} syntagm to describe only those light-induced effects which ideally do not modify from a thermodynamic point of view the temperature of the system. 
        \newline\indent 
 The \textit{transient magnetisation} terminology has been previously used in the literature to define the after-effect of a given laser-matter interaction, which can include the individual or combined effect of thermal or non-thermal phenomena. 
        This umbrella term is generally expressed mathematically as a time-dependent, vector quantity $\delta\bm{m}(t)$ which is used to describe laser-induced changes of the (reduced) magnetisation vector $\bm{m}$ that may or may not outlive the timescale of the excitation. In the work of John \textit{et al.} \cite{John2017MagnetisationPulses}, the origin of the transient $\delta\bm{m}(t)$ magnetisation is assumed to be the non-thermal mechanism of the inverse Faraday effect (IFE), a phenomenon widely investigated in many classes of materials as for example metals \cite{Hertel2006TheoryMetals,Kurkin2008TransientMagnetization,Popova2011TheoryExperiments,Battiato2014QuantumEffect,Berritta2016AbMetals,Freimuth2016Laser-inducedFerromagnets}, plasmonic systems \cite{Smolyaninov2005Plasmon-InducedNanostructures, Hurst2018MagneticEffect} as well as non-traditional materials such as molecular magnets  \cite{Tokman2009InverseMagnets} and magnetic ionic liquids \cite{Jin2011FemtosecondbmimFeCl4}. In a recent \textit{ab-initio} study it was shown that the absorption of circularly polarised light can further induce a cumulative, helicity-dependent magnetisation component in ferromagnets \cite{Scheid2019AbFerromagnets}. Unlike the IFE, the latter mechanism is of dissipative nature and scales linearly both with the laser pulse intensity and with time, being also argued it becomes dominant in the little explored ultra violet regime \cite{Scheid2021AbRegime}. In a different fashion, THz driven phonon excitation has been shown capable of inducing an ultra-fast, first-order phase transition in the magnetic insulator DyFeO$_3$ \cite{Afanasiev2021UltrafastPhonons}. The use of a sub-picosecond, mid-infrared electric field pulse in resonance with an optical phonon mode leads to the appearance of a macroscopic transient magnetisation in the system which accompanies the internal coherent spin reorientation from an antiferromagnetic to an weakly ferromagnetic state.
        Similarly, THz excitation of optical phonons in the antiferromagnetic difluoride CoF$_2$, followed by subsequent lattice dynamics in conjunction with transverse or longitudinal piezomagnetism gives rise to a transient net magnetisation \cite{Disa2020PolarizingField,Formisano2022Laser-inducedCoF2}. In addition, the study of magneto-optical phenomena might benefit from the advancements in the quantum optics field as the ultra-strong, light-matter coupling regime is being explored \cite{FriskKockum2019UltrastrongMatter}. For example, the adjacent topic of cavity magnonics offers the promise of enhanced magneto-optical fields with tailored chirality at desired wavelengths \cite{Hubener2020EngineeringCavities,Barman2021TheRoadmap,Li2022ADynamics}, which might augment the degree of control as well as the amplitude of an optically induced transient magnetisation component.  
        \newline\indent  
Typical numerical models of laser-induced magnetisation dynamics describe non-thermal phenomena more often via transverse and precession dynamics or relaxation mechanisms \cite{Hansteen2005FemtosecondFilms,Atoneche2010LargeGarnet,Nemec2012ExperimentalTorque,Huisman2016FemtosecondHeterostructures,Choi2017Optical-helicity-drivenFerromagnets}, with any optically induced longitudinal magnetisation changes being usually neglected due to their smaller amplitudes or much faster equilibration times.   Modifications of the magnetisation vector length are ultimately introduced as a result of the thermal effect of the laser pulse excitation in stochastic atomistic modelling \cite{Kazantseva2009LinearTemperature,Ostler2012UltrafastFerrimagnet,Ellis2016All-opticalDichroism} or by employing the Landau-Lifshitz-Bloch (LLB) equation \cite{Ramsay2015OpticalSemiconductorb, John2017MagnetisationPulses,Raposo2020MicromagneticDichroism}.  Furthermore, the IFE effect is often taken into account as a local field of very large amplitude of the order of $10~$T and even above \cite{Vahaplar2009UltrafastState,Vahaplar2012All-opticalModeling,Ellis2016All-opticalDichroism,Raposo2020MicromagneticDichroism,Longman2021Kilo-TeslaBeams}.
In previous micromagnetic works concerning the investigation of laser-induced domain-wall (DW) dynamics,  the light-matter interaction primarily leads to the appearance of temperature gradients or spin-wave (SW) excitations which ultimately drive DWs via entropic and magnonic torques \cite{Hinzke2011DomainEffectb,Schlickeiser2014RoleGradients,Moretti2017DomainGradients}. Generally in such situations, thermally induced transverse and precessional relaxation processes dominate the dynamics on the $ps-ns$ timescale, where spatial non-uniformities in the anisotropy or exchange stiffness parameters become relevant. 
        \newline\indent
 Concerning \textit{ab-initio} models, it has been argued that the IFE produces both magnetisation torque \cite{Freimuth2016Laser-inducedFerromagnets} as well as the  modification of spin and orbital magnetic moment \cite{Battiato2014QuantumEffect,Berritta2016AbMetals}. In an attempt to reflect the results of the \textit{ab-initio} theory, in relation to electrically induced Rashba Edelstein
effect in Mn$_2$Au,  recent atomistic spin-resolved models  include explicitly the presence of the orbital magnetic moment, and its interaction with the spin moment \cite{Selzer2022Current-inducedPrinciples}. 
  \newline\indent
In this work, we investigate the possibility to convert a transient, non-thermal, magnetisation contribution  into a transverse DW motion in a ferromagnetic system. 
As pointed out by Zhang \textit{et al.} \cite{Zhang2019Energy-EfficientTorque}, a femtosecond laser pulse excitation can help reduce the energy cost of current induced DW dynamics in perpendicularly magnetised wires. In their work, it is shown that the presence of a helicity-dependent, optical effect can reduce by $50\%$ the threshold density current of a spin-orbit torque driving mechanism in ultra thin Co/Ni/Co films, motivating from a technological point of view the potential importance of our study. 
\newline\indent
      Thus, the pure non-thermal route of driving magnetisation dynamics via longitudinal magnetisation changes remains little or completely unexplored.  Owing to the diversity of light-induced phenomena which can lead to the appearance of a transient magnetisation contribution, we approach the problem of DW dynamics without reference to a particular effect. The key idea of our study relies on the assumption that a non-thermal $\delta \bm{m}(t)$ contribution manifests as a longitudinal distortion in the magnetic texture. 
 The fundamental question which we try to answer here is the following: could pure longitudinal relaxation processes pass angular momentum to the transverse dynamics and lead to translational DW motion?
        In order to model this mechanism we make use of the  micromagnetic LLB equation which naturally allows for the description of longitudinal relaxation processes. 
        \newline\indent
    In the immediate section following this introductory part, we describe the theoretical background of our micromagnetic approach.
    The third section of the article is  divided in two parts. Firstly, in subsection "A" we approach the problem of DW dynamics in a 1D model.
 It is assumed the induced transient magnetisation acts instantaneously on the magnetic texture, all while disregarding any heating effects which might arise during a laser pulse excitation. The aim of this model is to present in a clear picture the mechanism of converting a longitudinal deformation of the magnetic texture into a DW motion based on the appearance of a non-thermal magnetisation gradient. 
 A more realistic full micromagnetic analysis of DW dynamics is employed in the second part "B" of the same results section. Here, we investigate the dynamics of an out-of-plane (OOP) DW in a thin magnetic stripe, also taking into account the presence of heating and the magnetostatic effects. 
 The DW dynamics are  investigated as a function of the laser fluence and amplitude of the induced magnetisation.
 Finally, we show how in the presence of the interfacial Dzyaloshinskii-Moryia interaction (DMI), it is possible to enhance the DW velocity and the maximum displacement achieved in a single-pulse excitation. The final section is reserved for conclusions and comments on the outlook of the developed model.

\section{Numerical model}
We consider a generic micromagnetic model describing a
ferromagnetic sample discretized in a lattice of $N$ identical cubic elements with lateral spacing $\Delta$, 
governed by the LLB equation \cite{Garanin1991GeneralizedFerromagnet,Garanin1997Fokker-PlanckFerromagnets,Chubykalo-Fesenko2006DynamicTemperature,Atxitia2007MicromagneticEquation}. Importantly, this formalism is valid at high temperatures and does not conserve the magnetisation length, thus, allowing modelling of longitudinal dynamics. The LLB equation reads:   
\begin{align}
\frac{d\bm{m}_i}{dt}&=-\gamma\bm{m}_i\times\bm{H}^i_{eff}+\gamma\alpha_{||}\frac{(\bm{m}_i\cdot\bm{H}^i_{eff})  \bm{m}_i}{m_i^2}\\ &-\gamma\alpha_{\perp}\frac{\bm{m}_i\times(\bm{m}_i\times\bm{H}^i_{eff})}{m_i^2} \nonumber.
\end{align}
The unit vector $\bm{m}_i$ is defined as $\bm{m}_i=\bm{M}_i(T)/M_s(0)$ with $M_s(0)$ being the saturation magnetisation at $0~K$ while $\bm{M}_i(T)$ is the temperature dependent magnetisation vector. The constant $\gamma$ denotes the electron gyromagnetic ratio, while $\alpha_{||}$ and $\alpha_{\perp}$ are the dimensionless longitudinal and transverse damping parameters defined as:
\begin{align}
\alpha_{||}=\lambda\frac{2T}{3T_c}, ~~~\alpha_{\perp}=\lambda\
\begin{cases}
(1-\frac{T}{3T_c})~~&T\lesssim T_c,\\
\frac{2T}{3T_c}~~~~~~~~~~&T\gtrsim T_c. 
\end{cases}
\end{align}
The proportionality factor $\lambda$ is a  measure of the intrinsic spin-flip scattering events which defines the coupling strength between the spin degrees of freedom and the thermal bath. The damping parameters $\alpha_{||}$ and $\alpha_{\perp}$ are directly proportional to the $T/T_c$ ratio where $T$ is the bath temperature and $T_c$ is the Curie point of the ferromagnetic sample.  The dynamics of each macrospin \textit{i} are governed by the total effective field $\bm{H}^i_{eff}$, defined in our case as:
\begin{align}
\label{eq:eff_field}
\bm{H}^i_{eff}&=\bm{H}^i_{ex} + \bm{H}^i_{ani} +\bm{H}^i_{dem}
\\ &+ \begin{cases}
\frac{1}{2\tilde{\chi}_{||}}\left(1-\frac{m^2_i}{m^2_e}\right)\bm{m}_i&T\lesssim T_c,\\
-\frac{1}{\tilde{\chi}_{||}}\left(1+\frac{3T_c}{5(T-T_c)}m_i^2\right)\bm{m}_i&T\gtrsim T_c. \nonumber 
\end{cases}
\end{align}
The temperature dependent $m_e=M_e(T)/M_s(0)$ value is the macrospin vector length at thermal equilibrium, obtained here within the mean-field approximation (MFA) by solving the self-consistent Curie-Weiss equation. Assuming a classical spin system ($S\rightarrow \infty $) this evaluates to: $m_e=L(\beta J_0m_e)$, where $L$ is the Langevin function, $J_0=3k_BT_C$ expresses the strength of the Heisenberg exchange coupling and $\beta=1/k_BT$ is a measure of the thermal field. The $\tilde{\chi}_{||}$ term denotes the reduced longitudinal susceptibility which in the MFA evaluates to:
\begin{equation}
\tilde{\chi}_{||}(T)=
\begin{cases}
\frac{\mu_{s}\beta L'}{1-\beta J_0L'}~~~&T\lesssim T_c, \\
\frac{\mu_{s}T_c}{J_0(T-T_c)}~~~&T\gtrsim T_c,
\end{cases}
\end{equation}
where $L'$ represents the derivative of the Langevin function with respect to the argument $x=\beta J_0m_e$ and the $\mu_s$ constant denotes the atomic magnetic moment. 
\newline\indent
The exchange field $\bm{H}^i_{ex}$ field term accounts for the micromagnetic exchange interaction between macrospins, being fundamentally defined and numerically approximated in the following manner \cite{Atxitia2007MicromagneticEquation}: 

\begin{equation}
\bm{H}^i_{ex}=\frac{2A(T)}{M_s(0)m_e^2}\nabla^2\bm{m}_i \simeq \frac{2A(T)}{M_s(0)m_e^2\Delta^2}\sum_{j=1}^{n_i}{(\bm{m}_j-\bm{m}_i)},\end{equation}
with the summation taking into account all the $n_i$ neighbours of each individual macrospin vector $\bm{m}_i$ and the $A(T)$ parameter indicating the temperature-dependent exchange stiffness.
\newline\indent
Given a Cartesian $Oxyz$ frame of reference, we consider  an uni-axial magnetocrystalline anisotropy with the easy-axis (EA) taken along the $Oz$ direction. The anisotropy field $\bm{H}^i_{ani}$ is thus defined as:
\begin{equation}
\bm{H}^i_{ani}=-\frac{(m^i_x\bm{e}_x+m^i_y\bm{e}_y)}{\tilde{\chi}_{\perp}}
\end{equation}
For practical reasons, the reduced transverse susceptibility $\tilde{\chi}_{\perp}$ is linked to a temperature dependent, anisotropy constant $K(T)$ via the relationship $\tilde{\chi}_{\perp}=[M_s(0)m_e]^2/2K(T)=M_s(T)^2/2K(T)$.
\newline\indent
The demagnetising field $\bm{H}^i_{dem}$ is
 expressed as a discrete convolution sum of the demagnetising tensor $\bm{N}(\bm{r}_i-\bm{r}_j)$ and the reduced magnetisation vector $\bm{m}_j$, taking into account the contribution of all macro-cells in the system:
\begin{equation}
\bm{H}^i_{dem}=-\mu_0M_s(0)\sum_jN(\bm{r}_i-\bm{r}_j)\cdot\bm{m}_j.
\end{equation}
Since the demagnetising tensor depends only on the relative positions $\bm{r}_{i,j}$ of the cells, its calculation is done only once at the start of each simulation employing the method of Newell \textit{et al.} \cite{Newell1993AMagnetization} The overall calculation of the $\bm{H}^i_{dem}$ field is carried out using the traditional approach based on the use of the Fast Fourier Transform (FFT) algorithm. This field contribution is only present for full micromagnetic calculations in a thin film geometry discussed in subsection "III.B".
\newline\indent
The last term in Eq. \eqref{eq:eff_field} constitutes the so-called longitudinal field $\bm{H}^i_{lon}$, a measure of the competition between the atomic spin ordering and the disorder induced by the thermal bath, which ultimately dictates the length of the macroscopic vector $\bm{m}$.
\newline\indent
The material parameters we use to model the ferromagnetic sample correspond to a generic Co system and can be consulted in Table \ref{table:MatParams}. Their temperature dependence is obtained making use of the equilibrium magnetisation $m_e$ discussed earlier in the context of the MFA. Thus, the uni-axial anisotropy constant is assumed to follow the Callen-Callen scaling law $K(T)\propto m_e^3$ \cite{Callen1966TheLawb}, while for the exchange stiffness parameter we initially consider the $A(T)\propto m_e^2$ dependence \cite{Atxitia2007MicromagneticEquation}. The latter will be revised in the case of the more realistic micromagnetic model used to describe DW dynamics in a stripe geometry. If otherwise not specified, the microscopic damping parameter is set to $\lambda=0.1$.
\begin{table}[H]
\centering
\begin{ruledtabular}
\begin{tabular}{cc}
Parameter & Value\\
\colrule
$M_s(0)$ &$1400~kA/m$ \cite{Ivanov2016SingleVortices}\\
$A(0)$ & $10~pJ/m$\\ 
$K(0)$ & $0.45~MJ/m^3$ \cite{Ivanov2016SingleVortices}\\
$T_c$ & $1480~K$ \cite{Moreno2016Temperature-dependentCo}\\
\end{tabular}
\end{ruledtabular}
\caption{Generic Co material parameters. The exchange stiffness parameter has been chosen smaller than the typical values found in literature \cite{Ivanov2016SingleVortices,Moreno2016Temperature-dependentCo} in order to decrease the $0~K$ DW width.}
\label{table:MatParams}
\end{table}

\section{Results}
In this section we apply the micromagnetic model described previously to the problem of DW dynamics in a ferromagnetic sample, assuming the presence of a transient, non-thermal magnetisation contribution $\delta \bm{m}(t)$ which longitudinally deforms the magnetic texture. Throughout this work, the $\delta \bm{m}(t)$ term will only be assumed to act along the $Oz$ direction.  The time-dependence of the aforementioned magnetisation term is disregarded initially in the case of the 1D model discussed in subsection "A", but later introduced in the case of the thin film geometry modelled in subsection "B". For these reasons, in the following pages we will be using either the $\delta m_z$ or $\delta m_z(t)$ notations when referring to the induced magnetisation component.

\subsection{1D model}
\begin{figure}[!ht]
    \centering \includegraphics[width=1.0\linewidth]{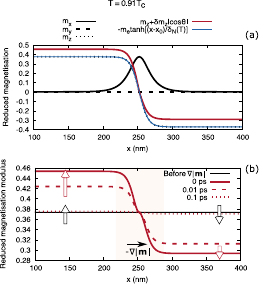}
    \caption{(a) An $180^0$ N\'eel DW configuration equilibrated in a $500~nm$ chain system at $T=0.91T_c$. By fitting the $m_z$ magnetisation profile we extract a wall width of $\delta_N(T)=12.62~nm$ which compares well with the analytical value $\delta(T)=12.64~nm$ given by Eq. \eqref{analytic_LLB_db}. (b) The effect of the longitudinal magnetisation contribution introduced via Eq. \eqref{IFE_chain} on the modulus of each macro-spin vector in the chain. The black straight line denotes the situation before the deformation is introduced, wherein all spins are characterised by the same modulus $|\bm{m}|$. Depending on their orientation with respect to the EA, the spins will either elongate or contract their length - graphically exemplified for two edge spins - leading to a magnetisation gradient $\nabla \bm{|m|}$ across the DW as it can be seen in the highlighted region, obtaining the initial configuration at $t=0~ps$. Subsequently, $\nabla \bm{|m|}$ will rapidly vanish on a \textit{fs} time-scale as suggested by the snapshots at $t=~0.01~ps$ and $t=~0.1~ps$.  }
    \label{fig:IFE_effect_chain}
\end{figure}

Let us consider first a spin chain of $500~nm$ in length which contains  at a constant temperature close  to the Curie point  $T=0.91T_c$ an $180^0$ N\'eel wall in the $Oxz$ plane - see Fig. \ref{fig:IFE_effect_chain}(a). 
To verify that we obtain a correct static N\'eel DW configuration within our model, we compare the numerically extracted wall width against the temperature-dependent expression given by \cite{Chen2019Landau-Lifshitz-BlochAntiferromagnets}:
\begin{equation}
\delta(T)  = \sqrt{\frac{2\tilde{\chi}_{\perp}A(T)}{M_sm_e^2}} = \sqrt{\frac{A(T)}{K(T)m_e}},
\label{analytic_LLB_db}
\end{equation}
Numerically we obtain the DW width fitting the $m_z$ magnetisation profile along the chain using the following equation:
\begin{equation}
m_z=-m_e\tanh[(x-x_0)/\delta_N (T))],
\end{equation}
where $x$ denotes the individual position of the spins and $x_0$ is the center of the wall. In our case, $\delta_N (T)$ evaluates to $12.62~nm$ at $T=0.91T_c$, in good agreement with the analytically obtained value of $\delta(T)=12.64~nm$. Although no magnetostatic interaction is taken into account in this spin chain study, we took two extra precautionary measures in trying to assure the reliability of
our results: a) the discretisation size is chosen in such a way as to be smaller than the characteristic exchange length given by $l_{ex}=\sqrt{\frac{2A(T)}{\mu_0M_s(T)^2}} = 2.85~nm$ \cite{Abo2013DefinitionLength}. The $l_{ex}$ parameter is independent of $T$ due to the scaling law of the temperature dependent exchange stiffness taken as $A(T)\sim m_e^{2}$. b) Although in the bulk of our results we used $\Delta=2.5~nm$, the calculations have also been carried out for $\Delta=1~nm$ producing no relevant qualitative nor quantitative differences in the results. 

Starting from the equilibrium DW configuration displayed in Fig. \ref{fig:IFE_effect_chain}(a), we introduce a magnetisation contribution which alters the macrospins' vector component along the EA direction in the following manner:
\begin{equation}
m_z ^{*i} = m_z^{i}+ \delta m_z|\cos\theta|,
\label{IFE_chain}
\end{equation}
\begin{figure*}[!ht]
    \centering
    \includegraphics[width=1.0\linewidth]{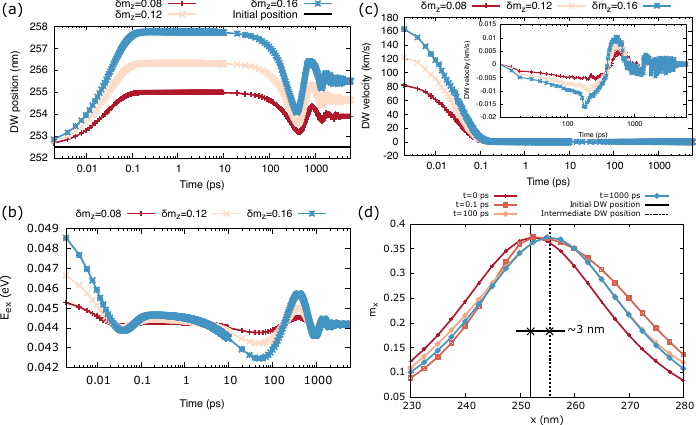}
    \caption{(a) DW displacement in time for different non-thermal, transient magnetisation contributions along $Oz$. The black solid line indicates the initial position of the wall. The DW position is tracked based on the center of mass method: $DW_{pos}=\sum_i m_x^ix^i/\sum_i m_x^i$. (b) The time variation of the micromagnetic exchange energy stored in the chain for all three different scenarios. (c) Instantaneous DW velocity extracted as the time derivative of the data presented in subfigure (a). The inset is adjusted to fit the points beyond the $10~ps$ time mark. (d) Time dependence of the N\'eel wall magnetisation profile along $Ox$ during the dynamics induced by a transient magnetisation of amplitude $\delta m_z=0.16$. The solid vertical line defines the initial DW position while the dotted line marks an intermediate state at which the wall finds itself approximately $3~nm$ away from the starting point according to Eq. \eqref{DW_pos}. The x axes in subplots (a) (b) and (c) are logarithmic.
    } 
    \label{fig:Figure2}
\end{figure*}
where $m_z^{i}$ and $m_z^{*i}$ are the $z$ magnetisation components of each macro-spin before and after the system is deformed; $\delta m_z$ is the amplitude of the longitudinal magnetisation change, modulated by the angle $\theta$ in turn defined by the relative orientation of the macro-spin vector with respect to the EA.  Eq. \eqref{IFE_chain} reflects the fact that our assumed non-thermal phenomenon as in the case of the IFE, depends on the angle between the laser polarisation and the magnetisation direction \cite{Freimuth2016Laser-inducedFerromagnets}. Thus, we consider that for a laser polarised perpendicularly to the local spin direction there is no modification of the $m_z$ component, i.e. $|\delta m_z|=0$, when they are perpendicular to each other.  

The LLB model allows one to go beyond the rigid macrospin approximation of standard micromagnetism ($|\bm{m}|=1|$). The longitudinal magnetisation change introduced in Eq. \eqref{IFE_chain} will distinctly modify the length of each spin vector, which in thermodynamic equilibrium conditions should be uniform across the chain and  readily available solving the self-consistent Curie-Weiss equation discussed previously in section "II".  In Fig. \ref{fig:IFE_effect_chain}(b) we evaluate the change in the local magnetisation modulus $|\bm{m}|$ across our spin chain due to the magnetisation contribution introduced along $Oz$. As it can be seen in the initial configuration displayed at $t=0~ps$, depending on their relative orientation with respect to the EA and the strength of the deformation, the spin vectors will either elongate or contract their magnetisation length leading to a $\nabla|\bm{m}|$ gradient across the DW.
\newline \indent
Note that a similar magnetisation gradient might occur during an ultrafast laser pulse excitation of a magnetic system leading to 
a non-uniform temperature profile in the sample. This can arise either due to the spatial profile of the laser pulse itself or due to a differential absorption  of circularly polarized light for the spin-up and spin-down expectation values on account of the magnetic circular dichroism effect \cite{Khorsand2012RoleRecording,Ellis2016All-opticalDichroism,Quessab2019ResolvingSwitching}. Ultimately, all these effects translate into the appearance of the "hot"/"cold" regions, where locally  the magnetic parameters $M_s$,$K$, $A$ are uneven through their temperature dependence. 

A pure optical route to induce a magnetisation gradient might be based on the IFE \cite{Battiato2014QuantumEffect,Berritta2016AbMetals,Freimuth2016Laser-inducedFerromagnets}. In this case, the temperature across the sample may remain uniform during excitation as presented in our model. However, since the strength of the magneto-optical response in the presence of the IFE depends on the relative orientation between the light propagation axis and the local magnetic vector, magnetisation gradients are expected to arise in non-coherent spin systems as for example in a DW configuration. Thus, although we study this effect at constant temperature, we shall use in the next paragraphs the "hot"/"cold" terminology to describe the regions of "small"/"large" magnetisation. Moreover, in similar fashion to the spin-Seebeck driven motion \cite{Hinzke2011DomainEffectb}, we report the displacement of our N\'eel DW towards the "hot" region of the system, that is the region where the macro-spin vectors have reduced their magnetisation length. In Fig. \ref{fig:Figure2}(a), one can track the DW motion in time for three different transient magnetisation contributions. The position of the wall is tracked by identifying its center of mass from the $m_x$ magnetisation profile using the following relationship \cite{Tatara2020CollectiveWall}: 
\begin{equation}
DW_{pos}=\sum_i m_x^ix^i/\sum_i m_x^i,
\label{DW_pos}
\end{equation}
where $x^i$ represents a macrospins' position along the chain. 
\newline\indent
In all situations investigated, the dynamic response can be separated in three distinct regions:\newline \indent
I) Firstly, a rapid displacement is obtained on the $fs$ timescale. Governed by longitudinal relaxation effects, the induced change in magnetisation is converted into a transverse motion of the DW in the direction of a smaller magnetisation length. During this process, the longitudinal field will revert the spin vectors' length back to the initial value before the transient magnetisation is introduced, thus neutralising the observed $\nabla|\bm{m}|$  gradient. In Fig. \ref{fig:IFE_effect_chain}(b) one can see how does $|\bm{m}|$ vary across the chain at different moments in time for the smallest deformation considered. Interestingly, already at the $0.1~ps$ mark the magnetisation gradient nearly vanishes and the spin vectors regain their original lengths. 
\newline \indent
II) No more longitudinal relaxation processes take place in the absence of $\nabla|\bm{m}|$. Furthermore, since the dynamics timescale is yet too short for any relevant transverse or precessional torques to act, the DW practically preserves its acquired position for several $ps$.
\newline \indent
III) Beyond 10 ps the remaining excess energy induced by the presence of the transient magnetisation is invested in much slower transverse and precessional relaxation processes which  promote an oscillating behaviour of the DW on the $ns$ timescale until its final equilibrium position is reached. 
\newline\indent
As we mention above, the direction of motion is towards the so-called "hot" region, as it also takes place in the DW motion driven by temperature gradients.
Previously, Schlickeiser \textit{et al.} \cite{Schlickeiser2014RoleGradients} have explained the motion of DWs in the presence of temperature gradients based on the existence of the so-called entropic torque which induces dynamics in a direction given by the $-\nabla A(T)$ gradient. Interestingly, in their analytical treatment of the DW dynamics developed in the framework of the LLB equation, they include also a driving mechanism due to the magnetisation gradient $-\nabla |\bm{m}|$ term - see equations 8 and 9 found in the Supplementary Material of \cite{Schlickeiser2014RoleGradients} - but conclude its effect is small in comparison to the former $-\nabla A$ contribution. In their situation, the dynamics were investigated on a longer time-scale where any longitudinal relaxation processes have already taken place.
\newline\indent
In order to understand the  driving mechanism,  we evaluate the time evolution of the exchange energy, using the following numerical approximation:
\begin{equation}
E_{ex} = A(T)V\sum_i\sum_j\left(\frac{\bm{n}_j-\bm{n}_i}{\Delta^2}\right)^2,
\end{equation}
where the counters $i$,$j$ loop over all the macro-spins in the system and their individual neighbours respectively and $V=\Delta^3$ is the volume of the cubic macro-cell. In standard micromagnetics the macrospin's vector length is always conserved ($|\bm{m}|=1$). To recover a similar definition, we normalise the $\bm{m}$ vector to the equilibrium magnetisation value $m_e$ and define $E_{ex}$ making use of the variable $\bm{n}=\bm{m}/m_e=\bm{M}(T)/M_s(T)$, where $|\bm{n}|\neq 1$. 
In Fig. \ref{fig:Figure2}(b) we  represent the time variation of the micromagnetic exchange energy $E_{ex}$ stored in the chain in all three situations. Comparing the initial and final states, one can appreciate qualitatively the amount of exchange energy introduced in the system due to the elongation/contraction of the spins. Thus, the DW distortion produces a large torque on neighboring spins through their elongation/contraction.
Beyond the $10~ps$ time threshold, the oscillating behaviour described earlier is also clearly observed in the exchange energy. In this timescale, however, $|\bm{n}|= 1$.
\newline\indent
A further mention here refers to the timescale and huge instantaneous velocity acquired by the DW during its initial displacement - see Fig. \ref{fig:Figure2}(c). Indeed, at this timescale the change of $\bm{m}(t)$ is governed by the longitudinal field $\bm{H}_{lon}$ - consequence of the relaxation process governed by  the internal Heisenberg interactions between an ensemble of atomic spins. On average the disorder present at the atomic level will be translated at the macroscopic scale as a change in the instantaneous magnetisation $\bm{m}(t)$, a representation of the competition between internal exchange fields and thermal fluctuations. The Heisenberg exchange coupling is the strongest known interaction in any magnetic system giving rise to ultra-fast dynamics on the $fs$ timescale. For our chosen system this internal exchange field $J_0/\mu_{at}$ evaluates to approximately $1.68 \times 10^4~T$
for a simple cubic cell with the size of $a=0.25~nm$.
As evidenced in subplot (b) of Fig. \ref{fig:IFE_effect_chain}, the induced magnetisation gradient vanishes around the $0.1~ps$ time threshold. Up to this point dominant longitudinal relaxation processes will take place under large $\bm{H}_{lon}$ fields giving rise to the enormous instantaneous velocities seen in Fig. \ref{fig:Figure2}(c). These large velocity values arise due to the ultra-fast timescale imposed by the longitudinal field. Any displacement of just a few $nm$ on the $fs$ timescale will lead to DW velocities in the $km/s$ domain. The analysis of the magnetisation profile ($m_x$ component) in Fig. \ref{fig:Figure2}(d) shows that the displacements of the DW center of mass results from the accommodation of the DW  profile to the equilibrium one on the timescale of longitudinal relaxation.  The snapshot at $t=0.1~ps$ presents the largest shape deviation from the initial configuration corresponding to the maximum DW displacement of approximately $5~nm$ recorded in Fig. \ref{fig:Figure2}(a).

\begin{figure}[t]
    \centering
    \includegraphics[width=0.7\linewidth]{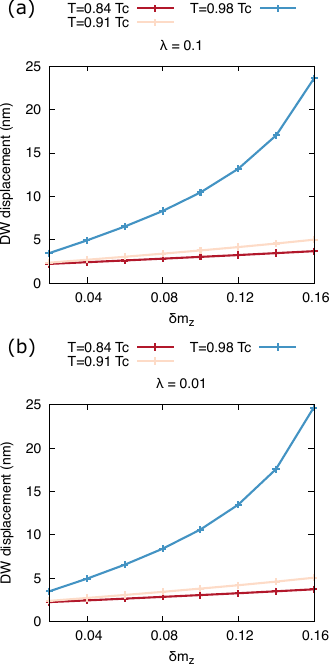}
    \caption{Final DW displacement achieved after complete equilibration as a function of the temperature $T$ and the amplitude $\delta m_z$ of the induced magnetisation for two different microscopic damping values (a) $\lambda=0.1$ and (b) $\lambda=0.01$. } 
    \label{fig:Figure3}
\end{figure}

As discussed earlier, the excess energy introduced in the system will be invested on the $ps-ns$ timescale in transverse and precession like torques which will lead finally to smaller velocities since the dynamics are governed then by the much smaller anisotropy $\bm{H}_{ani}$ and the micromagnetic exchange $\bm{H}_{ex}$ fields with $|\bm{n}|= 1$. An oscillating behaviour of the DW position is observed in the subsequent time snapshots taken at $t=100~ps$ and $t=1000~ps$. At this latter time threshold, a clear shift in the $m_x$ curve peak with respect to the initial state can be easily identified as direct evidence of the wall motion we have claimed. The DW will display onwards a back and forth motion  until final equilibration is reached. The oscillations mentioned so far become more obvious turning our attention to the $m_y$ magnetisation profile discussed in the Supplemental Material in Fig. S1 \cite{SeeGeometry}. Lastly, evidence of the DW motion can also be extracted from the $m_z$ magnetisation profile discussed in the same figure in the Supplemental \cite{SeeGeometry}. 
\newline \indent
To inquire the possibility for augmented DW dynamics, we investigate the displacement of the N\'eel wall as a function the applied temperature $T$ and the amplitude $\delta m_z$ of the induced transient magnetisation for two different values of the microscopic damping $\lambda$. 
For any given set of parameters, we run the LLB dynamics for several $ns$ extracting the final DW displacement after complete equilibration is reached. 
The result of this parameter sweep can be seen in Fig. \ref{fig:Figure3}. 
For any transient magnetisation amplitude $\delta m_z$ and irrespective of $\lambda$, the final displacement will be increased as we get closer to the Curie point. Interestingly, a smaller microscopic damping of $\lambda=0.01$ affects the end result to a very small extent, leading to a slightly larger final displacement at $T=0.98Tc$ and $\delta m_z=0.16$. 
The enhancement in displacement achieved when the temperature is increased can be explained referring ourselves solely to the longitudinal relaxation processes. 
The LLB equation allows  to define a longitudinal relaxation time as
\cite{Atxitia2007MicromagneticEquation,Nieves2014QuantumCase}:
\begin{equation}
\tau_{||}=\frac{\tilde{\chi_{||}}}{\gamma\alpha_{||}}=\frac{3\tilde{\chi_{||}}}{2\gamma}\frac{T_c}{\lambda T}
\label{longitudinal_rel}
\end{equation}
A general effect characteristic of second order phase transitions \cite{Chen1994Spin-dynamicsFerromagnet,Chubykalo-Fesenko2006DynamicTemperature}, the longitudinal relaxation time experiences a critical slowing down approaching the Curie temperature $T_c$. 
The main factor responsible for the overall increase of the $\tau_{||}$ relaxation time   is the divergence of the parallel susceptibility $\tilde{\chi}_{||}$ near the phase transition point, which dominates the $1/\lambda T$ dependence resulting
from the longitudinal damping parameter $\alpha_{||}$. 
In our case we can turn the "slow" longitudinal dynamics at elevated temperatures to our advantage: the key driving mechanism in this study is the magnetisation gradient $\nabla |\bm{m}|$ we induce across the DW through a longitudinal deformation of the system. As seen in Fig. \ref{fig:IFE_effect_chain}(b) and \ref{fig:Figure2}(a), as long as $\nabla |\bm{m}|$ is preserved the wall will continue in its displacement until temporarily reaching a halt  when $\nabla |\bm{m}| = 0$. 
Thus, increasing the life-time of $\nabla|\bm{m}|$ will  lead to larger DW displacements overall.
 The behavior of $\tau_{||}$ also explains a slight increase of the DW displacement for a smaller damping value.
\newline\indent
The relaxation or life-time of the induced magnetisation is governed here by the thermal longitudinal dynamics and consequently slows down near the Curie point. On the contrary, this effect could be parameterised by a stand-alone relaxation parameter which could result in a magnetisation gradient $\nabla|\bm{m}|$ that outlives the thermal longitudinal dynamics of the studied optical effect thus leading to significant DW displacements even away from $T_c$. 
 \newline\indent
To summarise, in this subsection we demonstrate the possibility to convert a non-thermal, transient magnetisation contribution followed by longitudinal relaxation processes into a subsequent transverse DW motion in a ferromagnetic system. The fundamental mechanism at the origin of the effect is the appearance of a magnetisation gradient $\nabla|\bm{m}|$ along the DW which allows for a displacement towards the "hot" region, that is the area of small magnetisation. The total distance covered by the DW is dependent on the life-time of the induced gradient as well the amplitude of the magnetisation contribution. Increasing the temperature of the bath $T$, one can delay the relaxation of the macro-spin vectors towards their equilibrium lengths allowing for a prolonged displacement of the wall. 
\subsection{Full micromagnetic model: stripe system}
\begin{figure*}[t]
    \centering
    \includegraphics[width=1.0\linewidth]{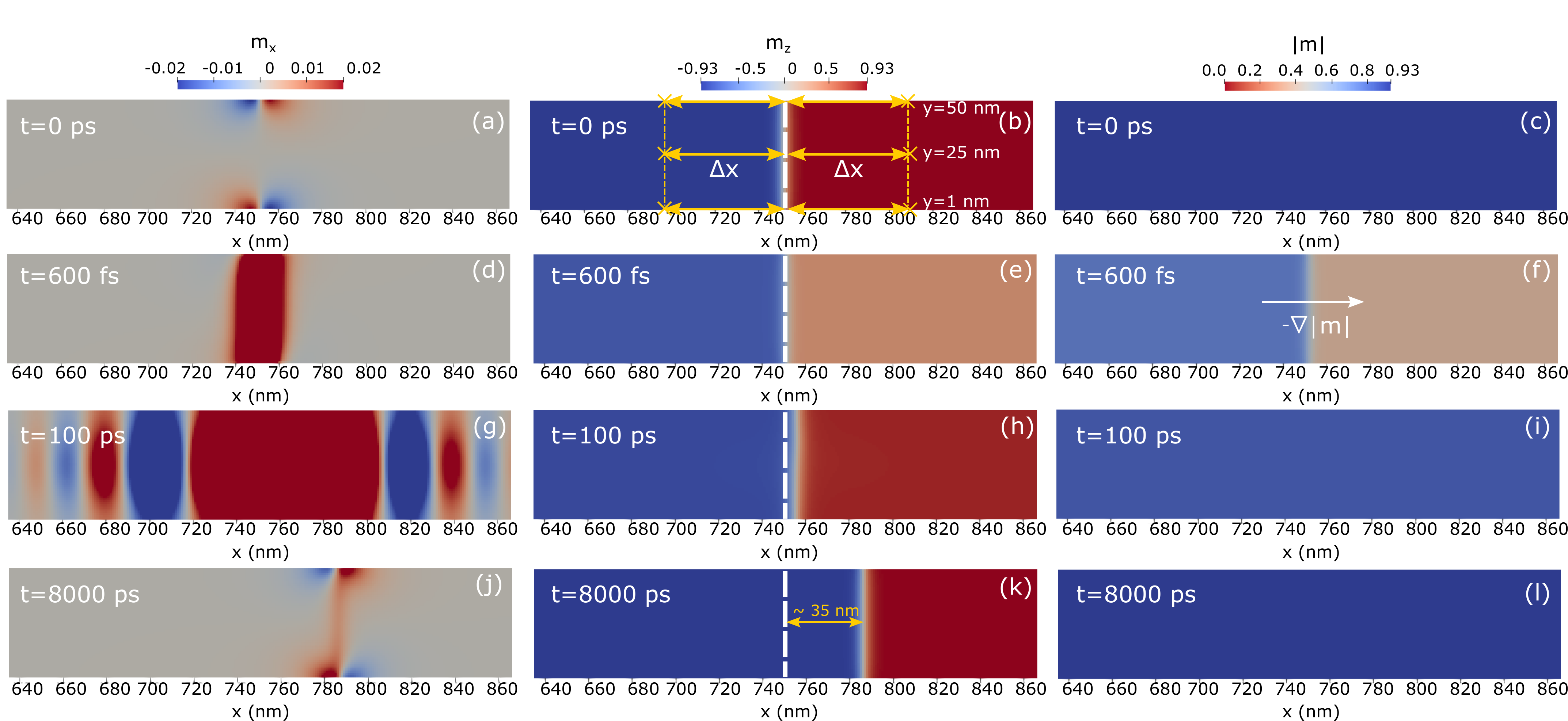}
    \caption{DW dynamics under the effect of a spatially uniform, ultra-fast laser pulse heating ($F=3~mJ/cm^2$, $t_p=200~fs$, $t_0=3t_p=600~fs$) and a transient magnetisation component of amplitude $\delta m_z^0=0.2$ induced along the $Oz$ direction - see equations \eqref{lpulse_power} and \eqref{dmz_stripe}. From left to right we display the $m_x$, $m_z$ and $|\bm{m}|$ magnetisation components along the thin film. For better visualisation we crop the system approximately between $x=640~nm$ and $x=860~nm$. The starting position of the wall is found around $x^0_{DW}=750~nm$, the middle region of the stripe. Subplots (a), (b) and (c) display the initial configuration of the sample at $t=0~ps$. Subplots (d), (e) and (f) showcase the magnetisation at the exact time when the laser power reaches a peak amplitude ($t= t_0=3t_p=600~fs$).
    In subplots (g), (h) and (i) we display the magnetic configuration at $t=100~ps$. The short-lived transient magnetisation induced on the $fs$ time-scale leads to the generation of SWs which propagate on the $ps-ns$ time-scale - see subplot (g). As seen in (i), the spin vectors recover their original magnetisation length and the $\nabla|\bm{m}|$ gradient  disappears. Finally, subplots (j), (k) and (l) showcase the state of the system at $8000~ps$, shortly before final equilibrium is reached. At this point, the DW has displaced approximately $35~nm$ away from its initial position as indicated in (k). The white dashed lines in (b), (e), (h) and (k) display the initial DW position.  In subplot (b) we indicate six different points (crossed marks) 
    at which the SW signal is analysed.
    }
    \label{fig:Figure7}
\end{figure*}

Moving away from the simple chain system discussed in the previous subsection, we further inquire the possibility for DW dynamics in a more complex situation by taking into account also the heating produced by the laser. We consider a
 stripe geometry, introducing also the magnetostatic contribution to the total effective field. 
Given the current perpendicular magnetic recording paradigm, an OOP anisotropy is preferred.  For the material parameters  presented in Table \ref{table:MatParams}, the competition between the OOP anisotropy and the magnetostatic interactions produces  a spin-reorientation transition below room temperature. 
To circumvent this, we augment the magnetocrystalline anisotropy of the system by increasing the $K(0)$ constant to $2.25~MJ/m^3$. 
\newline\indent
We consider a $1500~nm\times50~nm\times1~nm$ stripe system discretised in cubic cells of lateral size of $\Delta=1~nm$. While in the chain model the exchange length parameter $l_{ex}=\sqrt{\frac{2A(T)}{\mu_0M_s(T)^2}}$ was temperature independent due to the assumed MFA $A(T)\sim m_e^2$ scaling law, here we use  a more realistic scaling ($A(T)\sim m_e^{1.8}$) \cite{Moreno2016Temperature-dependentCo} which leads to an increasing $l_{ex}$ as a function of temperature starting from the $0~K$ value of $2.85~nm$, assuring ourselves the discretisation size $\Delta$ is properly chosen. 
At room temperature ($T=300~K$), the  DW configuration is of a Bloch type, as seen in Fig. \ref{fig:Figure7} at $t=0$. 
Starting from this initial configuration, we consider a spatially uniform ($\nabla T = 0$) laser-pulse heating of the thin film sample on the $fs$ time-scale. This rapid heating of the system is described by the two-temperature model (TTM) \cite{Kaganov1957RelaxationLattice,Lifshits1960OnRadiation,Anisimov1974ElectronPulses} which couples the electron and phonon baths via:
\begin{align}
C_e \frac{dT_e}{dt}&=-G_{ep}(T_e-T_{ph})+P(t)-C_e\frac{T_e-T_{room}}{\tau_{th}},\\
C_{ph} \frac{dT_{ph}}{dt}&=G_{ep}(T_e-T_{ph}),
\end{align}
where $C_e$ and $C_{ph}$ are the electron and phonon volumetric heat capacities and $G_{ep}$ is a measure of the coupling between the two baths. Working within the free electron approximation, the specific heat $C_e$ is taken as $C_e=\gamma_eT_e$, where $\gamma_e$ is a material dependent proportionality factor. Assuming $T_{ph}$ to be larger than the Debye temperature, $C_{ph}$ will be considered constant. A simple Newton cooling law is employed to model the heat transfer with the external medium ($T_{room}$) at a rate given by $\tau_{th}$. The time dependent laser pulse power is given by:
\begin{equation}
P(t)=\frac{A_{ab}F}{1.0645t_p2\sqrt{\ln{2}}\delta_{opt}}\exp\left[-\left(\frac{t-t_0}{t_p}\right)^2\right],
\label{lpulse_power}
\end{equation}
where $A_{ab}$ is an absorption coefficient, $F$ is the laser fluence at full width half maximum (FWHM), $\delta_{opt}$ is the optical penetration depth, $t_p$ is the pulse duration and $t_0$ denotes the moment of time when the laser pulse power reaches its peak amplitude. The numerical constants arise due to the FWHM relationship between the laser fluence $F$ and intensity $I$: $F=(\tau/2)\sqrt{\pi/\ln{2}}I\approx1.0645\tau I$, where $\tau=2t_p\sqrt{\ln{2}}$. Here, $t_p$ is set at $200~fs$. The TTM parameters we are employing are listed in Table \ref{table:2TM_parameters}. The $\gamma_e$, $C_{e}$ and $G_{ep}$ values correspond to a generic Co sample and have been extracted from \cite{Koopmans2009ExplainingDemagnetization,Chimata2012MicroscopicDynamics} while the $\tau_{th}$ parameter has been set to a standard value of $50~ps$. The absorption coefficient is set to a small value of $25\%$ similar to the work reported in \cite{John2017MagnetisationPulses} and $\delta_{opt}$ is set equal to the thin film thickness of $1~nm$.

\begin{table}[H]
\centering
\begin{ruledtabular}
\begin{tabular}{cc}
Parameter & Value\\
\midrule
$\gamma_e$ &$5.53\times10^3~Jm^{-3}K^{-2}$ \\
$C_{ph}$ & $2.07\times10^6~Jm^{-3}K^{-1}$ \\ 
$G_{ep}$ & $4.05\times10^{18}~Js^{-1}m^{-3}K^{-1}$\\
$\tau_{th}$ & $50~ps$\\
\end{tabular}
\end{ruledtabular}
\caption{TTM parameters for a generic Co sample as extracted from \cite{Koopmans2009ExplainingDemagnetization,Chimata2012MicroscopicDynamics}.}
\label{table:2TM_parameters}
\end{table}
\begin{figure}[!ht]
    \centering
    \includegraphics[width=1.0\linewidth]{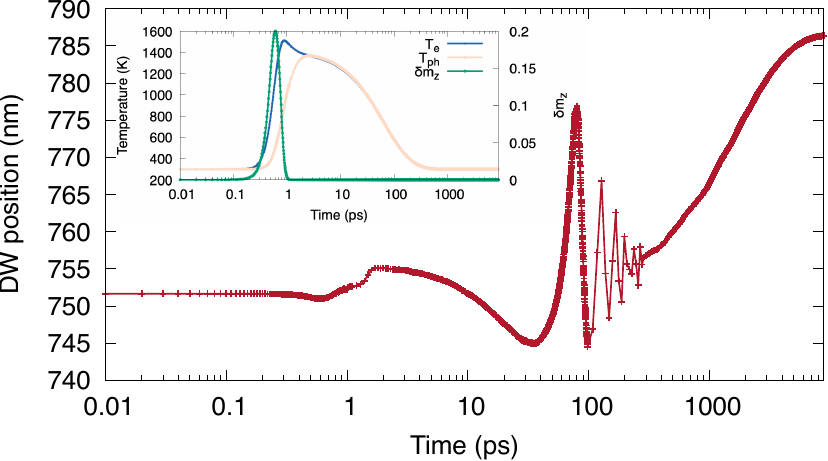}
    \caption{Center of mass DW dynamics extracted along the bottom-edge of the thin film system ($y=1~nm$) in Fig. \ref{fig:Figure7} under the effect of a spatially uniform, ultra-fast laser pulse heating ($F=3~mJ/cm^2$, $t_p=200~fs$, $t_0=3t_p=600~fs$) and a transient magnetisation component of amplitude $\delta m_z^0=0.2$ induced along the $Oz$ direction. The inset figure displays the electron and phonon temperature dynamics as well as the $\delta m_z$ variation in time. }
    \label{fig:figureX}
\end{figure}

In a similar fashion to John \textit{et al. }\cite{John2017MagnetisationPulses}, we further assume the laser-matter interaction gives rise to a transient magnetisation in the system along the $\bm{k}$ vector direction, parallel in this case to the EA ($Oz$) of the sample. We model the laser-induced magnetisation in the following manner:
\begin{equation}
m_z ^{i*} = m_z^i+ \delta m_z^0\exp{\left[-\left(\frac{t-t_0}{t_p}\right)^2\right]}= m_z^i+ \delta m_z(t)
\label{dmz_stripe}
\end{equation}
where $m_z^i$ is the equilibrium $z$ magnetisation component of each macro-spin in the system and $\delta m_z(t)$ is the time-dependent, laser-induced, non-thermal contribution of amplitude $\delta m_z^0$, assumed uniform across the sample. Unlike the chain model we discussed previously, we have dropped for simplicity the $|\cos\theta|$ angle dependence - see Eq. \eqref{IFE_chain} -  without changing the qualitative behaviour of the investigated DW dynamics. The temporal profile of the transient, laser induced magnetisation $\delta m_z(t)$ is assumed to precisely follow the time dependence of the laser pulse power given in Eq. \eqref{lpulse_power}. 
\begin{figure*}[t]
    \centering
    \includegraphics[width=1.0\linewidth]{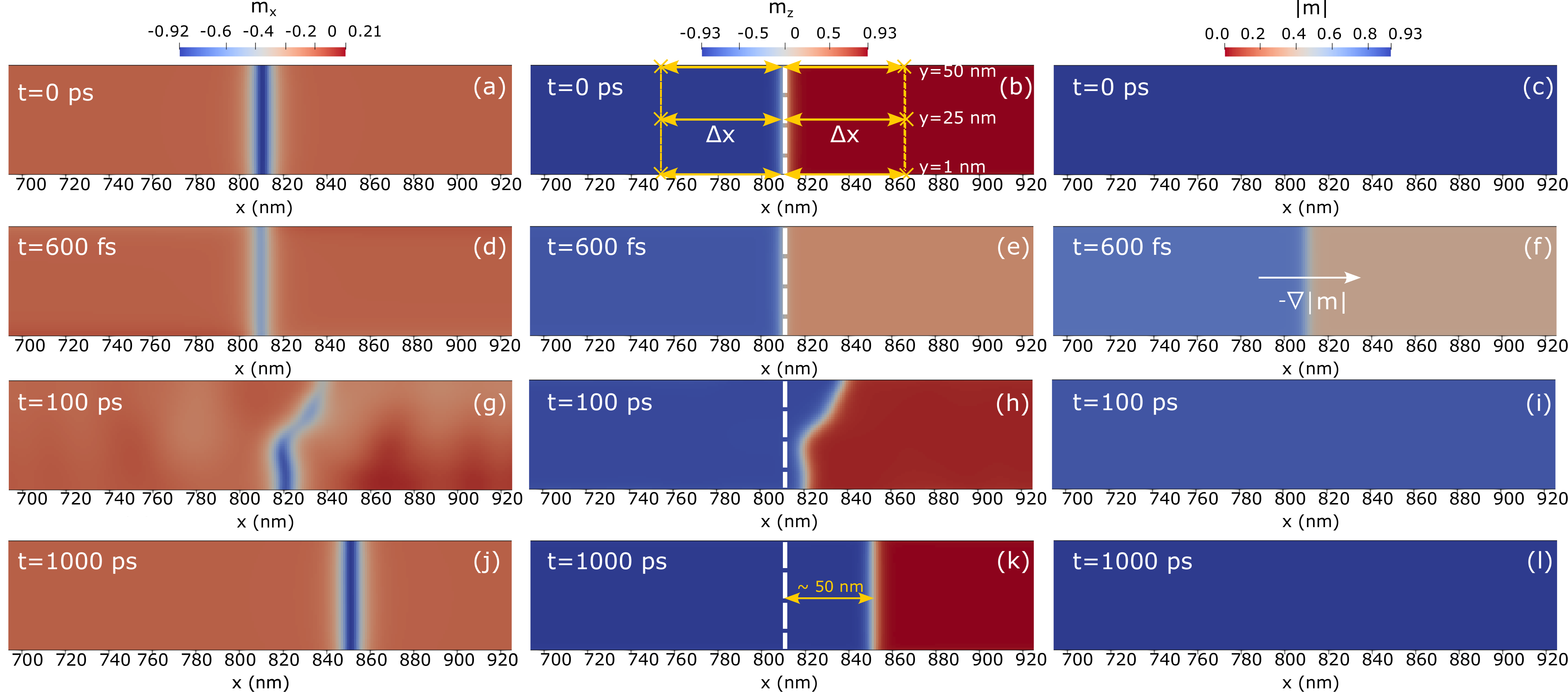}
    \caption{The numerical experiment of Fig. \ref{fig:Figure7} is repeated in the presence of an interfacial DMI. 
    From left to right we display the $m_x$, $m_z$ and $|\bm{m}|$ magnetisation components along the stripe where for better visualisation we have cropped the edges of the system. 
    Subfigures (d), (e) and (f) correspond to the moment of time when the laser pulse power reaches its peak amplitude ($t_0=t_p=600~fs$). 
    Figures (g), (h) and (i) display the magnetisation configuration at $t=100~ps$. 
    In figures (j), (k) and (l) we showcase the state of the system shortly before final equilibrium is reached. In the presence of the DMI, the DW dynamics are enhanced: in just $1000~ps$ the total displacement achieved is equal to approximately $50~nm$. The white dashed lines in (b), (e), (h) and (k) indicate the initial DW position. In subplot (b) we indicate six different points (crossed marks) at which the SW spectrum is analysed.
        }
    \label{fig:Figure8}
\end{figure*} 
\newline \indent
Fig. \ref{fig:Figure7} presents an example of our results considering for the purpose of visualisation a large transient magnetisation component of  $\delta m_z^0=0.2$. The optically induced dynamics are calculated
 using a pulse of fluence $F=3.0~mJ/cm^2$ and duration $t_p=200~fs$. The laser power is set to reach its peak value at $t_0=3t_p=600~fs$. 
 In subplots (d), (e) and (f) of Fig. \ref{fig:Figure7},
  we display the state of the system at $t=600~fs$, the exact moment of time when the laser pulse power reaches its peak amplitude. Due to the presence of the transient magnetisation $\delta m_z(t)$ induced on the time-scale of the ultra-fast heating, a $\nabla|\bm{m}|$ gradient is observed in subplot (f). As a consequence of the longitudinal dynamics, a positive $\delta m_z(t)$ contribution elongates the vector length of the down spins ("cold" region on the left side of the stripe) while it contracts the length of up spins ("hot" region on the right side of the stripe). 
\newline\indent 
Comparatively to the chain case,  both non-thermal and thermal longitudinal relaxation are now present.
Moreover, thermal longitudinal effects are acting on a larger timescale (up to  $ns$) due to the elevated electron temperatures compared to the non-thermal ones which are limited to a $fs$-$ps$ time-scale  as seen in the inset of Fig. \ref{fig:figureX}. 
Due to the dynamical behavior of the transient magnetisation which adapts to the intensity of the laser pulse on the timescale larger than the longitudinal relaxation time,
the DW displacement on the $fs$ timescale is almost null and can be attributed more to  a small spatial re-arrangement.   
A well visible DW motion can be observed within several $ps$ from the start of the pulse - see Fig. \ref{fig:figureX}.  The corresponding velocities are in the range of $1-10$ $km/s$, being large, but smaller than the ones observed in the chain case.
Interestingly, similar to the 1D case, after this  initial displacement, the DW is not moving around the $10~ps$ time stamp  since by then all non-thermal longitudinal effects have stopped and the transverse motion requires a larger time frame to develop.
\newline\indent
More surprisingly is that a much larger and oscillating DW displacement is visible at even larger timescale, see $t=100~ps$, when the spin-electron temperature and magnetisation magnitude have already reached the almost equilibrium value. Importantly,
 at the $t=100~ps$ time mark we observe the presence of SWs propagating along the thin film sample as seen in subplot (g) of Fig. \ref{fig:Figure7}. The emission of these SWs is due to the adjustment of the DW profile to the equilibrium value 
 due to the initial energy input introduced by the transient magnetisation contribution with the excess of the energy  emitted in form of SWs.
 \newline\indent
 Finally, we record the state of the system at $t=8~ns$ shortly before final equilibrirum is reached. The magnetisation profile slowly returns to its original shape as it can be seen comparing the first and last three sets of plots in Fig. \ref{fig:Figure7}. At this point, the DW has displaced approximately $35~nm$ away from its initial position as indicated in subfigure (k). Thus, in similar fashion to the chain model presented earlier, a transient magnetisation induced in the system followed by longitudinal relaxation of the macro-spin vectors on the $fs$ timescale leads to a DW displacement on the $ps-ns$ timescale towards the "hot" region of the sample - in this case denoting the area where the spin vectors have decreased in length keeping in mind that no temperature gradients arise in our simulation. The induced transient magnetisation gives rise to the emission and propagation of SWs (analogous to the oscillations discussed in Fig. \ref{fig:Figure2}) which drive the DW long after the $\nabla|\bm{m}|$ gradient has vanished.
\newline \indent
DW dynamics have been widely investigated both theoretically and experimentally also in the presence of the interfacial DMI \cite{Dzyaloshinsky1958AAntiferromagnetics,Moriya1960AnisotropicFerromagnetism}. While this antisymmetric exchange contribution has been shown detrimental to the field driven DW dynamics of in-plane magnetised ferromagnet/heavy-metal nano-wires \cite{Brandao2017UnderstandingNanowires}, in perpendicularly magnetised systems the DMI is seen capable of reducing the Walker breakdown field thus enhancing the DW velocities \cite{Thiaville2012DynamicsFilms,Yoshimura2015Soliton-likeDzyaloshinskiiMoriyainteraction}. On the other hand, a magnon-driven DW motion in conjunction with DMI induced linear angular momentum transfer and in the presence of an easy-plane anisotropy is shown to be more efficient in driving DWs compared to the angular momentum transfer mechanism \cite{Wang2015Magnon-drivenInteraction}. Thus, depending on the geometry of the system and the driving force, an interfacial DMI can have different effects on the dynamics of DWs. In what follows, we consider the very same experiment graphically described in Fig. \ref{fig:Figure7}, only now we include the presence of an interfacial DMI contribution which can arise for example due to the coupling to a heavy-metal layer such as Pt. Thus, we add an additional term in the total effective field acting on each macro-spin in the following manner \cite{Cortes-Ortuno2018ProposalInteraction,Lepadatu2020EmergenceDemagnetization}:
\begin{align}
\bm{H}_{DMI}=&-\frac{2D(T)}{M_s(0)m_e}\left[\nabla m_z - (\nabla\cdot\bm{m})\hat{\bm{z}}\right]
\label{DMI_field}
\end{align}
The temperature scaling of the DMI constant is set to: $D(T)=D(0)m_e^2$ (MFA), where $D(0)=2~mJ/m^2$ is taken at $0~K$. In Fig. \ref{fig:Figure8}, we display the time-evolution of our system under the action of the same laser-pulse excitation and transient magnetisation $\delta m_z(t)$ as in the previous case. Due to the presence of the DMI interaction, the $300~K$ equilibrium DW profile used as initial state at $t=0~ps$ has now transitioned to a N\'eel configuration as seen in subfigures (a), (b), and (c). 
In (d), (e) and (f) we once more display the state of the system exactly at the moment of time when the laser power reaches a peak amplitude ($t_0=3t_p=600~fs$). As seen in (f),  as a consequence of the induced $\delta m_z(t)$ contribution, a magnetisation gradient is formed across the wall. In comparison to the no-DMI case, the subsequent dynamics on the $ps-ns$ timescale result in an asymmetric DW profile as displayed in subplots (g), (h) and (i) at $t=100~ps$. The DW dynamics are accompanied also in this case by SW emission and propagation along the thin film sample as it can be seen for example in the $m_x$ in (g). 
Finally, in (j), (k) and (l) we display the state of the system just before final equilibrium is reached and the full recovery of the initial DW magnetisation profile is obtained. Interestingly, the observed dynamics are enhanced in the presence of the interfacial DMI: equilibrium is reached around the $1000~ps$ threshold while the final displacement we obtain increases to approximately $50~nm$. 
\begin{figure}[t]
    \centering
    \includegraphics[width=0.7\linewidth]{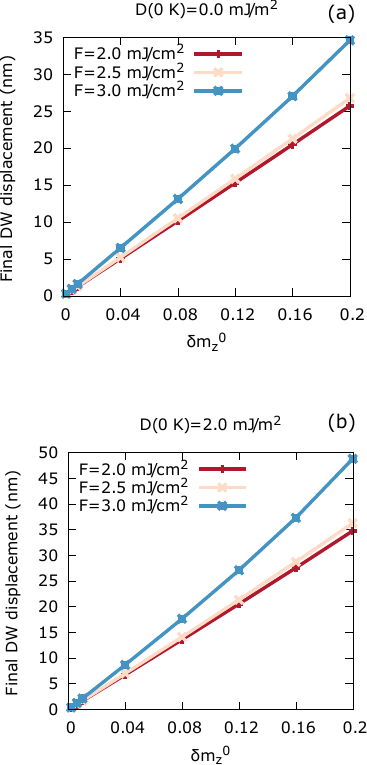}
    \caption{Final DW displacement achieved under the action of an ultra-fast, uniform laser pulse heating and a transient magnetisation induced along $Oz$. The displacement is graphically represented against the laser pulse fluence $F$ and the amplitude $\delta m_z^0$ of the induced magnetisation. The parameter sweep is considered both in the absence (a) as well as in the presence (b) of an interfacial DMI contribution - see Eq.
    \eqref{DMI_field}. In both situations, the microscoping damping parameter is set to $\lambda=0.1$.} 
    \label{fig:Figure9}
\end{figure}
\newline\indent
Fig. \ref{fig:Figure9} presents the final displacement as a function of the maximum induced magnetisation for various laser pulse fluences. 
Irrespective of the presence/absence of DMI, the total distance covered by the DW increases linearly as a function of $\delta m_z^0$, a similar behaviour to the chain model results obtained at temperatures $T=0.84~T_c$ and $T=0.91~T_c$ as seen in Fig. \ref{fig:Figure3}. However, in our stripe simulations we do not obtain a parabolic-like variation of the displacement with respect the $\delta m_z^0$ variable as seen in the chain case at $T=0.98~T_c$.
More interestingly, in the presence of the interfacial DMI contribution, the total distance  is larger compared to the no-DMI situation for any ($F$, $\delta m_z^0$) pair of parameters, an effect which graphically is more visible beyond the $\delta m_z^0=0.04$ limit. 

\begin{figure*}[t]
    \centering
    \includegraphics[width=1.0\linewidth]{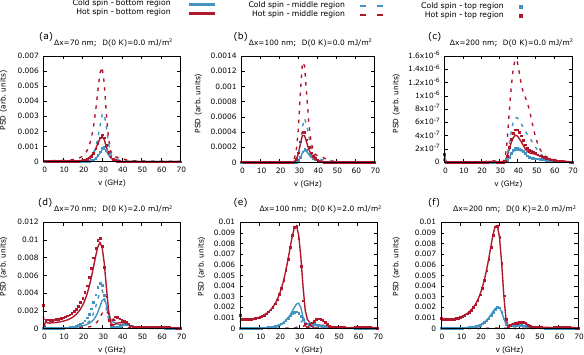}
    \caption{
    Spectral analysis of the FFT power extracted from the time-dependent $m_x(t)$ signal at six independent macro-cell sites at the bottom edge ($y=1~nm$), middle-region ($y=25~nm$) and top edge ($y=50~nm$) of the stripe system situated at a variable distance $\Delta x$ away from the initial DW position $x^0_{DW}$ as graphically represented in subplot (b) of both Fig.
    \ref{fig:Figure7} and \ref{fig:Figure8}. The analysis is employed both in the absence - top row - as well as presence - bottom row - of the interfacial DMI exchange contribution ($D(0~K)=2.0~  mJ/cm^2$). We consider three $\Delta x$ values: $\pm 70~nm$, $\pm 100~nm$ and $\pm 200~nm$  (subplots seen from left to right) such that the DW is always confined between the ($x^0_{DW},x^0_{DW}+\Delta x$) boundaries during its displacement. In each of the subplots, the data graphically represented in red (blue) corresponds to the hot (cold) region. We remind the reader the "hot/cold" terminology refers to regions of small/large magnetisation, in our work being generated in the absence of any temperature gradients. The bottom edge signal is represented by solid lines, the middle region signal by dashed lines and finally the top edge corresponding data by squared points. 
    }
    \label{fig:Figure10}
\end{figure*}

Unlike the chain model discussed earlier where an instantaneous magnetisation gradient leads to a rapid DW displacement under dominant longitudinal relaxation processes on the $fs-ps$ time-scale, in the case of a more realistic stripe modelling, we observed  a significant SW emission. To expand our understanding of the observed DW dynamics and to understand the role of the interfacial DMI contribution, we analysed the SW spectrum at pair points in the "cold" and "hot" regions taken at an equal distance with respect to the initial DW position along  the bottom edge of the stripe ($y=1~nm$), in the  middle region ($y=25~nm$) as well as the top edge ($y=50~nm$)  - see subplot (b) in both Fig. \ref{fig:Figure7} and \ref{fig:Figure8}. At each of these points, we extract the $m_x(t)$ time-dependent magnetisation component along $Ox$ and perform the FFT analyzing  the Power Spectral Density (PSD). 
In Fig. \ref{fig:Figure10}, we present the extracted PSD both in the absence and presence of the interfacial DMI contribution.
\newline \indent
As it is expected, PSD is different at the nanostripe edges and in the middle, due to the influence of magnetostatic interactions. Considering first the no-DMI case as seen in subplots (a), (b) and (c), we observe a dominant PSD peak, extracted from the $m_x(t)$ signal corresponding to the "hot" macro-spins situated along the middle region of the stripe in comparison to their "cold" counter-parts as well as any signal analysed at the edges of the thin film. If we correlate the PSD to the energy carried by left $\bm{j}_m^l$ and right $\bm{j}_m^r$ propagating magnon-spin currents, the motion of the wall on the $ps-ns$ time scale where any $\nabla|\bm{m}|$ gradient has already vanished may be explained by a net magnon current $\bm{j}_m=\bm{j}_m^r+\bm{j}_m^l$ which propagates from the "hot" to the "cold" area along a dominant channel corresponding to the middle-region of the stripe. For the conservation of the net angular momentum, the DW needs to displace opposite to the direction of $\bm{j}_m$ \cite{Yan2011All-magnonicPropagation,Hinzke2011DomainEffectb,Kovalev2012ThermomagnonicIOPscience,Yan2013AngularFerromagnets,Kim2015Landau-LifshitzTorque}, that is from the "cold" to the "hot" region. We further observe the amplitude of the PSD signal decreasing as a function of $\Delta x$, a signature of the SW damping as the propagation approaches the edges of the system. Interestingly, the frequency distribution of the generated magnons increases in width and shifts to larger $\nu$ values as we probe the SW signal further away from the initial DW position $x^0_{DW}$, an effect whose origins remain to be investigated. 
Lastly, the SWs analysed at the bottom and top edges of the thin film sample display a  closely matching power spectrum, evidenced for all considered $\Delta x$ values. 
\newline\indent
With the introduction of the interfacial DMI, the DW dynamics are enhanced as seen both by the time-dependent analysis graphically employed in Fig. \ref{fig:Figure7} and \ref{fig:Figure8} as well as the final displacement investigation of Fig. \ref{fig:Figure9}. As evidenced by our spectral analysis of the SW signal, the addition of the antisymmetric exchange leads to several effects: firstly, while the net magnonic spin current preserves its direction of motion from the "hot" to the "cold" area, the dominant propagation channel can no longer be attributed to the middle-region of the stripe but to the top and bottom edges as evidenced in subplots (d), (e) and (f) of Fig. \ref{fig:Figure10}. This edge channeling effect in in the presence of the DMI has been previously reported in \cite{Garcia-Sanchez2014NonreciprocalInteractionb}. Secondly, the magnitude of the PSD is much larger than for the non-DMI case suggesting augmented magnon currents which could account for the enhanced DW dynamics we observe. Thirdly.
the frequency distribution of the generated magnon population increases in the presence of the DMI; unlike the previous case, we no longer observe an evident frequency shift and widening of the calculated PSD as we modify the $\Delta x$ variable. Finally, the analysed SWs display a less pronounced damping  in space as evidenced by the small change in the peak amplitude of the PSD as we move away from the initial $x^0_{DW}$ DW position. 
\newline\indent
To avoid repetition, we will reserve the conclusions dedicated to this subsection for the next and final part of the article.

\section{Conclusions}
The main goal of our work has been the demonstration of a new DW driving mechanism  and not an exhaustive analysis of its efficiency or direct comparison to other displacement methods. A fully quantitative analysis has been avoided also in light of the generality of the effect as not one, but multiple magneto-optical phenomena have been shown capable of inducing a transient $\delta \bm{m}(t)$ magnetisation contribution as discussed in the introductory part \cite{John2017MagnetisationPulses,Scheid2019AbFerromagnets,Disa2020PolarizingField,Afanasiev2021UltrafastPhonons,Scheid2021AbRegime,Formisano2022Laser-inducedCoF2}.  The rigourous investigation of acquired DW velocities as a function of parameters such as the laser pulse width or fluence can be investigated in a future work with exact reference to a given magneto-optical phenomenon at origin. Nonetheless, making use of the inherent longitudinal dynamics described by the LLB equation, we demonstrate the possibility to convert a transient, non-thermal magnetisation contribution followed by a subsequent longitudinal relaxation of the macro-spins into a transverse DW motion. 
\newline\indent
First of all, we approached the problem in a simpler, chain macro-spin model where no heating effects have been considered nor the presence of the magnetostatic interaction has been taken into account. Nonetheless, we showed that by virtue of a longitudinal deformation of the magnetic texture induced by a transient, magnetisation contribution it is possible to displace a ferromagnetic DW. The mechanism is based on the appearance of a non-thermal magnetisation gradient $\nabla|\bm{m}|$ which enables the motion of the DW towards the "hot" region corresponding to an area of small magnetisation. It has been shown the displacement primarily takes places on a very fast $fs$ timescale where longitudinal relaxation processes are dominant. In our analysis, we have seen the distance covered by the DW is proportional to the life-time of the induced gradient as well as the amplitude of the transient magnetisation. The role of the precession and transverse relaxation mechanisms in the acquired displacement is reduced to some DW oscillations on the $ns$ timescale in this chain model as the DW loses its main drive on the longer $ps$-$ns$ timescale where the $\nabla|\bm{m}|$ has already vanished.
\newline\indent
Secondly, we verified the viability of the suggested driving mechanism in a more realistic micromagnetic study in which we nucleate an OOP DW configuration in a stripe geometry. In this case, we expand the complexity of the model by taking into account the presence of the magnetostatic interactions as well as consider the system which no longer finds itself in thermal equilibrium but it is uniformly heated across its area ($\nabla T=0$) using an ultra-fast laser pulse excitation described by the TTM. Furthermore, we considered the induced transient magnetisation contribution that follows the temporal profile of the pump pulse moving away from the instantaneous-like behaviour assumed in the chain model. While we once again identified the appearance of an optically induced magnetisation gradient which favours the DW displacement towards the "hot" region, the overall driving mechanism displaces several different features. The presence of the transient magnetisation contribution in conjunction with the uniform laser heating effect give rise to emission and propagation of SWs which drive the DW long after the $\nabla|\bm{m}|$ gradient has vanished. Interestingly, the observed motion can be enhanced in the presence of an interfacial DMI contribution irrespective of the laser pulse fluence or strength of the induced magnetisation contribution. A PSD analysis revealed the existence of a net magnon spin-current, propagating from the "hot" to the "cold" region, opposite to the direction of the wall displacement. The addition of the antisymmetric exchange interaction gives rise to a SW edge channeling effect \cite{Garcia-Sanchez2014NonreciprocalInteractionb} while it overall increases the density of the net magnon spin current, thus accounting for the enhanced DW dynamics.
\newline\indent
A laser-induced transient magnetisation equivalent to our $\delta m_z^0$ parameter was calculated for example in the case of L1$_0$ FePt in the work of John \textit{et al.} \cite{John2017MagnetisationPulses} on account of the previously established quantum theory of the IFE \cite{Battiato2014QuantumEffect}. The $\delta m_z^0$ contribution at a photon energy of $1.55~eV$  and laser intensity of $68~GW/cm^2$ was found to be $-7.1\%M_s(0~K)$ or $-3.45\%M_s(0~K)$ for a left ($\sigma-$) and right ($\sigma+$) circularly polarised laser pulse. These values, in light of our simulations, would produce a visible DW displacement even for one laser pulse. In a Co sample, Berritta \textit{et al.} \cite{Berritta2016AbMetals} calculated in the same approach the IFE induced magnetisation for an identical photon energy but smaller laser intensity of $10~GW/cm^2$, obtaining contributions of $-4.8\times10^{-3}~\mu_B/$at. vol. and $-13\times10^{-3}~\mu_B/$at. vol. for a $\sigma+$ and $\sigma-$ polarisation respectively. Assuming a cubic atomic volume of lateral size $a=0.25~nm$  \cite{Evans2014AtomisticNanomaterials} and considering the saturation magnetisation of our Co sample $M_s(0~K)=1400~kA/m$, the latter \textit{ab initio} result evaluates to approximately $-0.2\%M_s(0~K)$ and $-0.55\%M_s(0~K)$. In our model this small transient magnetisation leads to insignificant displacements. Since the IFE scales linearly with the laser pulse intensity $I$, it is expected a larger intensity of $68~GW/cm^2$ as used in the aforementioned FePt work would enhance the effect in a Co sample. While we increase the amplitude of the induced transient magnetisation up to $20\%$ of the zero-Kelvin $M_s$, our work suggests that a contribution in the range of $4\%-8\%$ in the presence of the interfacial DMI is enough to obtain a DW displacement between $10~nm - 18~nm$. To further increase the strength of the effect, one could envision the use of multiple laser pulse excitations or the investigation of large Verdet constants materials which display a better suitability for the magneto-optical coupling based on the IFE \cite{Pershan1966TheoreticalPhenomena,Belotelov2010InverseHeterostructures} reminding the reader once more that a transient magnetisation can be attributed to a broad range of light-induced phenomena and not solely the IFE.
Additionally, it remains to be seen how taking into account an absorption based contribution to the induced magnetisation as discussed by Scheid \textit{et al.} \cite{Scheid2019AbFerromagnets,Scheid2021AbRegime} can further enhance the dynamics. 
Finally, noncollinear spiral systems have been reported 
to contribute significantly to the magnitude of optically induced effects \cite{Freimuth2021Laser-inducedSpirals}.

\begin{acknowledgments}
This project has received funding from the European Union’s Horizon 2020 research and innovation program under the Marie Skłodowska-Curie ITN COMRAD (grant agreement No 861300).
 The authors acknowledge financial support by the grant PID2019-108075RB-C31  funded by Ministry of Science and Innovation of Spain MCIN/AEI/ 10.13039/501100011033 and  the grant 023AEP034 funded by Spanish National Research Council (CSIC).
\end{acknowledgments}




\cleardoublepage
\phantomsection
\bibliography{IFE.bib}

\begin{thebibliography}{90}%
\makeatletter
\providecommand \@ifxundefined [1]{%
 \@ifx{#1\undefined}
}%
\providecommand \@ifnum [1]{%
 \ifnum #1\expandafter \@firstoftwo
 \else \expandafter \@secondoftwo
 \fi
}%
\providecommand \@ifx [1]{%
 \ifx #1\expandafter \@firstoftwo
 \else \expandafter \@secondoftwo
 \fi
}%
\providecommand \natexlab [1]{#1}%
\providecommand \enquote  [1]{``#1''}%
\providecommand \bibnamefont  [1]{#1}%
\providecommand \bibfnamefont [1]{#1}%
\providecommand \citenamefont [1]{#1}%
\providecommand \href@noop [0]{\@secondoftwo}%
\providecommand \href [0]{\begingroup \@sanitize@url \@href}%
\providecommand \@href[1]{\@@startlink{#1}\@@href}%
\providecommand \@@href[1]{\endgroup#1\@@endlink}%
\providecommand \@sanitize@url [0]{\catcode `\\12\catcode `\$12\catcode `\&12\catcode `\#12\catcode `\^12\catcode `\_12\catcode `\%12\relax}%
\providecommand \@@startlink[1]{}%
\providecommand \@@endlink[0]{}%
\providecommand \url  [0]{\begingroup\@sanitize@url \@url }%
\providecommand \@url [1]{\endgroup\@href {#1}{\urlprefix }}%
\providecommand \urlprefix  [0]{URL }%
\providecommand \Eprint [0]{\href }%
\providecommand \doibase [0]{https://doi.org/}%
\providecommand \selectlanguage [0]{\@gobble}%
\providecommand \bibinfo  [0]{\@secondoftwo}%
\providecommand \bibfield  [0]{\@secondoftwo}%
\providecommand \translation [1]{[#1]}%
\providecommand \BibitemOpen [0]{}%
\providecommand \bibitemStop [0]{}%
\providecommand \bibitemNoStop [0]{.\EOS\space}%
\providecommand \EOS [0]{\spacefactor3000\relax}%
\providecommand \BibitemShut  [1]{\csname bibitem#1\endcsname}%
\let\auto@bib@innerbib\@empty
\bibitem [{\citenamefont {Beaurepaire}\ \emph {et~al.}(1996)\citenamefont {Beaurepaire}, \citenamefont {Merle}, \citenamefont {Daunois},\ and\ \citenamefont {Bigot}}]{Beaurepaire1996UltrafastNickel}%
  \BibitemOpen
  \bibfield  {author} {\bibinfo {author} {\bibfnamefont {E.}~\bibnamefont {Beaurepaire}}, \bibinfo {author} {\bibfnamefont {J.~C.}\ \bibnamefont {Merle}}, \bibinfo {author} {\bibfnamefont {A.}~\bibnamefont {Daunois}},\ and\ \bibinfo {author} {\bibfnamefont {J.~Y.}\ \bibnamefont {Bigot}},\ }\bibfield  {title} {\bibinfo {title} {{Ultrafast Spin Dynamics in Ferromagnetic Nickel}},\ }\href {https://doi.org/10.1103/PhysRevLett.76.4250} {\bibfield  {journal} {\bibinfo  {journal} {Physical Review Letters}\ }\textbf {\bibinfo {volume} {76}},\ \bibinfo {pages} {4250} (\bibinfo {year} {1996})}\BibitemShut {NoStop}%
\bibitem [{\citenamefont {Koopmans}\ \emph {et~al.}(2000)\citenamefont {Koopmans}, \citenamefont {Van~Kampen}, \citenamefont {Kohlhepp},\ and\ \citenamefont {De~Jonge}}]{Koopmans2000UltrafastOptics}%
  \BibitemOpen
  \bibfield  {author} {\bibinfo {author} {\bibfnamefont {B.}~\bibnamefont {Koopmans}}, \bibinfo {author} {\bibfnamefont {M.}~\bibnamefont {Van~Kampen}}, \bibinfo {author} {\bibfnamefont {J.~T.}\ \bibnamefont {Kohlhepp}},\ and\ \bibinfo {author} {\bibfnamefont {W.~J.}\ \bibnamefont {De~Jonge}},\ }\bibfield  {title} {\bibinfo {title} {{Ultrafast Magneto-Optics in Nickel: Magnetism or Optics?}},\ }\href {https://doi.org/10.1103/PhysRevLett.85.844} {\bibfield  {journal} {\bibinfo  {journal} {Physical Review Letters}\ }\textbf {\bibinfo {volume} {85}},\ \bibinfo {pages} {844} (\bibinfo {year} {2000})}\BibitemShut {NoStop}%
\bibitem [{\citenamefont {Hohlfeld}\ \emph {et~al.}(2001)\citenamefont {Hohlfeld}, \citenamefont {Gerrits}, \citenamefont {Bilderbeek}, \citenamefont {Rasing}, \citenamefont {Awano},\ and\ \citenamefont {Ohta}}]{Hohlfeld2001FastPulses}%
  \BibitemOpen
  \bibfield  {author} {\bibinfo {author} {\bibfnamefont {J.}~\bibnamefont {Hohlfeld}}, \bibinfo {author} {\bibfnamefont {T.}~\bibnamefont {Gerrits}}, \bibinfo {author} {\bibfnamefont {M.}~\bibnamefont {Bilderbeek}}, \bibinfo {author} {\bibfnamefont {T.}~\bibnamefont {Rasing}}, \bibinfo {author} {\bibfnamefont {H.}~\bibnamefont {Awano}},\ and\ \bibinfo {author} {\bibfnamefont {N.}~\bibnamefont {Ohta}},\ }\bibfield  {title} {\bibinfo {title} {{Fast magnetization reversal of GdFeCo induced by femtosecond laser pulses}},\ }\href {https://doi.org/10.1103/PhysRevB.65.012413} {\bibfield  {journal} {\bibinfo  {journal} {Physical Review B}\ }\textbf {\bibinfo {volume} {65}},\ \bibinfo {pages} {012413} (\bibinfo {year} {2001})}\BibitemShut {NoStop}%
\bibitem [{\citenamefont {Stanciu}\ \emph {et~al.}(2006)\citenamefont {Stanciu}, \citenamefont {Kimel}, \citenamefont {Hansteen}, \citenamefont {Tsukamoto}, \citenamefont {Itoh}, \citenamefont {Kirilyuk},\ and\ \citenamefont {Rasing}}]{Stanciu2006UltrafastCompensation}%
  \BibitemOpen
  \bibfield  {author} {\bibinfo {author} {\bibfnamefont {C.~D.}\ \bibnamefont {Stanciu}}, \bibinfo {author} {\bibfnamefont {A.~V.}\ \bibnamefont {Kimel}}, \bibinfo {author} {\bibfnamefont {F.}~\bibnamefont {Hansteen}}, \bibinfo {author} {\bibfnamefont {A.}~\bibnamefont {Tsukamoto}}, \bibinfo {author} {\bibfnamefont {A.}~\bibnamefont {Itoh}}, \bibinfo {author} {\bibfnamefont {A.}~\bibnamefont {Kirilyuk}},\ and\ \bibinfo {author} {\bibfnamefont {T.}~\bibnamefont {Rasing}},\ }\bibfield  {title} {\bibinfo {title} {{Ultrafast spin dynamics across compensation points in ferrimagnetic GdFeCo: The role of angular momentum compensation}},\ }\href {https://doi.org/10.1103/PHYSREVB.73.220402/FIGURES/3/MEDIUM} {\bibfield  {journal} {\bibinfo  {journal} {Physical Review B - Condensed Matter and Materials Physics}\ }\textbf {\bibinfo {volume} {73}},\ \bibinfo {pages} {220402} (\bibinfo {year} {2006})}\BibitemShut {NoStop}%
\bibitem [{\citenamefont {Stanciu}\ \emph {et~al.}(2007)\citenamefont {Stanciu}, \citenamefont {Hansteen}, \citenamefont {Kimel}, \citenamefont {Kirilyuk}, \citenamefont {Tsukamoto}, \citenamefont {Itoh},\ and\ \citenamefont {Rasing}}]{Stanciu2007All-opticalLight}%
  \BibitemOpen
  \bibfield  {author} {\bibinfo {author} {\bibfnamefont {C.~D.}\ \bibnamefont {Stanciu}}, \bibinfo {author} {\bibfnamefont {F.}~\bibnamefont {Hansteen}}, \bibinfo {author} {\bibfnamefont {A.~V.}\ \bibnamefont {Kimel}}, \bibinfo {author} {\bibfnamefont {A.}~\bibnamefont {Kirilyuk}}, \bibinfo {author} {\bibfnamefont {A.}~\bibnamefont {Tsukamoto}}, \bibinfo {author} {\bibfnamefont {A.}~\bibnamefont {Itoh}},\ and\ \bibinfo {author} {\bibfnamefont {T.}~\bibnamefont {Rasing}},\ }\bibfield  {title} {\bibinfo {title} {{All-optical magnetic recording with circularly polarized light}},\ }\href {https://doi.org/10.1103/PHYSREVLETT.99.047601/FIGURES/4/MEDIUM} {\bibfield  {journal} {\bibinfo  {journal} {Physical Review Letters}\ }\textbf {\bibinfo {volume} {99}},\ \bibinfo {pages} {047601} (\bibinfo {year} {2007})}\BibitemShut {NoStop}%
\bibitem [{\citenamefont {Radu}\ \emph {et~al.}(2011)\citenamefont {Radu}, \citenamefont {Vahaplar}, \citenamefont {Stamm}, \citenamefont {Kachel}, \citenamefont {Pontius}, \citenamefont {D{\"{u}}rr}, \citenamefont {Ostler}, \citenamefont {Barker}, \citenamefont {Evans}, \citenamefont {Chantrell}, \citenamefont {Tsukamoto}, \citenamefont {Itoh}, \citenamefont {Kirilyuk}, \citenamefont {Rasing},\ and\ \citenamefont {Kimel}}]{Radu2011TransientSpins}%
  \BibitemOpen
  \bibfield  {author} {\bibinfo {author} {\bibfnamefont {I.}~\bibnamefont {Radu}}, \bibinfo {author} {\bibfnamefont {K.}~\bibnamefont {Vahaplar}}, \bibinfo {author} {\bibfnamefont {C.}~\bibnamefont {Stamm}}, \bibinfo {author} {\bibfnamefont {T.}~\bibnamefont {Kachel}}, \bibinfo {author} {\bibfnamefont {N.}~\bibnamefont {Pontius}}, \bibinfo {author} {\bibfnamefont {H.~A.}\ \bibnamefont {D{\"{u}}rr}}, \bibinfo {author} {\bibfnamefont {T.~A.}\ \bibnamefont {Ostler}}, \bibinfo {author} {\bibfnamefont {J.}~\bibnamefont {Barker}}, \bibinfo {author} {\bibfnamefont {R.~F.}\ \bibnamefont {Evans}}, \bibinfo {author} {\bibfnamefont {R.~W.}\ \bibnamefont {Chantrell}}, \bibinfo {author} {\bibfnamefont {A.}~\bibnamefont {Tsukamoto}}, \bibinfo {author} {\bibfnamefont {A.}~\bibnamefont {Itoh}}, \bibinfo {author} {\bibfnamefont {A.}~\bibnamefont {Kirilyuk}}, \bibinfo {author} {\bibfnamefont {T.}~\bibnamefont {Rasing}},\ and\ \bibinfo {author} {\bibfnamefont {A.~V.}\ \bibnamefont {Kimel}},\ }\bibfield  {title} {\bibinfo
  {title} {{Transient ferromagnetic-like state mediating ultrafast reversal of antiferromagnetically coupled spins}},\ }\href {https://doi.org/10.1038/nature09901} {\bibfield  {journal} {\bibinfo  {journal} {Nature 2011 472:7342}\ }\textbf {\bibinfo {volume} {472}},\ \bibinfo {pages} {205} (\bibinfo {year} {2011})}\BibitemShut {NoStop}%
\bibitem [{\citenamefont {Eschenlohr}\ \emph {et~al.}(2013)\citenamefont {Eschenlohr}, \citenamefont {Battiato}, \citenamefont {Maldonado}, \citenamefont {Pontius}, \citenamefont {Kachel}, \citenamefont {Holldack}, \citenamefont {Mitzner}, \citenamefont {F{\"{o}}hlisch}, \citenamefont {Oppeneer},\ and\ \citenamefont {Stamm}}]{Eschenlohr2013UltrafastDemagnetization}%
  \BibitemOpen
  \bibfield  {author} {\bibinfo {author} {\bibfnamefont {A.}~\bibnamefont {Eschenlohr}}, \bibinfo {author} {\bibfnamefont {M.}~\bibnamefont {Battiato}}, \bibinfo {author} {\bibfnamefont {P.}~\bibnamefont {Maldonado}}, \bibinfo {author} {\bibfnamefont {N.}~\bibnamefont {Pontius}}, \bibinfo {author} {\bibfnamefont {T.}~\bibnamefont {Kachel}}, \bibinfo {author} {\bibfnamefont {K.}~\bibnamefont {Holldack}}, \bibinfo {author} {\bibfnamefont {R.}~\bibnamefont {Mitzner}}, \bibinfo {author} {\bibfnamefont {A.}~\bibnamefont {F{\"{o}}hlisch}}, \bibinfo {author} {\bibfnamefont {P.~M.}\ \bibnamefont {Oppeneer}},\ and\ \bibinfo {author} {\bibfnamefont {C.}~\bibnamefont {Stamm}},\ }\bibfield  {title} {\bibinfo {title} {{Ultrafast spin transport as key to femtosecond demagnetization}},\ }\href {https://doi.org/10.1038/nmat3546} {\bibfield  {journal} {\bibinfo  {journal} {Nature Materials 2013 12:4}\ }\textbf {\bibinfo {volume} {12}},\ \bibinfo {pages} {332} (\bibinfo {year} {2013})}\BibitemShut {NoStop}%
\bibitem [{\citenamefont {John}\ \emph {et~al.}(2017)\citenamefont {John}, \citenamefont {Berritta}, \citenamefont {Hinzke}, \citenamefont {M{\"{u}}ller}, \citenamefont {Santos}, \citenamefont {Ulrichs}, \citenamefont {Nieves}, \citenamefont {Walowski}, \citenamefont {Mondal}, \citenamefont {Chubykalo-Fesenko}, \citenamefont {McCord}, \citenamefont {Oppeneer}, \citenamefont {Nowak},\ and\ \citenamefont {M{\"{u}}nzenberg}}]{John2017MagnetisationPulses}%
  \BibitemOpen
  \bibfield  {author} {\bibinfo {author} {\bibfnamefont {R.}~\bibnamefont {John}}, \bibinfo {author} {\bibfnamefont {M.}~\bibnamefont {Berritta}}, \bibinfo {author} {\bibfnamefont {D.}~\bibnamefont {Hinzke}}, \bibinfo {author} {\bibfnamefont {C.}~\bibnamefont {M{\"{u}}ller}}, \bibinfo {author} {\bibfnamefont {T.}~\bibnamefont {Santos}}, \bibinfo {author} {\bibfnamefont {H.}~\bibnamefont {Ulrichs}}, \bibinfo {author} {\bibfnamefont {P.}~\bibnamefont {Nieves}}, \bibinfo {author} {\bibfnamefont {J.}~\bibnamefont {Walowski}}, \bibinfo {author} {\bibfnamefont {R.}~\bibnamefont {Mondal}}, \bibinfo {author} {\bibfnamefont {O.}~\bibnamefont {Chubykalo-Fesenko}}, \bibinfo {author} {\bibfnamefont {J.}~\bibnamefont {McCord}}, \bibinfo {author} {\bibfnamefont {P.~M.}\ \bibnamefont {Oppeneer}}, \bibinfo {author} {\bibfnamefont {U.}~\bibnamefont {Nowak}},\ and\ \bibinfo {author} {\bibfnamefont {M.}~\bibnamefont {M{\"{u}}nzenberg}},\ }\bibfield  {title} {\bibinfo {title} {{Magnetisation switching of FePt nanoparticle
  recording medium by femtosecond laser pulses}},\ }\href {https://doi.org/10.1038/s41598-017-04167-w} {\bibfield  {journal} {\bibinfo  {journal} {Scientific Reports 2017 7:1}\ }\textbf {\bibinfo {volume} {7}},\ \bibinfo {pages} {1} (\bibinfo {year} {2017})}\BibitemShut {NoStop}%
\bibitem [{\citenamefont {Kimel}\ \emph {et~al.}(2002)\citenamefont {Kimel}, \citenamefont {Pisarev}, \citenamefont {Hohlfeld},\ and\ \citenamefont {Rasing}}]{Kimel2002UltrafastCoupling}%
  \BibitemOpen
  \bibfield  {author} {\bibinfo {author} {\bibfnamefont {A.~V.}\ \bibnamefont {Kimel}}, \bibinfo {author} {\bibfnamefont {R.~V.}\ \bibnamefont {Pisarev}}, \bibinfo {author} {\bibfnamefont {J.}~\bibnamefont {Hohlfeld}},\ and\ \bibinfo {author} {\bibfnamefont {T.}~\bibnamefont {Rasing}},\ }\bibfield  {title} {\bibinfo {title} {{Ultrafast Quenching of the Antiferromagnetic Order in [Formula presented]: Direct Optical Probing of the Phonon-Magnon Coupling}},\ }\href {https://doi.org/10.1103/PHYSREVLETT.89.287401/FIGURES/4/MEDIUM} {\bibfield  {journal} {\bibinfo  {journal} {Physical Review Letters}\ }\textbf {\bibinfo {volume} {89}},\ \bibinfo {pages} {287401} (\bibinfo {year} {2002})}\BibitemShut {NoStop}%
\bibitem [{\citenamefont {Hansteen}\ \emph {et~al.}(2005)\citenamefont {Hansteen}, \citenamefont {Kimel}, \citenamefont {Kirilyuk},\ and\ \citenamefont {Rasing}}]{Hansteen2005FemtosecondFilms}%
  \BibitemOpen
  \bibfield  {author} {\bibinfo {author} {\bibfnamefont {F.}~\bibnamefont {Hansteen}}, \bibinfo {author} {\bibfnamefont {A.}~\bibnamefont {Kimel}}, \bibinfo {author} {\bibfnamefont {A.}~\bibnamefont {Kirilyuk}},\ and\ \bibinfo {author} {\bibfnamefont {T.}~\bibnamefont {Rasing}},\ }\bibfield  {title} {\bibinfo {title} {{Femtosecond photomagnetic switching of spins in ferrimagnetic garnet films}},\ }\href {https://doi.org/10.1103/PHYSREVLETT.95.047402/FIGURES/3/MEDIUM} {\bibfield  {journal} {\bibinfo  {journal} {Physical Review Letters}\ }\textbf {\bibinfo {volume} {95}},\ \bibinfo {pages} {047402} (\bibinfo {year} {2005})}\BibitemShut {NoStop}%
\bibitem [{\citenamefont {Atoneche}\ \emph {et~al.}(2010)\citenamefont {Atoneche}, \citenamefont {Kalashnikova}, \citenamefont {Kimel}, \citenamefont {Stupakiewicz}, \citenamefont {Maziewski}, \citenamefont {Kirilyuk},\ and\ \citenamefont {Rasing}}]{Atoneche2010LargeGarnet}%
  \BibitemOpen
  \bibfield  {author} {\bibinfo {author} {\bibfnamefont {F.}~\bibnamefont {Atoneche}}, \bibinfo {author} {\bibfnamefont {A.~M.}\ \bibnamefont {Kalashnikova}}, \bibinfo {author} {\bibfnamefont {A.~V.}\ \bibnamefont {Kimel}}, \bibinfo {author} {\bibfnamefont {A.}~\bibnamefont {Stupakiewicz}}, \bibinfo {author} {\bibfnamefont {A.}~\bibnamefont {Maziewski}}, \bibinfo {author} {\bibfnamefont {A.}~\bibnamefont {Kirilyuk}},\ and\ \bibinfo {author} {\bibfnamefont {T.}~\bibnamefont {Rasing}},\ }\bibfield  {title} {\bibinfo {title} {{Large ultrafast photoinduced magnetic anisotropy in a cobalt-substituted yttrium iron garnet}},\ }\href {https://doi.org/10.1103/PHYSREVB.81.214440/FIGURES/7/MEDIUM} {\bibfield  {journal} {\bibinfo  {journal} {Physical Review B - Condensed Matter and Materials Physics}\ }\textbf {\bibinfo {volume} {81}},\ \bibinfo {pages} {214440} (\bibinfo {year} {2010})}\BibitemShut {NoStop}%
\bibitem [{\citenamefont {Stupakiewicz}\ \emph {et~al.}(2017)\citenamefont {Stupakiewicz}, \citenamefont {Szerenos}, \citenamefont {Afanasiev}, \citenamefont {Kirilyuk},\ and\ \citenamefont {Kimel}}]{Stupakiewicz2017UltrafastMedium}%
  \BibitemOpen
  \bibfield  {author} {\bibinfo {author} {\bibfnamefont {A.}~\bibnamefont {Stupakiewicz}}, \bibinfo {author} {\bibfnamefont {K.}~\bibnamefont {Szerenos}}, \bibinfo {author} {\bibfnamefont {D.}~\bibnamefont {Afanasiev}}, \bibinfo {author} {\bibfnamefont {A.}~\bibnamefont {Kirilyuk}},\ and\ \bibinfo {author} {\bibfnamefont {A.~V.}\ \bibnamefont {Kimel}},\ }\bibfield  {title} {\bibinfo {title} {{Ultrafast nonthermal photo-magnetic recording in a transparent medium}},\ }\href {https://doi.org/10.1038/nature20807} {\bibfield  {journal} {\bibinfo  {journal} {Nature 2017 542:7639}\ }\textbf {\bibinfo {volume} {542}},\ \bibinfo {pages} {71} (\bibinfo {year} {2017})}\BibitemShut {NoStop}%
\bibitem [{\citenamefont {Stupakiewicz}\ \emph {et~al.}(2019)\citenamefont {Stupakiewicz}, \citenamefont {Szerenos}, \citenamefont {Davydova}, \citenamefont {Zvezdin}, \citenamefont {Zvezdin}, \citenamefont {Kirilyuk},\ and\ \citenamefont {Kimel}}]{Stupakiewicz2019SelectionGarnet}%
  \BibitemOpen
  \bibfield  {author} {\bibinfo {author} {\bibfnamefont {A.}~\bibnamefont {Stupakiewicz}}, \bibinfo {author} {\bibfnamefont {K.}~\bibnamefont {Szerenos}}, \bibinfo {author} {\bibfnamefont {M.~D.}\ \bibnamefont {Davydova}}, \bibinfo {author} {\bibfnamefont {K.~A.}\ \bibnamefont {Zvezdin}}, \bibinfo {author} {\bibfnamefont {A.~K.}\ \bibnamefont {Zvezdin}}, \bibinfo {author} {\bibfnamefont {A.}~\bibnamefont {Kirilyuk}},\ and\ \bibinfo {author} {\bibfnamefont {A.~V.}\ \bibnamefont {Kimel}},\ }\bibfield  {title} {\bibinfo {title} {{Selection rules for all-optical magnetic recording in iron garnet}},\ }\href {https://doi.org/10.1038/s41467-019-08458-w} {\bibfield  {journal} {\bibinfo  {journal} {Nature Communications 2019 10:1}\ }\textbf {\bibinfo {volume} {10}},\ \bibinfo {pages} {1} (\bibinfo {year} {2019})}\BibitemShut {NoStop}%
\bibitem [{\citenamefont {Kojima}\ \emph {et~al.}(2003)\citenamefont {Kojima}, \citenamefont {Shimano}, \citenamefont {Hashimoto}, \citenamefont {Katsumoto}, \citenamefont {Iye},\ and\ \citenamefont {Kuwata-Gonokami}}]{Kojima2003ObservationMeasurementsb}%
  \BibitemOpen
  \bibfield  {author} {\bibinfo {author} {\bibfnamefont {E.}~\bibnamefont {Kojima}}, \bibinfo {author} {\bibfnamefont {R.}~\bibnamefont {Shimano}}, \bibinfo {author} {\bibfnamefont {Y.}~\bibnamefont {Hashimoto}}, \bibinfo {author} {\bibfnamefont {S.}~\bibnamefont {Katsumoto}}, \bibinfo {author} {\bibfnamefont {Y.}~\bibnamefont {Iye}},\ and\ \bibinfo {author} {\bibfnamefont {M.}~\bibnamefont {Kuwata-Gonokami}},\ }\bibfield  {title} {\bibinfo {title} {{Observation of the spin-charge thermal isolation of ferromagnetic Ga0.94Mn0.06As by time-resolved magneto-optical measurements}},\ }\href {https://doi.org/10.1103/PhysRevB.68.193203} {\bibfield  {journal} {\bibinfo  {journal} {Physical Review B}\ }\textbf {\bibinfo {volume} {68}},\ \bibinfo {pages} {193203} (\bibinfo {year} {2003})}\BibitemShut {NoStop}%
\bibitem [{\citenamefont {Wang}\ \emph {et~al.}(2007)\citenamefont {Wang}, \citenamefont {Ren}, \citenamefont {Liu}, \citenamefont {Furdyna}, \citenamefont {Grimsditch},\ and\ \citenamefont {Merlin}}]{Wang2007Light-inducedModes}%
  \BibitemOpen
  \bibfield  {author} {\bibinfo {author} {\bibfnamefont {D.~M.}\ \bibnamefont {Wang}}, \bibinfo {author} {\bibfnamefont {Y.~H.}\ \bibnamefont {Ren}}, \bibinfo {author} {\bibfnamefont {X.}~\bibnamefont {Liu}}, \bibinfo {author} {\bibfnamefont {J.~K.}\ \bibnamefont {Furdyna}}, \bibinfo {author} {\bibfnamefont {M.}~\bibnamefont {Grimsditch}},\ and\ \bibinfo {author} {\bibfnamefont {R.}~\bibnamefont {Merlin}},\ }\bibfield  {title} {\bibinfo {title} {{Light-induced magnetic precession in (Ga,Mn)As slabs: Hybrid standing-wave Damon-Eshbach modes}},\ }\href {https://doi.org/10.1103/PHYSREVB.75.233308/FIGURES/4/MEDIUM} {\bibfield  {journal} {\bibinfo  {journal} {Physical Review B - Condensed Matter and Materials Physics}\ }\textbf {\bibinfo {volume} {75}},\ \bibinfo {pages} {233308} (\bibinfo {year} {2007})}\BibitemShut {NoStop}%
\bibitem [{\citenamefont {N{\v{e}}mec}\ \emph {et~al.}(2012)\citenamefont {N{\v{e}}mec}, \citenamefont {Rozkotov{\'{a}}}, \citenamefont {Tesa{\v{r}}ov{\'{a}}}, \citenamefont {Troj{\'{a}}nek}, \citenamefont {De~Ranieri}, \citenamefont {Olejn{\'{i}}k}, \citenamefont {Zemen}, \citenamefont {Nov{\'{a}}k}, \citenamefont {Cukr}, \citenamefont {Mal{\'{y}}},\ and\ \citenamefont {Jungwirth}}]{Nemec2012ExperimentalTorque}%
  \BibitemOpen
  \bibfield  {author} {\bibinfo {author} {\bibfnamefont {P.}~\bibnamefont {N{\v{e}}mec}}, \bibinfo {author} {\bibfnamefont {E.}~\bibnamefont {Rozkotov{\'{a}}}}, \bibinfo {author} {\bibfnamefont {N.}~\bibnamefont {Tesa{\v{r}}ov{\'{a}}}}, \bibinfo {author} {\bibfnamefont {F.}~\bibnamefont {Troj{\'{a}}nek}}, \bibinfo {author} {\bibfnamefont {E.}~\bibnamefont {De~Ranieri}}, \bibinfo {author} {\bibfnamefont {K.}~\bibnamefont {Olejn{\'{i}}k}}, \bibinfo {author} {\bibfnamefont {J.}~\bibnamefont {Zemen}}, \bibinfo {author} {\bibfnamefont {V.}~\bibnamefont {Nov{\'{a}}k}}, \bibinfo {author} {\bibfnamefont {M.}~\bibnamefont {Cukr}}, \bibinfo {author} {\bibfnamefont {P.}~\bibnamefont {Mal{\'{y}}}},\ and\ \bibinfo {author} {\bibfnamefont {T.}~\bibnamefont {Jungwirth}},\ }\bibfield  {title} {\bibinfo {title} {{Experimental observation of the optical spin transfer torque}},\ }\href {https://doi.org/10.1038/nphys2279} {\bibfield  {journal} {\bibinfo  {journal} {Nature Physics 2012 8:5}\ }\textbf {\bibinfo {volume} {8}},\
  \bibinfo {pages} {411} (\bibinfo {year} {2012})}\BibitemShut {NoStop}%
\bibitem [{\citenamefont {Ramsay}\ \emph {et~al.}(2015)\citenamefont {Ramsay}, \citenamefont {Roy}, \citenamefont {Haigh}, \citenamefont {Otxoa}, \citenamefont {Irvine}, \citenamefont {Janda}, \citenamefont {Campion}, \citenamefont {Gallagher},\ and\ \citenamefont {Wunderlich}}]{Ramsay2015OpticalSemiconductorb}%
  \BibitemOpen
  \bibfield  {author} {\bibinfo {author} {\bibfnamefont {A.~J.}\ \bibnamefont {Ramsay}}, \bibinfo {author} {\bibfnamefont {P.~E.}\ \bibnamefont {Roy}}, \bibinfo {author} {\bibfnamefont {J.~E.}\ \bibnamefont {Haigh}}, \bibinfo {author} {\bibfnamefont {R.~M.}\ \bibnamefont {Otxoa}}, \bibinfo {author} {\bibfnamefont {A.~C.}\ \bibnamefont {Irvine}}, \bibinfo {author} {\bibfnamefont {T.}~\bibnamefont {Janda}}, \bibinfo {author} {\bibfnamefont {R.~P.}\ \bibnamefont {Campion}}, \bibinfo {author} {\bibfnamefont {B.~L.}\ \bibnamefont {Gallagher}},\ and\ \bibinfo {author} {\bibfnamefont {J.}~\bibnamefont {Wunderlich}},\ }\bibfield  {title} {\bibinfo {title} {{Optical Spin-Transfer-Torque-Driven Domain-Wall Motion in a Ferromagnetic Semiconductor}},\ }\href {https://journals.aps.org/prl/abstract/10.1103/PhysRevLett.114.067202} {\bibfield  {journal} {\bibinfo  {journal} {Phys. Rev. Lett.}\ }\textbf {\bibinfo {volume} {114}} (\bibinfo {year} {2015})}\BibitemShut {NoStop}%
\bibitem [{\citenamefont {Ishii}\ \emph {et~al.}(2018)\citenamefont {Ishii}, \citenamefont {Yamakawa}, \citenamefont {Kanaki}, \citenamefont {Miyamoto}, \citenamefont {Kida}, \citenamefont {Okamoto}, \citenamefont {Tanaka},\ and\ \citenamefont {Ohya}}]{Ishii2018UltrafastFilm}%
  \BibitemOpen
  \bibfield  {author} {\bibinfo {author} {\bibfnamefont {T.}~\bibnamefont {Ishii}}, \bibinfo {author} {\bibfnamefont {H.}~\bibnamefont {Yamakawa}}, \bibinfo {author} {\bibfnamefont {T.}~\bibnamefont {Kanaki}}, \bibinfo {author} {\bibfnamefont {T.}~\bibnamefont {Miyamoto}}, \bibinfo {author} {\bibfnamefont {N.}~\bibnamefont {Kida}}, \bibinfo {author} {\bibfnamefont {H.}~\bibnamefont {Okamoto}}, \bibinfo {author} {\bibfnamefont {M.}~\bibnamefont {Tanaka}},\ and\ \bibinfo {author} {\bibfnamefont {S.}~\bibnamefont {Ohya}},\ }\bibfield  {title} {\bibinfo {title} {{Ultrafast magnetization modulation induced by the electric field component of a terahertz pulse in a ferromagnetic-semiconductor thin film}},\ }\href {https://doi.org/10.1038/s41598-018-25266-2} {\bibfield  {journal} {\bibinfo  {journal} {Scientific Reports 2018 8:1}\ }\textbf {\bibinfo {volume} {8}},\ \bibinfo {pages} {1} (\bibinfo {year} {2018})}\BibitemShut {NoStop}%
\bibitem [{\citenamefont {Kirilyuk}\ \emph {et~al.}(2010)\citenamefont {Kirilyuk}, \citenamefont {Kimel},\ and\ \citenamefont {Rasing}}]{Kirilyuk2010UltrafastOrder}%
  \BibitemOpen
  \bibfield  {author} {\bibinfo {author} {\bibfnamefont {A.}~\bibnamefont {Kirilyuk}}, \bibinfo {author} {\bibfnamefont {A.~V.}\ \bibnamefont {Kimel}},\ and\ \bibinfo {author} {\bibfnamefont {T.}~\bibnamefont {Rasing}},\ }\bibfield  {title} {\bibinfo {title} {{Ultrafast optical manipulation of magnetic order}},\ }\href {https://doi.org/10.1103/REVMODPHYS.82.2731/FIGURES/51/MEDIUM} {\bibfield  {journal} {\bibinfo  {journal} {Reviews of Modern Physics}\ }\textbf {\bibinfo {volume} {82}},\ \bibinfo {pages} {2731} (\bibinfo {year} {2010})}\BibitemShut {NoStop}%
\bibitem [{\citenamefont {Ostler}\ \emph {et~al.}(2012)\citenamefont {Ostler}, \citenamefont {Barker}, \citenamefont {Evans}, \citenamefont {Chantrell}, \citenamefont {Atxitia}, \citenamefont {Chubykalo-Fesenko}, \citenamefont {El~Moussaoui}, \citenamefont {Le~Guyader}, \citenamefont {Mengotti}, \citenamefont {Heyderman}, \citenamefont {Nolting}, \citenamefont {Tsukamoto}, \citenamefont {Itoh}, \citenamefont {Afanasiev}, \citenamefont {Ivanov}, \citenamefont {Kalashnikova}, \citenamefont {Vahaplar}, \citenamefont {Mentink}, \citenamefont {Kirilyuk}, \citenamefont {Rasing},\ and\ \citenamefont {Kimel}}]{Ostler2012UltrafastFerrimagnet}%
  \BibitemOpen
  \bibfield  {author} {\bibinfo {author} {\bibfnamefont {T.~A.}\ \bibnamefont {Ostler}}, \bibinfo {author} {\bibfnamefont {J.}~\bibnamefont {Barker}}, \bibinfo {author} {\bibfnamefont {R.~F.}\ \bibnamefont {Evans}}, \bibinfo {author} {\bibfnamefont {R.~W.}\ \bibnamefont {Chantrell}}, \bibinfo {author} {\bibfnamefont {U.}~\bibnamefont {Atxitia}}, \bibinfo {author} {\bibfnamefont {O.}~\bibnamefont {Chubykalo-Fesenko}}, \bibinfo {author} {\bibfnamefont {S.}~\bibnamefont {El~Moussaoui}}, \bibinfo {author} {\bibfnamefont {L.}~\bibnamefont {Le~Guyader}}, \bibinfo {author} {\bibfnamefont {E.}~\bibnamefont {Mengotti}}, \bibinfo {author} {\bibfnamefont {L.~J.}\ \bibnamefont {Heyderman}}, \bibinfo {author} {\bibfnamefont {F.}~\bibnamefont {Nolting}}, \bibinfo {author} {\bibfnamefont {A.}~\bibnamefont {Tsukamoto}}, \bibinfo {author} {\bibfnamefont {A.}~\bibnamefont {Itoh}}, \bibinfo {author} {\bibfnamefont {D.}~\bibnamefont {Afanasiev}}, \bibinfo {author} {\bibfnamefont {B.~A.}\ \bibnamefont {Ivanov}}, \bibinfo {author}
  {\bibfnamefont {A.~M.}\ \bibnamefont {Kalashnikova}}, \bibinfo {author} {\bibfnamefont {K.}~\bibnamefont {Vahaplar}}, \bibinfo {author} {\bibfnamefont {J.}~\bibnamefont {Mentink}}, \bibinfo {author} {\bibfnamefont {A.}~\bibnamefont {Kirilyuk}}, \bibinfo {author} {\bibfnamefont {T.}~\bibnamefont {Rasing}},\ and\ \bibinfo {author} {\bibfnamefont {A.~V.}\ \bibnamefont {Kimel}},\ }\bibfield  {title} {\bibinfo {title} {{Ultrafast heating as a sufficient stimulus for magnetization reversal in a ferrimagnet}},\ }\href {https://doi.org/10.1038/ncomms1666} {\bibfield  {journal} {\bibinfo  {journal} {Nature Communications 2012 3:1}\ }\textbf {\bibinfo {volume} {3}},\ \bibinfo {pages} {1} (\bibinfo {year} {2012})}\BibitemShut {NoStop}%
\bibitem [{\citenamefont {Disa}\ \emph {et~al.}(2020)\citenamefont {Disa}, \citenamefont {Fechner}, \citenamefont {Nova}, \citenamefont {Liu}, \citenamefont {F{\"{o}}rst}, \citenamefont {Prabhakaran}, \citenamefont {Radaelli},\ and\ \citenamefont {Cavalleri}}]{Disa2020PolarizingField}%
  \BibitemOpen
  \bibfield  {author} {\bibinfo {author} {\bibfnamefont {A.~S.}\ \bibnamefont {Disa}}, \bibinfo {author} {\bibfnamefont {M.}~\bibnamefont {Fechner}}, \bibinfo {author} {\bibfnamefont {T.~F.}\ \bibnamefont {Nova}}, \bibinfo {author} {\bibfnamefont {B.}~\bibnamefont {Liu}}, \bibinfo {author} {\bibfnamefont {M.}~\bibnamefont {F{\"{o}}rst}}, \bibinfo {author} {\bibfnamefont {D.}~\bibnamefont {Prabhakaran}}, \bibinfo {author} {\bibfnamefont {P.~G.}\ \bibnamefont {Radaelli}},\ and\ \bibinfo {author} {\bibfnamefont {A.}~\bibnamefont {Cavalleri}},\ }\bibfield  {title} {\bibinfo {title} {{Polarizing an antiferromagnet by optical engineering of the crystal field}},\ }\href {https://doi.org/10.1038/s41567-020-0936-3} {\bibfield  {journal} {\bibinfo  {journal} {Nature Physics 2020 16:9}\ }\textbf {\bibinfo {volume} {16}},\ \bibinfo {pages} {937} (\bibinfo {year} {2020})}\BibitemShut {NoStop}%
\bibitem [{\citenamefont {Afanasiev}\ \emph {et~al.}(2021)\citenamefont {Afanasiev}, \citenamefont {Hortensius}, \citenamefont {Ivanov}, \citenamefont {Sasani}, \citenamefont {Bousquet}, \citenamefont {Blanter}, \citenamefont {Mikhaylovskiy}, \citenamefont {Kimel},\ and\ \citenamefont {Caviglia}}]{Afanasiev2021UltrafastPhonons}%
  \BibitemOpen
  \bibfield  {author} {\bibinfo {author} {\bibfnamefont {D.}~\bibnamefont {Afanasiev}}, \bibinfo {author} {\bibfnamefont {J.~R.}\ \bibnamefont {Hortensius}}, \bibinfo {author} {\bibfnamefont {B.~A.}\ \bibnamefont {Ivanov}}, \bibinfo {author} {\bibfnamefont {A.}~\bibnamefont {Sasani}}, \bibinfo {author} {\bibfnamefont {E.}~\bibnamefont {Bousquet}}, \bibinfo {author} {\bibfnamefont {Y.~M.}\ \bibnamefont {Blanter}}, \bibinfo {author} {\bibfnamefont {R.~V.}\ \bibnamefont {Mikhaylovskiy}}, \bibinfo {author} {\bibfnamefont {A.~V.}\ \bibnamefont {Kimel}},\ and\ \bibinfo {author} {\bibfnamefont {A.~D.}\ \bibnamefont {Caviglia}},\ }\bibfield  {title} {\bibinfo {title} {{Ultrafast control of magnetic interactions via light-driven phonons}},\ }\href {https://doi.org/10.1038/s41563-021-00922-7} {\bibfield  {journal} {\bibinfo  {journal} {Nature Materials 2021 20:5}\ }\textbf {\bibinfo {volume} {20}},\ \bibinfo {pages} {607} (\bibinfo {year} {2021})}\BibitemShut {NoStop}%
\bibitem [{\citenamefont {Formisano}\ \emph {et~al.}(2022)\citenamefont {Formisano}, \citenamefont {Dubrovin}, \citenamefont {Pisarev}, \citenamefont {Kalashnikova},\ and\ \citenamefont {Kimel}}]{Formisano2022Laser-inducedCoF2}%
  \BibitemOpen
  \bibfield  {author} {\bibinfo {author} {\bibfnamefont {F.}~\bibnamefont {Formisano}}, \bibinfo {author} {\bibfnamefont {R.~M.}\ \bibnamefont {Dubrovin}}, \bibinfo {author} {\bibfnamefont {R.~V.}\ \bibnamefont {Pisarev}}, \bibinfo {author} {\bibfnamefont {A.~M.}\ \bibnamefont {Kalashnikova}},\ and\ \bibinfo {author} {\bibfnamefont {A.~V.}\ \bibnamefont {Kimel}},\ }\bibfield  {title} {\bibinfo {title} {{Laser-induced THz magnetism of antiferromagnetic CoF2}},\ }\href {https://doi.org/10.1088/1361-648X/AC5C20} {\bibfield  {journal} {\bibinfo  {journal} {Journal of Physics: Condensed Matter}\ }\textbf {\bibinfo {volume} {34}},\ \bibinfo {pages} {225801} (\bibinfo {year} {2022})}\BibitemShut {NoStop}%
\bibitem [{\citenamefont {Hertel}(2006)}]{Hertel2006TheoryMetals}%
  \BibitemOpen
  \bibfield  {author} {\bibinfo {author} {\bibfnamefont {R.}~\bibnamefont {Hertel}},\ }\bibfield  {title} {\bibinfo {title} {{Theory of the inverse Faraday effect in metals}},\ }\href {https://doi.org/10.1016/J.JMMM.2005.10.225} {\bibfield  {journal} {\bibinfo  {journal} {Journal of Magnetism and Magnetic Materials}\ }\textbf {\bibinfo {volume} {303}},\ \bibinfo {pages} {L1} (\bibinfo {year} {2006})}\BibitemShut {NoStop}%
\bibitem [{\citenamefont {Kurkin}\ \emph {et~al.}(2008)\citenamefont {Kurkin}, \citenamefont {Bakulina},\ and\ \citenamefont {Pisarev}}]{Kurkin2008TransientMagnetization}%
  \BibitemOpen
  \bibfield  {author} {\bibinfo {author} {\bibfnamefont {M.~I.}\ \bibnamefont {Kurkin}}, \bibinfo {author} {\bibfnamefont {N.~B.}\ \bibnamefont {Bakulina}},\ and\ \bibinfo {author} {\bibfnamefont {R.~V.}\ \bibnamefont {Pisarev}},\ }\bibfield  {title} {\bibinfo {title} {{Transient inverse Faraday effect and ultrafast optical switching of magnetization}},\ }\href {https://doi.org/10.1103/PHYSREVB.78.134430/FIGURES/1/MEDIUM} {\bibfield  {journal} {\bibinfo  {journal} {Physical Review B - Condensed Matter and Materials Physics}\ }\textbf {\bibinfo {volume} {78}},\ \bibinfo {pages} {134430} (\bibinfo {year} {2008})}\BibitemShut {NoStop}%
\bibitem [{\citenamefont {Popova}\ \emph {et~al.}(2011)\citenamefont {Popova}, \citenamefont {Bringer},\ and\ \citenamefont {Bl{\"{u}}gel}}]{Popova2011TheoryExperiments}%
  \BibitemOpen
  \bibfield  {author} {\bibinfo {author} {\bibfnamefont {D.}~\bibnamefont {Popova}}, \bibinfo {author} {\bibfnamefont {A.}~\bibnamefont {Bringer}},\ and\ \bibinfo {author} {\bibfnamefont {S.}~\bibnamefont {Bl{\"{u}}gel}},\ }\bibfield  {title} {\bibinfo {title} {{Theory of the inverse Faraday effect in view of ultrafast magnetization experiments}},\ }\href {https://doi.org/10.1103/PHYSREVB.84.214421/FIGURES/5/MEDIUM} {\bibfield  {journal} {\bibinfo  {journal} {Physical Review B - Condensed Matter and Materials Physics}\ }\textbf {\bibinfo {volume} {84}},\ \bibinfo {pages} {214421} (\bibinfo {year} {2011})}\BibitemShut {NoStop}%
\bibitem [{\citenamefont {Battiato}\ \emph {et~al.}(2014)\citenamefont {Battiato}, \citenamefont {Barbalinardo},\ and\ \citenamefont {Oppeneer}}]{Battiato2014QuantumEffect}%
  \BibitemOpen
  \bibfield  {author} {\bibinfo {author} {\bibfnamefont {M.}~\bibnamefont {Battiato}}, \bibinfo {author} {\bibfnamefont {G.}~\bibnamefont {Barbalinardo}},\ and\ \bibinfo {author} {\bibfnamefont {P.~M.}\ \bibnamefont {Oppeneer}},\ }\bibfield  {title} {\bibinfo {title} {{Quantum theory of the inverse Faraday effect}},\ }\href {https://doi.org/10.1103/PHYSREVB.89.014413/FIGURES/2/MEDIUM} {\bibfield  {journal} {\bibinfo  {journal} {Physical Review B - Condensed Matter and Materials Physics}\ }\textbf {\bibinfo {volume} {89}},\ \bibinfo {pages} {014413} (\bibinfo {year} {2014})}\BibitemShut {NoStop}%
\bibitem [{\citenamefont {Berritta}\ \emph {et~al.}(2016)\citenamefont {Berritta}, \citenamefont {Mondal}, \citenamefont {Carva},\ and\ \citenamefont {Oppeneer}}]{Berritta2016AbMetals}%
  \BibitemOpen
  \bibfield  {author} {\bibinfo {author} {\bibfnamefont {M.}~\bibnamefont {Berritta}}, \bibinfo {author} {\bibfnamefont {R.}~\bibnamefont {Mondal}}, \bibinfo {author} {\bibfnamefont {K.}~\bibnamefont {Carva}},\ and\ \bibinfo {author} {\bibfnamefont {P.~M.}\ \bibnamefont {Oppeneer}},\ }\bibfield  {title} {\bibinfo {title} {{Ab Initio Theory of Coherent Laser-Induced Magnetization in Metals}},\ }\href {https://doi.org/10.1103/PHYSREVLETT.117.137203/FIGURES/5/MEDIUM} {\bibfield  {journal} {\bibinfo  {journal} {Physical Review Letters}\ }\textbf {\bibinfo {volume} {117}},\ \bibinfo {pages} {137203} (\bibinfo {year} {2016})}\BibitemShut {NoStop}%
\bibitem [{\citenamefont {Freimuth}\ \emph {et~al.}(2016)\citenamefont {Freimuth}, \citenamefont {Bl{\"{u}}gel},\ and\ \citenamefont {Mokrousov}}]{Freimuth2016Laser-inducedFerromagnets}%
  \BibitemOpen
  \bibfield  {author} {\bibinfo {author} {\bibfnamefont {F.}~\bibnamefont {Freimuth}}, \bibinfo {author} {\bibfnamefont {S.}~\bibnamefont {Bl{\"{u}}gel}},\ and\ \bibinfo {author} {\bibfnamefont {Y.}~\bibnamefont {Mokrousov}},\ }\bibfield  {title} {\bibinfo {title} {{Laser-induced torques in metallic ferromagnets}},\ }\href {https://doi.org/10.1103/PHYSREVB.94.144432/FIGURES/5/MEDIUM} {\bibfield  {journal} {\bibinfo  {journal} {Physical Review B}\ }\textbf {\bibinfo {volume} {94}},\ \bibinfo {pages} {144432} (\bibinfo {year} {2016})}\BibitemShut {NoStop}%
\bibitem [{\citenamefont {Smolyaninov}\ \emph {et~al.}(2005)\citenamefont {Smolyaninov}, \citenamefont {Davis}, \citenamefont {Smolyaninova}, \citenamefont {Schaefer}, \citenamefont {Elliott},\ and\ \citenamefont {Zayats}}]{Smolyaninov2005Plasmon-InducedNanostructures}%
  \BibitemOpen
  \bibfield  {author} {\bibinfo {author} {\bibfnamefont {I.~I.}\ \bibnamefont {Smolyaninov}}, \bibinfo {author} {\bibfnamefont {C.~C.}\ \bibnamefont {Davis}}, \bibinfo {author} {\bibfnamefont {V.~N.}\ \bibnamefont {Smolyaninova}}, \bibinfo {author} {\bibfnamefont {D.}~\bibnamefont {Schaefer}}, \bibinfo {author} {\bibfnamefont {J.}~\bibnamefont {Elliott}},\ and\ \bibinfo {author} {\bibfnamefont {A.~V.}\ \bibnamefont {Zayats}},\ }\bibfield  {title} {\bibinfo {title} {{Plasmon-Induced magnetization of metallic nanostructures}},\ }\href {https://doi.org/10.1103/PHYSREVB.71.035425/FIGURES/9/MEDIUM} {\bibfield  {journal} {\bibinfo  {journal} {Physical Review B - Condensed Matter and Materials Physics}\ }\textbf {\bibinfo {volume} {71}},\ \bibinfo {pages} {035425} (\bibinfo {year} {2005})}\BibitemShut {NoStop}%
\bibitem [{\citenamefont {Hurst}\ \emph {et~al.}(2018)\citenamefont {Hurst}, \citenamefont {Oppeneer}, \citenamefont {Manfredi},\ and\ \citenamefont {Hervieux}}]{Hurst2018MagneticEffect}%
  \BibitemOpen
  \bibfield  {author} {\bibinfo {author} {\bibfnamefont {J.}~\bibnamefont {Hurst}}, \bibinfo {author} {\bibfnamefont {P.~M.}\ \bibnamefont {Oppeneer}}, \bibinfo {author} {\bibfnamefont {G.}~\bibnamefont {Manfredi}},\ and\ \bibinfo {author} {\bibfnamefont {P.~A.}\ \bibnamefont {Hervieux}},\ }\bibfield  {title} {\bibinfo {title} {{Magnetic moment generation in small gold nanoparticles via the plasmonic inverse Faraday effect}},\ }\href {https://doi.org/10.1103/PHYSREVB.98.134439/FIGURES/11/MEDIUM} {\bibfield  {journal} {\bibinfo  {journal} {Physical Review B}\ }\textbf {\bibinfo {volume} {94}},\ \bibinfo {pages} {134439} (\bibinfo {year} {2018})}\BibitemShut {NoStop}%
\bibitem [{\citenamefont {Tokman}\ and\ \citenamefont {Shvetsov}(2009)}]{Tokman2009InverseMagnets}%
  \BibitemOpen
  \bibfield  {author} {\bibinfo {author} {\bibfnamefont {I.~D.}\ \bibnamefont {Tokman}}\ and\ \bibinfo {author} {\bibfnamefont {A.~V.}\ \bibnamefont {Shvetsov}},\ }\bibfield  {title} {\bibinfo {title} {{Inverse Faraday effect in crystals of molecular magnets}},\ }\href {https://doi.org/10.3103/S1062873809010092} {\bibfield  {journal} {\bibinfo  {journal} {Bulletin of the Russian Academy of Sciences: Physics 2009 73:1}\ }\textbf {\bibinfo {volume} {73}},\ \bibinfo {pages} {28} (\bibinfo {year} {2009})}\BibitemShut {NoStop}%
\bibitem [{\citenamefont {Jin}\ \emph {et~al.}(2011)\citenamefont {Jin}, \citenamefont {Ma}, \citenamefont {Li}, \citenamefont {Ma}, \citenamefont {Wang},\ and\ \citenamefont {Zhao}}]{Jin2011FemtosecondbmimFeCl4}%
  \BibitemOpen
  \bibfield  {author} {\bibinfo {author} {\bibfnamefont {Z.}~\bibnamefont {Jin}}, \bibinfo {author} {\bibfnamefont {H.}~\bibnamefont {Ma}}, \bibinfo {author} {\bibfnamefont {D.}~\bibnamefont {Li}}, \bibinfo {author} {\bibfnamefont {G.}~\bibnamefont {Ma}}, \bibinfo {author} {\bibfnamefont {M.}~\bibnamefont {Wang}},\ and\ \bibinfo {author} {\bibfnamefont {C.}~\bibnamefont {Zhao}},\ }\bibfield  {title} {\bibinfo {title} {{Femtosecond inverse Faraday effect in magnetic ionic liquid [bmim]FeCl4}},\ }\href {https://doi.org/10.1063/1.3574442} {\bibfield  {journal} {\bibinfo  {journal} {Journal of Applied Physics}\ }\textbf {\bibinfo {volume} {109}},\ \bibinfo {pages} {073109} (\bibinfo {year} {2011})}\BibitemShut {NoStop}%
\bibitem [{\citenamefont {Scheid}\ \emph {et~al.}(2019)\citenamefont {Scheid}, \citenamefont {Malinowski}, \citenamefont {Mangin},\ and\ \citenamefont {Leb{\`{e}}gue}}]{Scheid2019AbFerromagnets}%
  \BibitemOpen
  \bibfield  {author} {\bibinfo {author} {\bibfnamefont {P.}~\bibnamefont {Scheid}}, \bibinfo {author} {\bibfnamefont {G.}~\bibnamefont {Malinowski}}, \bibinfo {author} {\bibfnamefont {S.}~\bibnamefont {Mangin}},\ and\ \bibinfo {author} {\bibfnamefont {S.}~\bibnamefont {Leb{\`{e}}gue}},\ }\bibfield  {title} {\bibinfo {title} {{Ab initio theory of magnetization induced by light absorption in ferromagnets}},\ }\href {https://doi.org/10.1103/PHYSREVB.100.214402/FIGURES/4/MEDIUM} {\bibfield  {journal} {\bibinfo  {journal} {Physical Review B}\ }\textbf {\bibinfo {volume} {100}},\ \bibinfo {pages} {214402} (\bibinfo {year} {2019})}\BibitemShut {NoStop}%
\bibitem [{\citenamefont {Scheid}\ \emph {et~al.}(2021)\citenamefont {Scheid}, \citenamefont {Sharma}, \citenamefont {Malinowski}, \citenamefont {Mangin},\ and\ \citenamefont {Leb{\`{e}}gue}}]{Scheid2021AbRegime}%
  \BibitemOpen
  \bibfield  {author} {\bibinfo {author} {\bibfnamefont {P.}~\bibnamefont {Scheid}}, \bibinfo {author} {\bibfnamefont {S.}~\bibnamefont {Sharma}}, \bibinfo {author} {\bibfnamefont {G.}~\bibnamefont {Malinowski}}, \bibinfo {author} {\bibfnamefont {S.}~\bibnamefont {Mangin}},\ and\ \bibinfo {author} {\bibfnamefont {S.}~\bibnamefont {Leb{\`{e}}gue}},\ }\bibfield  {title} {\bibinfo {title} {{Ab Initio Study of Helicity-Dependent Light-Induced Demagnetization: From the Optical Regime to the Extreme Ultraviolet Regime}},\ }\href {https://doi.org/10.1021/ACS.NANOLETT.0C04166/ASSET/IMAGES/MEDIUM/NL0C04166{\_}0005.GIF} {\bibfield  {journal} {\bibinfo  {journal} {Nano Letters}\ }\textbf {\bibinfo {volume} {21}},\ \bibinfo {pages} {1943} (\bibinfo {year} {2021})}\BibitemShut {NoStop}%
\bibitem [{\citenamefont {Frisk~Kockum}\ \emph {et~al.}(2019)\citenamefont {Frisk~Kockum}, \citenamefont {Miranowicz}, \citenamefont {De~Liberato}, \citenamefont {Savasta},\ and\ \citenamefont {Nori}}]{FriskKockum2019UltrastrongMatter}%
  \BibitemOpen
  \bibfield  {author} {\bibinfo {author} {\bibfnamefont {A.}~\bibnamefont {Frisk~Kockum}}, \bibinfo {author} {\bibfnamefont {A.}~\bibnamefont {Miranowicz}}, \bibinfo {author} {\bibfnamefont {S.}~\bibnamefont {De~Liberato}}, \bibinfo {author} {\bibfnamefont {S.}~\bibnamefont {Savasta}},\ and\ \bibinfo {author} {\bibfnamefont {F.}~\bibnamefont {Nori}},\ }\bibfield  {title} {\bibinfo {title} {{Ultrastrong coupling between light and matter}},\ }\href {https://doi.org/10.1038/s42254-018-0006-2} {\bibfield  {journal} {\bibinfo  {journal} {Nature Reviews Physics 2019 1:1}\ }\textbf {\bibinfo {volume} {1}},\ \bibinfo {pages} {19} (\bibinfo {year} {2019})}\BibitemShut {NoStop}%
\bibitem [{\citenamefont {H{\"{u}}bener}\ \emph {et~al.}(2020)\citenamefont {H{\"{u}}bener}, \citenamefont {De~Giovannini}, \citenamefont {Sch{\"{a}}fer}, \citenamefont {Andberger}, \citenamefont {Ruggenthaler}, \citenamefont {Faist},\ and\ \citenamefont {Rubio}}]{Hubener2020EngineeringCavities}%
  \BibitemOpen
  \bibfield  {author} {\bibinfo {author} {\bibfnamefont {H.}~\bibnamefont {H{\"{u}}bener}}, \bibinfo {author} {\bibfnamefont {U.}~\bibnamefont {De~Giovannini}}, \bibinfo {author} {\bibfnamefont {C.}~\bibnamefont {Sch{\"{a}}fer}}, \bibinfo {author} {\bibfnamefont {J.}~\bibnamefont {Andberger}}, \bibinfo {author} {\bibfnamefont {M.}~\bibnamefont {Ruggenthaler}}, \bibinfo {author} {\bibfnamefont {J.}~\bibnamefont {Faist}},\ and\ \bibinfo {author} {\bibfnamefont {A.}~\bibnamefont {Rubio}},\ }\bibfield  {title} {\bibinfo {title} {{Engineering quantum materials with chiral optical cavities}},\ }\href {https://doi.org/10.1038/s41563-020-00801-7} {\bibfield  {journal} {\bibinfo  {journal} {Nature Materials 2020 20:4}\ }\textbf {\bibinfo {volume} {20}},\ \bibinfo {pages} {438} (\bibinfo {year} {2020})}\BibitemShut {NoStop}%
\bibitem [{\citenamefont {Barman}\ \emph {et~al.}(2021)\citenamefont {Barman}, \citenamefont {Gubbiotti}, \citenamefont {Ladak}, \citenamefont {Adeyeye}, \citenamefont {Krawczyk}, \citenamefont {Grafe}, \citenamefont {Adelmann}, \citenamefont {Cotofana}, \citenamefont {Naeemi}, \citenamefont {Vasyuchka}, \citenamefont {Hillebrands}, \citenamefont {Nikitov}, \citenamefont {Yu}, \citenamefont {Grundler}, \citenamefont {Sadovnikov}, \citenamefont {Grachev}, \citenamefont {Sheshukova}, \citenamefont {Duquesne}, \citenamefont {Marangolo}, \citenamefont {Csaba}, \citenamefont {Porod}, \citenamefont {Demidov}, \citenamefont {Urazhdin}, \citenamefont {Demokritov}, \citenamefont {Albisetti}, \citenamefont {Petti}, \citenamefont {Bertacco}, \citenamefont {Schultheiss}, \citenamefont {Kruglyak}, \citenamefont {Poimanov}, \citenamefont {Sahoo}, \citenamefont {Sinha}, \citenamefont {Yang}, \citenamefont {M{\"{u}}nzenberg}, \citenamefont {Moriyama}, \citenamefont {Mizukami}, \citenamefont {Landeros}, \citenamefont
  {Gallardo}, \citenamefont {Carlotti}, \citenamefont {Kim}, \citenamefont {Stamps}, \citenamefont {Camley}, \citenamefont {Rana}, \citenamefont {Otani}, \citenamefont {Yu}, \citenamefont {Yu}, \citenamefont {Bauer}, \citenamefont {Back}, \citenamefont {Uhrig}, \citenamefont {Dobrovolskiy}, \citenamefont {Budinska}, \citenamefont {Qin}, \citenamefont {Van~Dijken}, \citenamefont {Chumak}, \citenamefont {Khitun}, \citenamefont {Nikonov}, \citenamefont {Young}, \citenamefont {Zingsem},\ and\ \citenamefont {Winklhofer}}]{Barman2021TheRoadmap}%
  \BibitemOpen
  \bibfield  {author} {\bibinfo {author} {\bibfnamefont {A.}~\bibnamefont {Barman}}, \bibinfo {author} {\bibfnamefont {G.}~\bibnamefont {Gubbiotti}}, \bibinfo {author} {\bibfnamefont {S.}~\bibnamefont {Ladak}}, \bibinfo {author} {\bibfnamefont {A.~O.}\ \bibnamefont {Adeyeye}}, \bibinfo {author} {\bibfnamefont {M.}~\bibnamefont {Krawczyk}}, \bibinfo {author} {\bibfnamefont {J.}~\bibnamefont {Grafe}}, \bibinfo {author} {\bibfnamefont {C.}~\bibnamefont {Adelmann}}, \bibinfo {author} {\bibfnamefont {S.}~\bibnamefont {Cotofana}}, \bibinfo {author} {\bibfnamefont {A.}~\bibnamefont {Naeemi}}, \bibinfo {author} {\bibfnamefont {V.~I.}\ \bibnamefont {Vasyuchka}}, \bibinfo {author} {\bibfnamefont {B.}~\bibnamefont {Hillebrands}}, \bibinfo {author} {\bibfnamefont {S.~A.}\ \bibnamefont {Nikitov}}, \bibinfo {author} {\bibfnamefont {H.}~\bibnamefont {Yu}}, \bibinfo {author} {\bibfnamefont {D.}~\bibnamefont {Grundler}}, \bibinfo {author} {\bibfnamefont {A.~V.}\ \bibnamefont {Sadovnikov}}, \bibinfo {author} {\bibfnamefont
  {A.~A.}\ \bibnamefont {Grachev}}, \bibinfo {author} {\bibfnamefont {S.~E.}\ \bibnamefont {Sheshukova}}, \bibinfo {author} {\bibfnamefont {J.~Y.}\ \bibnamefont {Duquesne}}, \bibinfo {author} {\bibfnamefont {M.}~\bibnamefont {Marangolo}}, \bibinfo {author} {\bibfnamefont {G.}~\bibnamefont {Csaba}}, \bibinfo {author} {\bibfnamefont {W.}~\bibnamefont {Porod}}, \bibinfo {author} {\bibfnamefont {V.~E.}\ \bibnamefont {Demidov}}, \bibinfo {author} {\bibfnamefont {S.}~\bibnamefont {Urazhdin}}, \bibinfo {author} {\bibfnamefont {S.~O.}\ \bibnamefont {Demokritov}}, \bibinfo {author} {\bibfnamefont {E.}~\bibnamefont {Albisetti}}, \bibinfo {author} {\bibfnamefont {D.}~\bibnamefont {Petti}}, \bibinfo {author} {\bibfnamefont {R.}~\bibnamefont {Bertacco}}, \bibinfo {author} {\bibfnamefont {H.}~\bibnamefont {Schultheiss}}, \bibinfo {author} {\bibfnamefont {V.~V.}\ \bibnamefont {Kruglyak}}, \bibinfo {author} {\bibfnamefont {V.~D.}\ \bibnamefont {Poimanov}}, \bibinfo {author} {\bibfnamefont {S.}~\bibnamefont {Sahoo}}, \bibinfo
  {author} {\bibfnamefont {J.}~\bibnamefont {Sinha}}, \bibinfo {author} {\bibfnamefont {H.}~\bibnamefont {Yang}}, \bibinfo {author} {\bibfnamefont {M.}~\bibnamefont {M{\"{u}}nzenberg}}, \bibinfo {author} {\bibfnamefont {T.}~\bibnamefont {Moriyama}}, \bibinfo {author} {\bibfnamefont {S.}~\bibnamefont {Mizukami}}, \bibinfo {author} {\bibfnamefont {P.}~\bibnamefont {Landeros}}, \bibinfo {author} {\bibfnamefont {R.~A.}\ \bibnamefont {Gallardo}}, \bibinfo {author} {\bibfnamefont {G.}~\bibnamefont {Carlotti}}, \bibinfo {author} {\bibfnamefont {J.~V.}\ \bibnamefont {Kim}}, \bibinfo {author} {\bibfnamefont {R.~L.}\ \bibnamefont {Stamps}}, \bibinfo {author} {\bibfnamefont {R.~E.}\ \bibnamefont {Camley}}, \bibinfo {author} {\bibfnamefont {B.}~\bibnamefont {Rana}}, \bibinfo {author} {\bibfnamefont {Y.}~\bibnamefont {Otani}}, \bibinfo {author} {\bibfnamefont {W.}~\bibnamefont {Yu}}, \bibinfo {author} {\bibfnamefont {T.}~\bibnamefont {Yu}}, \bibinfo {author} {\bibfnamefont {G.~E.}\ \bibnamefont {Bauer}}, \bibinfo {author}
  {\bibfnamefont {C.}~\bibnamefont {Back}}, \bibinfo {author} {\bibfnamefont {G.~S.}\ \bibnamefont {Uhrig}}, \bibinfo {author} {\bibfnamefont {O.~V.}\ \bibnamefont {Dobrovolskiy}}, \bibinfo {author} {\bibfnamefont {B.}~\bibnamefont {Budinska}}, \bibinfo {author} {\bibfnamefont {H.}~\bibnamefont {Qin}}, \bibinfo {author} {\bibfnamefont {S.}~\bibnamefont {Van~Dijken}}, \bibinfo {author} {\bibfnamefont {A.~V.}\ \bibnamefont {Chumak}}, \bibinfo {author} {\bibfnamefont {A.}~\bibnamefont {Khitun}}, \bibinfo {author} {\bibfnamefont {D.~E.}\ \bibnamefont {Nikonov}}, \bibinfo {author} {\bibfnamefont {I.~A.}\ \bibnamefont {Young}}, \bibinfo {author} {\bibfnamefont {B.~W.}\ \bibnamefont {Zingsem}},\ and\ \bibinfo {author} {\bibfnamefont {M.}~\bibnamefont {Winklhofer}},\ }\bibfield  {title} {\bibinfo {title} {{The 2021 Magnonics Roadmap}},\ }\href {https://doi.org/10.1088/1361-648X/ABEC1A} {\bibfield  {journal} {\bibinfo  {journal} {Journal of Physics: Condensed Matter}\ }\textbf {\bibinfo {volume} {33}},\ \bibinfo
  {pages} {413001} (\bibinfo {year} {2021})}\BibitemShut {NoStop}%
\bibitem [{\citenamefont {Li}\ \emph {et~al.}(2022)\citenamefont {Li}, \citenamefont {Yang}, \citenamefont {Mondal}, \citenamefont {Tzschaschel},\ and\ \citenamefont {Pal}}]{Li2022ADynamics}%
  \BibitemOpen
  \bibfield  {author} {\bibinfo {author} {\bibfnamefont {J.}~\bibnamefont {Li}}, \bibinfo {author} {\bibfnamefont {C.~J.}\ \bibnamefont {Yang}}, \bibinfo {author} {\bibfnamefont {R.}~\bibnamefont {Mondal}}, \bibinfo {author} {\bibfnamefont {C.}~\bibnamefont {Tzschaschel}},\ and\ \bibinfo {author} {\bibfnamefont {S.}~\bibnamefont {Pal}},\ }\bibfield  {title} {\bibinfo {title} {{A perspective on nonlinearities in coherent magnetization dynamics}},\ }\href {https://doi.org/10.1063/5.0075999/2832824} {\bibfield  {journal} {\bibinfo  {journal} {Applied Physics Letters}\ }\textbf {\bibinfo {volume} {120}},\ \bibinfo {pages} {50501} (\bibinfo {year} {2022})}\BibitemShut {NoStop}%
\bibitem [{\citenamefont {Huisman}\ \emph {et~al.}(2016)\citenamefont {Huisman}, \citenamefont {Mikhaylovskiy}, \citenamefont {Costa}, \citenamefont {Freimuth}, \citenamefont {Paz}, \citenamefont {Ventura}, \citenamefont {Freitas}, \citenamefont {Bl{\"{u}}gel}, \citenamefont {Mokrousov}, \citenamefont {Rasing},\ and\ \citenamefont {Kimel}}]{Huisman2016FemtosecondHeterostructures}%
  \BibitemOpen
  \bibfield  {author} {\bibinfo {author} {\bibfnamefont {T.~J.}\ \bibnamefont {Huisman}}, \bibinfo {author} {\bibfnamefont {R.~V.}\ \bibnamefont {Mikhaylovskiy}}, \bibinfo {author} {\bibfnamefont {J.~D.}\ \bibnamefont {Costa}}, \bibinfo {author} {\bibfnamefont {F.}~\bibnamefont {Freimuth}}, \bibinfo {author} {\bibfnamefont {E.}~\bibnamefont {Paz}}, \bibinfo {author} {\bibfnamefont {J.}~\bibnamefont {Ventura}}, \bibinfo {author} {\bibfnamefont {P.~P.}\ \bibnamefont {Freitas}}, \bibinfo {author} {\bibfnamefont {S.}~\bibnamefont {Bl{\"{u}}gel}}, \bibinfo {author} {\bibfnamefont {Y.}~\bibnamefont {Mokrousov}}, \bibinfo {author} {\bibfnamefont {T.}~\bibnamefont {Rasing}},\ and\ \bibinfo {author} {\bibfnamefont {A.~V.}\ \bibnamefont {Kimel}},\ }\bibfield  {title} {\bibinfo {title} {{Femtosecond control of electric currents in metallic ferromagnetic heterostructures}},\ }\href {https://doi.org/10.1038/nnano.2015.331} {\bibfield  {journal} {\bibinfo  {journal} {Nature Nanotechnology 2016 11:5}\ }\textbf {\bibinfo
  {volume} {11}},\ \bibinfo {pages} {455} (\bibinfo {year} {2016})}\BibitemShut {NoStop}%
\bibitem [{\citenamefont {Choi}\ \emph {et~al.}(2017)\citenamefont {Choi}, \citenamefont {Schleife},\ and\ \citenamefont {Cahill}}]{Choi2017Optical-helicity-drivenFerromagnets}%
  \BibitemOpen
  \bibfield  {author} {\bibinfo {author} {\bibfnamefont {G.~M.}\ \bibnamefont {Choi}}, \bibinfo {author} {\bibfnamefont {A.}~\bibnamefont {Schleife}},\ and\ \bibinfo {author} {\bibfnamefont {D.~G.}\ \bibnamefont {Cahill}},\ }\bibfield  {title} {\bibinfo {title} {{Optical-helicity-driven magnetization dynamics in metallic ferromagnets}},\ }\href {https://doi.org/10.1038/ncomms15085} {\bibfield  {journal} {\bibinfo  {journal} {Nature Communications 2017 8:1}\ }\textbf {\bibinfo {volume} {8}},\ \bibinfo {pages} {1} (\bibinfo {year} {2017})}\BibitemShut {NoStop}%
\bibitem [{\citenamefont {Kazantseva}\ \emph {et~al.}(2009)\citenamefont {Kazantseva}, \citenamefont {Hinzke}, \citenamefont {Chantrell},\ and\ \citenamefont {Nowak}}]{Kazantseva2009LinearTemperature}%
  \BibitemOpen
  \bibfield  {author} {\bibinfo {author} {\bibfnamefont {N.}~\bibnamefont {Kazantseva}}, \bibinfo {author} {\bibfnamefont {D.}~\bibnamefont {Hinzke}}, \bibinfo {author} {\bibfnamefont {R.~W.}\ \bibnamefont {Chantrell}},\ and\ \bibinfo {author} {\bibfnamefont {U.}~\bibnamefont {Nowak}},\ }\bibfield  {title} {\bibinfo {title} {{Linear and elliptical magnetization reversal close to the Curie temperature}},\ }\href {https://doi.org/10.1209/0295-5075/86/27006} {\bibfield  {journal} {\bibinfo  {journal} {Europhysics Letters}\ }\textbf {\bibinfo {volume} {86}},\ \bibinfo {pages} {27006} (\bibinfo {year} {2009})}\BibitemShut {NoStop}%
\bibitem [{\citenamefont {Ellis}\ \emph {et~al.}(2016)\citenamefont {Ellis}, \citenamefont {Fullerton},\ and\ \citenamefont {Chantrell}}]{Ellis2016All-opticalDichroism}%
  \BibitemOpen
  \bibfield  {author} {\bibinfo {author} {\bibfnamefont {M.~O.}\ \bibnamefont {Ellis}}, \bibinfo {author} {\bibfnamefont {E.~E.}\ \bibnamefont {Fullerton}},\ and\ \bibinfo {author} {\bibfnamefont {R.~W.}\ \bibnamefont {Chantrell}},\ }\bibfield  {title} {\bibinfo {title} {{All-optical switching in granular ferromagnets caused by magnetic circular dichroism}},\ }\href {https://doi.org/10.1038/srep30522} {\bibfield  {journal} {\bibinfo  {journal} {Scientific Reports 2016 6:1}\ }\textbf {\bibinfo {volume} {6}},\ \bibinfo {pages} {1} (\bibinfo {year} {2016})}\BibitemShut {NoStop}%
\bibitem [{\citenamefont {Raposo}\ \emph {et~al.}(2020)\citenamefont {Raposo}, \citenamefont {Guedas}, \citenamefont {Garc{\'{i}}a-S{\'{a}}nchez}, \citenamefont {Hern{\'{a}}ndez}, \citenamefont {Zazo},\ and\ \citenamefont {Mart{\'{i}}nez}}]{Raposo2020MicromagneticDichroism}%
  \BibitemOpen
  \bibfield  {author} {\bibinfo {author} {\bibfnamefont {V.}~\bibnamefont {Raposo}}, \bibinfo {author} {\bibfnamefont {R.}~\bibnamefont {Guedas}}, \bibinfo {author} {\bibfnamefont {F.}~\bibnamefont {Garc{\'{i}}a-S{\'{a}}nchez}}, \bibinfo {author} {\bibfnamefont {M.~A.}\ \bibnamefont {Hern{\'{a}}ndez}}, \bibinfo {author} {\bibfnamefont {M.}~\bibnamefont {Zazo}},\ and\ \bibinfo {author} {\bibfnamefont {E.}~\bibnamefont {Mart{\'{i}}nez}},\ }\bibfield  {title} {\bibinfo {title} {{Micromagnetic Modeling of All Optical Switching of Ferromagnetic Thin Films: The Role of Inverse Faraday Effect and Magnetic Circular Dichroism}},\ }\href {https://doi.org/10.3390/APP10041307} {\bibfield  {journal} {\bibinfo  {journal} {Applied Sciences 2020, Vol. 10, Page 1307}\ }\textbf {\bibinfo {volume} {10}},\ \bibinfo {pages} {1307} (\bibinfo {year} {2020})}\BibitemShut {NoStop}%
\bibitem [{\citenamefont {Vahaplar}\ \emph {et~al.}(2009)\citenamefont {Vahaplar}, \citenamefont {Kalashnikova}, \citenamefont {Kimel}, \citenamefont {Hinzke}, \citenamefont {Nowak}, \citenamefont {Chantrell}, \citenamefont {Tsukamoto}, \citenamefont {Itoh}, \citenamefont {Kirilyuk},\ and\ \citenamefont {Rasing}}]{Vahaplar2009UltrafastState}%
  \BibitemOpen
  \bibfield  {author} {\bibinfo {author} {\bibfnamefont {K.}~\bibnamefont {Vahaplar}}, \bibinfo {author} {\bibfnamefont {A.~M.}\ \bibnamefont {Kalashnikova}}, \bibinfo {author} {\bibfnamefont {A.~V.}\ \bibnamefont {Kimel}}, \bibinfo {author} {\bibfnamefont {D.}~\bibnamefont {Hinzke}}, \bibinfo {author} {\bibfnamefont {U.}~\bibnamefont {Nowak}}, \bibinfo {author} {\bibfnamefont {R.}~\bibnamefont {Chantrell}}, \bibinfo {author} {\bibfnamefont {A.}~\bibnamefont {Tsukamoto}}, \bibinfo {author} {\bibfnamefont {A.}~\bibnamefont {Itoh}}, \bibinfo {author} {\bibfnamefont {A.}~\bibnamefont {Kirilyuk}},\ and\ \bibinfo {author} {\bibfnamefont {T.}~\bibnamefont {Rasing}},\ }\bibfield  {title} {\bibinfo {title} {{Ultrafast Path for Optical Magnetization Reversal via a Strongly Nonequilibrium State}},\ }\href {https://doi.org/10.1103/PHYSREVLETT.103.117201/FIGURES/3/MEDIUM} {\bibfield  {journal} {\bibinfo  {journal} {Physical Review Letters}\ }\textbf {\bibinfo {volume} {103}},\ \bibinfo {pages} {117201} (\bibinfo {year}
  {2009})}\BibitemShut {NoStop}%
\bibitem [{\citenamefont {Vahaplar}\ \emph {et~al.}(2012)\citenamefont {Vahaplar}, \citenamefont {Kalashnikova}, \citenamefont {Kimel}, \citenamefont {Gerlach}, \citenamefont {Hinzke}, \citenamefont {Nowak}, \citenamefont {Chantrell}, \citenamefont {Tsukamoto}, \citenamefont {Itoh}, \citenamefont {Kirilyuk},\ and\ \citenamefont {Rasing}}]{Vahaplar2012All-opticalModeling}%
  \BibitemOpen
  \bibfield  {author} {\bibinfo {author} {\bibfnamefont {K.}~\bibnamefont {Vahaplar}}, \bibinfo {author} {\bibfnamefont {A.~M.}\ \bibnamefont {Kalashnikova}}, \bibinfo {author} {\bibfnamefont {A.~V.}\ \bibnamefont {Kimel}}, \bibinfo {author} {\bibfnamefont {S.}~\bibnamefont {Gerlach}}, \bibinfo {author} {\bibfnamefont {D.}~\bibnamefont {Hinzke}}, \bibinfo {author} {\bibfnamefont {U.}~\bibnamefont {Nowak}}, \bibinfo {author} {\bibfnamefont {R.}~\bibnamefont {Chantrell}}, \bibinfo {author} {\bibfnamefont {A.}~\bibnamefont {Tsukamoto}}, \bibinfo {author} {\bibfnamefont {A.}~\bibnamefont {Itoh}}, \bibinfo {author} {\bibfnamefont {A.}~\bibnamefont {Kirilyuk}},\ and\ \bibinfo {author} {\bibfnamefont {T.}~\bibnamefont {Rasing}},\ }\bibfield  {title} {\bibinfo {title} {{All-optical magnetization reversal by circularly polarized laser pulses: Experiment and multiscale modeling}},\ }\href {https://doi.org/10.1103/PHYSREVB.85.104402/FIGURES/12/MEDIUM} {\bibfield  {journal} {\bibinfo  {journal} {Physical Review B -
  Condensed Matter and Materials Physics}\ }\textbf {\bibinfo {volume} {85}},\ \bibinfo {pages} {104402} (\bibinfo {year} {2012})}\BibitemShut {NoStop}%
\bibitem [{\citenamefont {Longman}\ and\ \citenamefont {Fedosejevs}(2021)}]{Longman2021Kilo-TeslaBeams}%
  \BibitemOpen
  \bibfield  {author} {\bibinfo {author} {\bibfnamefont {A.}~\bibnamefont {Longman}}\ and\ \bibinfo {author} {\bibfnamefont {R.}~\bibnamefont {Fedosejevs}},\ }\bibfield  {title} {\bibinfo {title} {{Kilo-Tesla axial magnetic field generation with high intensity spin and orbital angular momentum beams}},\ }\href {https://doi.org/10.1103/PHYSREVRESEARCH.3.043180/FIGURES/7/MEDIUM} {\bibfield  {journal} {\bibinfo  {journal} {Physical Review Research}\ }\textbf {\bibinfo {volume} {3}},\ \bibinfo {pages} {043180} (\bibinfo {year} {2021})}\BibitemShut {NoStop}%
\bibitem [{\citenamefont {Hinzke}\ and\ \citenamefont {Nowak}(2011)}]{Hinzke2011DomainEffectb}%
  \BibitemOpen
  \bibfield  {author} {\bibinfo {author} {\bibfnamefont {D.}~\bibnamefont {Hinzke}}\ and\ \bibinfo {author} {\bibfnamefont {U.}~\bibnamefont {Nowak}},\ }\bibfield  {title} {\bibinfo {title} {{Domain wall motion by the magnonic spin seebeck effect}},\ }\href {https://doi.org/10.1103/PHYSREVLETT.107.027205/FIGURES/3/MEDIUM} {\bibfield  {journal} {\bibinfo  {journal} {Physical Review Letters}\ }\textbf {\bibinfo {volume} {107}},\ \bibinfo {pages} {027205} (\bibinfo {year} {2011})}\BibitemShut {NoStop}%
\bibitem [{\citenamefont {Schlickeiser}\ \emph {et~al.}(2014)\citenamefont {Schlickeiser}, \citenamefont {Ritzmann}, \citenamefont {Hinzke},\ and\ \citenamefont {Nowak}}]{Schlickeiser2014RoleGradients}%
  \BibitemOpen
  \bibfield  {author} {\bibinfo {author} {\bibfnamefont {F.}~\bibnamefont {Schlickeiser}}, \bibinfo {author} {\bibfnamefont {U.}~\bibnamefont {Ritzmann}}, \bibinfo {author} {\bibfnamefont {D.}~\bibnamefont {Hinzke}},\ and\ \bibinfo {author} {\bibfnamefont {U.}~\bibnamefont {Nowak}},\ }\bibfield  {title} {\bibinfo {title} {{Role of entropy in domain wall motion in thermal gradients}},\ }\href {https://doi.org/10.1103/PHYSREVLETT.113.097201/FIGURES/5/MEDIUM} {\bibfield  {journal} {\bibinfo  {journal} {Physical Review Letters}\ }\textbf {\bibinfo {volume} {113}},\ \bibinfo {pages} {097201} (\bibinfo {year} {2014})}\BibitemShut {NoStop}%
\bibitem [{\citenamefont {Moretti}\ \emph {et~al.}(2017)\citenamefont {Moretti}, \citenamefont {Raposo}, \citenamefont {Martinez},\ and\ \citenamefont {Lopez-Diaz}}]{Moretti2017DomainGradients}%
  \BibitemOpen
  \bibfield  {author} {\bibinfo {author} {\bibfnamefont {S.}~\bibnamefont {Moretti}}, \bibinfo {author} {\bibfnamefont {V.}~\bibnamefont {Raposo}}, \bibinfo {author} {\bibfnamefont {E.}~\bibnamefont {Martinez}},\ and\ \bibinfo {author} {\bibfnamefont {L.}~\bibnamefont {Lopez-Diaz}},\ }\bibfield  {title} {\bibinfo {title} {{Domain wall motion by localized temperature gradients}},\ }\href {https://doi.org/10.1103/PHYSREVB.95.064419/FIGURES/12/MEDIUM} {\bibfield  {journal} {\bibinfo  {journal} {Physical Review B}\ }\textbf {\bibinfo {volume} {95}},\ \bibinfo {pages} {064419} (\bibinfo {year} {2017})}\BibitemShut {NoStop}%
\bibitem [{\citenamefont {Selzer}\ \emph {et~al.}(2022)\citenamefont {Selzer}, \citenamefont {Salemi}, \citenamefont {De{\'{a}}k}, \citenamefont {Simon}, \citenamefont {Szunyogh}, \citenamefont {Oppeneer},\ and\ \citenamefont {Nowak}}]{Selzer2022Current-inducedPrinciples}%
  \BibitemOpen
  \bibfield  {author} {\bibinfo {author} {\bibfnamefont {S.}~\bibnamefont {Selzer}}, \bibinfo {author} {\bibfnamefont {L.}~\bibnamefont {Salemi}}, \bibinfo {author} {\bibfnamefont {A.}~\bibnamefont {De{\'{a}}k}}, \bibinfo {author} {\bibfnamefont {E.}~\bibnamefont {Simon}}, \bibinfo {author} {\bibfnamefont {L.}~\bibnamefont {Szunyogh}}, \bibinfo {author} {\bibfnamefont {P.~M.}\ \bibnamefont {Oppeneer}},\ and\ \bibinfo {author} {\bibfnamefont {U.}~\bibnamefont {Nowak}},\ }\bibfield  {title} {\bibinfo {title} {{Current-induced switching of antiferromagnetic order in Mn2Au from first principles}},\ }\href {https://doi.org/10.1103/PHYSREVB.105.174416/FIGURES/6/MEDIUM} {\bibfield  {journal} {\bibinfo  {journal} {Physical Review B}\ }\textbf {\bibinfo {volume} {105}},\ \bibinfo {pages} {174416} (\bibinfo {year} {2022})}\BibitemShut {NoStop}%
\bibitem [{\citenamefont {Zhang}\ \emph {et~al.}(2019)\citenamefont {Zhang}, \citenamefont {Xu}, \citenamefont {Zhao}, \citenamefont {Zhu}, \citenamefont {Lin}, \citenamefont {Hehn}, \citenamefont {Malinowski}, \citenamefont {Ravelosona},\ and\ \citenamefont {Mangin}}]{Zhang2019Energy-EfficientTorque}%
  \BibitemOpen
  \bibfield  {author} {\bibinfo {author} {\bibfnamefont {B.}~\bibnamefont {Zhang}}, \bibinfo {author} {\bibfnamefont {Y.}~\bibnamefont {Xu}}, \bibinfo {author} {\bibfnamefont {W.}~\bibnamefont {Zhao}}, \bibinfo {author} {\bibfnamefont {D.}~\bibnamefont {Zhu}}, \bibinfo {author} {\bibfnamefont {X.}~\bibnamefont {Lin}}, \bibinfo {author} {\bibfnamefont {M.}~\bibnamefont {Hehn}}, \bibinfo {author} {\bibfnamefont {G.}~\bibnamefont {Malinowski}}, \bibinfo {author} {\bibfnamefont {D.}~\bibnamefont {Ravelosona}},\ and\ \bibinfo {author} {\bibfnamefont {S.}~\bibnamefont {Mangin}},\ }\bibfield  {title} {\bibinfo {title} {{Energy-Efficient Domain-Wall Motion Governed by the Interplay of Helicity-Dependent Optical Effect and Spin-Orbit Torque}},\ }\href {https://doi.org/10.1103/PHYSREVAPPLIED.11.034001/FIGURES/5/MEDIUM} {\bibfield  {journal} {\bibinfo  {journal} {Physical Review Applied}\ }\textbf {\bibinfo {volume} {11}},\ \bibinfo {pages} {034001} (\bibinfo {year} {2019})}\BibitemShut {NoStop}%
\bibitem [{\citenamefont {Garanin}(1991)}]{Garanin1991GeneralizedFerromagnet}%
  \BibitemOpen
  \bibfield  {author} {\bibinfo {author} {\bibfnamefont {D.~A.}\ \bibnamefont {Garanin}},\ }\bibfield  {title} {\bibinfo {title} {{Generalized equation of motion for a ferromagnet}},\ }\href {https://doi.org/10.1016/0378-4371(91)90395-S} {\bibfield  {journal} {\bibinfo  {journal} {Physica A: Statistical Mechanics and its Applications}\ }\textbf {\bibinfo {volume} {172}},\ \bibinfo {pages} {470} (\bibinfo {year} {1991})}\BibitemShut {NoStop}%
\bibitem [{\citenamefont {Garanin}(1997)}]{Garanin1997Fokker-PlanckFerromagnets}%
  \BibitemOpen
  \bibfield  {author} {\bibinfo {author} {\bibfnamefont {D.}~\bibnamefont {Garanin}},\ }\bibfield  {title} {\bibinfo {title} {{Fokker-Planck and Landau-Lifshitz-Bloch equations for classical ferromagnets}},\ }\href {https://doi.org/10.1103/PhysRevB.55.3050} {\bibfield  {journal} {\bibinfo  {journal} {Physical Review B}\ }\textbf {\bibinfo {volume} {55}},\ \bibinfo {pages} {3050} (\bibinfo {year} {1997})}\BibitemShut {NoStop}%
\bibitem [{\citenamefont {Chubykalo-Fesenko}\ \emph {et~al.}(2006)\citenamefont {Chubykalo-Fesenko}, \citenamefont {Nowak}, \citenamefont {Chantrell},\ and\ \citenamefont {Garanin}}]{Chubykalo-Fesenko2006DynamicTemperature}%
  \BibitemOpen
  \bibfield  {author} {\bibinfo {author} {\bibfnamefont {O.}~\bibnamefont {Chubykalo-Fesenko}}, \bibinfo {author} {\bibfnamefont {U.}~\bibnamefont {Nowak}}, \bibinfo {author} {\bibfnamefont {R.~W.}\ \bibnamefont {Chantrell}},\ and\ \bibinfo {author} {\bibfnamefont {D.}~\bibnamefont {Garanin}},\ }\bibfield  {title} {\bibinfo {title} {{Dynamic approach for micromagnetics close to the Curie temperature}},\ }\href {https://doi.org/10.1103/PHYSREVB.74.094436/FIGURES/3/MEDIUM} {\bibfield  {journal} {\bibinfo  {journal} {Physical Review B - Condensed Matter and Materials Physics}\ }\textbf {\bibinfo {volume} {74}},\ \bibinfo {pages} {094436} (\bibinfo {year} {2006})}\BibitemShut {NoStop}%
\bibitem [{\citenamefont {Atxitia}\ \emph {et~al.}(2007)\citenamefont {Atxitia}, \citenamefont {Chubykalo-Fesenko}, \citenamefont {Kazantseva}, \citenamefont {Hinzke}, \citenamefont {Nowak},\ and\ \citenamefont {Chantrell}}]{Atxitia2007MicromagneticEquation}%
  \BibitemOpen
  \bibfield  {author} {\bibinfo {author} {\bibfnamefont {U.}~\bibnamefont {Atxitia}}, \bibinfo {author} {\bibfnamefont {O.}~\bibnamefont {Chubykalo-Fesenko}}, \bibinfo {author} {\bibfnamefont {N.}~\bibnamefont {Kazantseva}}, \bibinfo {author} {\bibfnamefont {D.}~\bibnamefont {Hinzke}}, \bibinfo {author} {\bibfnamefont {U.}~\bibnamefont {Nowak}},\ and\ \bibinfo {author} {\bibfnamefont {R.~W.}\ \bibnamefont {Chantrell}},\ }\bibfield  {title} {\bibinfo {title} {{Micromagnetic modeling of laser-induced magnetization dynamics using the Landau-Lifshitz-Bloch equation}},\ }\href {https://doi.org/10.1063/1.2822807} {\bibfield  {journal} {\bibinfo  {journal} {Applied Physics Letters}\ }\textbf {\bibinfo {volume} {91}},\ \bibinfo {pages} {232507} (\bibinfo {year} {2007})}\BibitemShut {NoStop}%
\bibitem [{\citenamefont {Newell}\ \emph {et~al.}(1993)\citenamefont {Newell}, \citenamefont {Williams},\ and\ \citenamefont {Dunlop}}]{Newell1993AMagnetization}%
  \BibitemOpen
  \bibfield  {author} {\bibinfo {author} {\bibfnamefont {A.~J.}\ \bibnamefont {Newell}}, \bibinfo {author} {\bibfnamefont {W.}~\bibnamefont {Williams}},\ and\ \bibinfo {author} {\bibfnamefont {D.~J.}\ \bibnamefont {Dunlop}},\ }\bibfield  {title} {\bibinfo {title} {{A generalization of the demagnetizing tensor for nonuniform magnetization}},\ }\href {https://doi.org/10.1029/93JB00694} {\bibfield  {journal} {\bibinfo  {journal} {Journal of Geophysical Research: Solid Earth}\ }\textbf {\bibinfo {volume} {98}},\ \bibinfo {pages} {9551} (\bibinfo {year} {1993})}\BibitemShut {NoStop}%
\bibitem [{\citenamefont {Callen}\ and\ \citenamefont {Callen}(1966)}]{Callen1966TheLawb}%
  \BibitemOpen
  \bibfield  {author} {\bibinfo {author} {\bibfnamefont {H.~B.}\ \bibnamefont {Callen}}\ and\ \bibinfo {author} {\bibfnamefont {E.}~\bibnamefont {Callen}},\ }\bibfield  {title} {\bibinfo {title} {{The present status of the temperature dependence of magnetocrystalline anisotropy, and the l(l+1)2 power law}},\ }\href {https://doi.org/10.1016/0022-3697(66)90012-6} {\bibfield  {journal} {\bibinfo  {journal} {Journal of Physics and Chemistry of Solids}\ }\textbf {\bibinfo {volume} {27}},\ \bibinfo {pages} {1271} (\bibinfo {year} {1966})}\BibitemShut {NoStop}%
\bibitem [{\citenamefont {Ivanov}\ \emph {et~al.}(2016)\citenamefont {Ivanov}, \citenamefont {Chuvilin}, \citenamefont {Vivas}, \citenamefont {Kosel}, \citenamefont {Chubykalo-Fesenko},\ and\ \citenamefont {V{\'{a}}zquez}}]{Ivanov2016SingleVortices}%
  \BibitemOpen
  \bibfield  {author} {\bibinfo {author} {\bibfnamefont {Y.~P.}\ \bibnamefont {Ivanov}}, \bibinfo {author} {\bibfnamefont {A.}~\bibnamefont {Chuvilin}}, \bibinfo {author} {\bibfnamefont {L.~G.}\ \bibnamefont {Vivas}}, \bibinfo {author} {\bibfnamefont {J.}~\bibnamefont {Kosel}}, \bibinfo {author} {\bibfnamefont {O.}~\bibnamefont {Chubykalo-Fesenko}},\ and\ \bibinfo {author} {\bibfnamefont {M.}~\bibnamefont {V{\'{a}}zquez}},\ }\bibfield  {title} {\bibinfo {title} {{Single crystalline cylindrical nanowires – toward dense 3D arrays of magnetic vortices}},\ }\href {https://doi.org/10.1038/srep23844} {\bibfield  {journal} {\bibinfo  {journal} {Scientific Reports 2016 6:1}\ }\textbf {\bibinfo {volume} {6}},\ \bibinfo {pages} {1} (\bibinfo {year} {2016})}\BibitemShut {NoStop}%
\bibitem [{\citenamefont {Moreno}\ \emph {et~al.}(2016)\citenamefont {Moreno}, \citenamefont {Evans}, \citenamefont {Khmelevskyi}, \citenamefont {Mu{\~{n}}oz}, \citenamefont {Chantrell},\ and\ \citenamefont {Chubykalo-Fesenko}}]{Moreno2016Temperature-dependentCo}%
  \BibitemOpen
  \bibfield  {author} {\bibinfo {author} {\bibfnamefont {R.}~\bibnamefont {Moreno}}, \bibinfo {author} {\bibfnamefont {R.~F.}\ \bibnamefont {Evans}}, \bibinfo {author} {\bibfnamefont {S.}~\bibnamefont {Khmelevskyi}}, \bibinfo {author} {\bibfnamefont {M.~C.}\ \bibnamefont {Mu{\~{n}}oz}}, \bibinfo {author} {\bibfnamefont {R.~W.}\ \bibnamefont {Chantrell}},\ and\ \bibinfo {author} {\bibfnamefont {O.}~\bibnamefont {Chubykalo-Fesenko}},\ }\bibfield  {title} {\bibinfo {title} {{Temperature-dependent exchange stiffness and domain wall width in Co}},\ }\href {https://doi.org/10.1103/PHYSREVB.94.104433/FIGURES/3/MEDIUM} {\bibfield  {journal} {\bibinfo  {journal} {Physical Review B}\ }\textbf {\bibinfo {volume} {94}},\ \bibinfo {pages} {104433} (\bibinfo {year} {2016})}\BibitemShut {NoStop}%
\bibitem [{\citenamefont {Chen}\ \emph {et~al.}(2019)\citenamefont {Chen}, \citenamefont {Yan}, \citenamefont {Qin},\ and\ \citenamefont {Liu}}]{Chen2019Landau-Lifshitz-BlochAntiferromagnets}%
  \BibitemOpen
  \bibfield  {author} {\bibinfo {author} {\bibfnamefont {Z.~Y.}\ \bibnamefont {Chen}}, \bibinfo {author} {\bibfnamefont {Z.~R.}\ \bibnamefont {Yan}}, \bibinfo {author} {\bibfnamefont {M.~H.}\ \bibnamefont {Qin}},\ and\ \bibinfo {author} {\bibfnamefont {J.~M.}\ \bibnamefont {Liu}},\ }\bibfield  {title} {\bibinfo {title} {{Landau-Lifshitz-Bloch equation for domain wall motion in antiferromagnets}},\ }\href {https://doi.org/10.1103/PHYSREVB.99.214436/FIGURES/4/MEDIUM} {\bibfield  {journal} {\bibinfo  {journal} {Physical Review B}\ }\textbf {\bibinfo {volume} {99}},\ \bibinfo {pages} {214436} (\bibinfo {year} {2019})}\BibitemShut {NoStop}%
\bibitem [{\citenamefont {Abo}\ \emph {et~al.}(2013)\citenamefont {Abo}, \citenamefont {Hong}, \citenamefont {Park}, \citenamefont {Lee}, \citenamefont {Lee},\ and\ \citenamefont {Choi}}]{Abo2013DefinitionLength}%
  \BibitemOpen
  \bibfield  {author} {\bibinfo {author} {\bibfnamefont {G.~S.}\ \bibnamefont {Abo}}, \bibinfo {author} {\bibfnamefont {Y.~K.}\ \bibnamefont {Hong}}, \bibinfo {author} {\bibfnamefont {J.}~\bibnamefont {Park}}, \bibinfo {author} {\bibfnamefont {J.}~\bibnamefont {Lee}}, \bibinfo {author} {\bibfnamefont {W.}~\bibnamefont {Lee}},\ and\ \bibinfo {author} {\bibfnamefont {B.~C.}\ \bibnamefont {Choi}},\ }\bibfield  {title} {\bibinfo {title} {{Definition of magnetic exchange length}},\ }\href {https://doi.org/10.1109/TMAG.2013.2258028} {\bibfield  {journal} {\bibinfo  {journal} {IEEE Transactions on Magnetics}\ }\textbf {\bibinfo {volume} {49}},\ \bibinfo {pages} {4937} (\bibinfo {year} {2013})}\BibitemShut {NoStop}%
\bibitem [{\citenamefont {Khorsand}\ \emph {et~al.}(2012)\citenamefont {Khorsand}, \citenamefont {Savoini}, \citenamefont {Kirilyuk}, \citenamefont {Kimel}, \citenamefont {Tsukamoto}, \citenamefont {Itoh},\ and\ \citenamefont {Rasing}}]{Khorsand2012RoleRecording}%
  \BibitemOpen
  \bibfield  {author} {\bibinfo {author} {\bibfnamefont {A.~R.}\ \bibnamefont {Khorsand}}, \bibinfo {author} {\bibfnamefont {M.}~\bibnamefont {Savoini}}, \bibinfo {author} {\bibfnamefont {A.}~\bibnamefont {Kirilyuk}}, \bibinfo {author} {\bibfnamefont {A.~V.}\ \bibnamefont {Kimel}}, \bibinfo {author} {\bibfnamefont {A.}~\bibnamefont {Tsukamoto}}, \bibinfo {author} {\bibfnamefont {A.}~\bibnamefont {Itoh}},\ and\ \bibinfo {author} {\bibfnamefont {T.}~\bibnamefont {Rasing}},\ }\bibfield  {title} {\bibinfo {title} {{Role of magnetic circular dichroism in all-optical magnetic recording}},\ }\href {https://doi.org/10.1103/PHYSREVLETT.108.127205/FIGURES/4/MEDIUM} {\bibfield  {journal} {\bibinfo  {journal} {Physical Review Letters}\ }\textbf {\bibinfo {volume} {108}},\ \bibinfo {pages} {127205} (\bibinfo {year} {2012})}\BibitemShut {NoStop}%
\bibitem [{\citenamefont {Quessab}\ \emph {et~al.}(2019)\citenamefont {Quessab}, \citenamefont {Deb}, \citenamefont {Gorchon}, \citenamefont {Hehn}, \citenamefont {Malinowski},\ and\ \citenamefont {Mangin}}]{Quessab2019ResolvingSwitching}%
  \BibitemOpen
  \bibfield  {author} {\bibinfo {author} {\bibfnamefont {Y.}~\bibnamefont {Quessab}}, \bibinfo {author} {\bibfnamefont {M.}~\bibnamefont {Deb}}, \bibinfo {author} {\bibfnamefont {J.}~\bibnamefont {Gorchon}}, \bibinfo {author} {\bibfnamefont {M.}~\bibnamefont {Hehn}}, \bibinfo {author} {\bibfnamefont {G.}~\bibnamefont {Malinowski}},\ and\ \bibinfo {author} {\bibfnamefont {S.}~\bibnamefont {Mangin}},\ }\bibfield  {title} {\bibinfo {title} {{Resolving the role of magnetic circular dichroism in multishot helicity-dependent all-optical switching}},\ }\href {https://doi.org/10.1103/PHYSREVB.100.024425/FIGURES/5/MEDIUM} {\bibfield  {journal} {\bibinfo  {journal} {Physical Review B}\ }\textbf {\bibinfo {volume} {100}},\ \bibinfo {pages} {024425} (\bibinfo {year} {2019})}\BibitemShut {NoStop}%
\bibitem [{\citenamefont {Tatara}\ and\ \citenamefont {Otxoa De~Zuazola}(2020)}]{Tatara2020CollectiveWall}%
  \BibitemOpen
  \bibfield  {author} {\bibinfo {author} {\bibfnamefont {G.}~\bibnamefont {Tatara}}\ and\ \bibinfo {author} {\bibfnamefont {R.~M.}\ \bibnamefont {Otxoa De~Zuazola}},\ }\bibfield  {title} {\bibinfo {title} {{Collective coordinate study of spin-wave emission from a dynamic domain wall}},\ }\href {https://doi.org/10.1103/PHYSREVB.101.224425/FIGURES/5/MEDIUM} {\bibfield  {journal} {\bibinfo  {journal} {Physical Review B}\ }\textbf {\bibinfo {volume} {101}},\ \bibinfo {pages} {224425} (\bibinfo {year} {2020})}\BibitemShut {NoStop}%
\bibitem [{See()}]{SeeGeometry}%
  \BibitemOpen
  \href@noop {} {\bibinfo {title} {{See Supplemental Material at [] for analysis of time-dependent DW profile in the chain geometry}}}\BibitemShut {NoStop}%
\bibitem [{\citenamefont {Nieves}\ \emph {et~al.}(2014)\citenamefont {Nieves}, \citenamefont {Serantes}, \citenamefont {Atxitia},\ and\ \citenamefont {Chubykalo-Fesenko}}]{Nieves2014QuantumCase}%
  \BibitemOpen
  \bibfield  {author} {\bibinfo {author} {\bibfnamefont {P.}~\bibnamefont {Nieves}}, \bibinfo {author} {\bibfnamefont {D.}~\bibnamefont {Serantes}}, \bibinfo {author} {\bibfnamefont {U.}~\bibnamefont {Atxitia}},\ and\ \bibinfo {author} {\bibfnamefont {O.}~\bibnamefont {Chubykalo-Fesenko}},\ }\bibfield  {title} {\bibinfo {title} {{Quantum Landau-Lifshitz-Bloch equation and its comparison with the classical case}},\ }\href {https://doi.org/10.1103/PHYSREVB.90.104428/FIGURES/7/MEDIUM} {\bibfield  {journal} {\bibinfo  {journal} {Physical Review B - Condensed Matter and Materials Physics}\ }\textbf {\bibinfo {volume} {90}},\ \bibinfo {pages} {104428} (\bibinfo {year} {2014})}\BibitemShut {NoStop}%
\bibitem [{\citenamefont {Chen}\ and\ \citenamefont {Landau}(1994)}]{Chen1994Spin-dynamicsFerromagnet}%
  \BibitemOpen
  \bibfield  {author} {\bibinfo {author} {\bibfnamefont {K.}~\bibnamefont {Chen}}\ and\ \bibinfo {author} {\bibfnamefont {D.~P.}\ \bibnamefont {Landau}},\ }\bibfield  {title} {\bibinfo {title} {{Spin-dynamics study of the dynamic critical behavior of the three-dimensional classical Heisenberg ferromagnet}},\ }\href {https://doi.org/10.1103/PhysRevB.49.3266} {\bibfield  {journal} {\bibinfo  {journal} {Physical Review B}\ }\textbf {\bibinfo {volume} {49}},\ \bibinfo {pages} {3266} (\bibinfo {year} {1994})}\BibitemShut {NoStop}%
\bibitem [{\citenamefont {Kaganov}\ \emph {et~al.}(1957)\citenamefont {Kaganov}, \citenamefont {Lifshitz},\ and\ \citenamefont {Tanatarov}}]{Kaganov1957RelaxationLattice}%
  \BibitemOpen
  \bibfield  {author} {\bibinfo {author} {\bibfnamefont {M.}~\bibnamefont {Kaganov}}, \bibinfo {author} {\bibfnamefont {I.}~\bibnamefont {Lifshitz}},\ and\ \bibinfo {author} {\bibfnamefont {L.}~\bibnamefont {Tanatarov}},\ }\bibfield  {title} {\bibinfo {title} {{Relaxation between Electrons and the Crystalline Lattice}},\ }\href {http://www.jetp.ras.ru/cgi-bin/e/index/e/4/2/p173?a=list} {\bibfield  {journal} {\bibinfo  {journal} {Journal of Experimental and Theoretical Physics}\ }\textbf {\bibinfo {volume} {4}} (\bibinfo {year} {1957})}\BibitemShut {NoStop}%
\bibitem [{\citenamefont {Lifshits}\ \emph {et~al.}(1960)\citenamefont {Lifshits}, \citenamefont {Kaganov},\ and\ \citenamefont {Tanatarov}}]{Lifshits1960OnRadiation}%
  \BibitemOpen
  \bibfield  {author} {\bibinfo {author} {\bibfnamefont {I.~M.}\ \bibnamefont {Lifshits}}, \bibinfo {author} {\bibfnamefont {M.~I.}\ \bibnamefont {Kaganov}},\ and\ \bibinfo {author} {\bibfnamefont {L.~V.}\ \bibnamefont {Tanatarov}},\ }\bibfield  {title} {\bibinfo {title} {{On the theory of the changes produced in metals by radiation}},\ }\href {https://doi.org/10.1007/BF01479732} {\bibfield  {journal} {\bibinfo  {journal} {The Soviet Journal of Atomic Energy 1960 6:4}\ }\textbf {\bibinfo {volume} {6}},\ \bibinfo {pages} {261} (\bibinfo {year} {1960})}\BibitemShut {NoStop}%
\bibitem [{\citenamefont {Anisimov}\ \emph {et~al.}(1974)\citenamefont {Anisimov}, \citenamefont {Kapeliovich}, \citenamefont {Perelman}, \citenamefont {Anisimov}, \citenamefont {Kapeliovich},\ and\ \citenamefont {Perelman}}]{Anisimov1974ElectronPulses}%
  \BibitemOpen
  \bibfield  {author} {\bibinfo {author} {\bibfnamefont {S.~I.}\ \bibnamefont {Anisimov}}, \bibinfo {author} {\bibfnamefont {B.~L.}\ \bibnamefont {Kapeliovich}}, \bibinfo {author} {\bibfnamefont {T.~L.}\ \bibnamefont {Perelman}}, \bibinfo {author} {\bibfnamefont {S.~I.}\ \bibnamefont {Anisimov}}, \bibinfo {author} {\bibfnamefont {B.~L.}\ \bibnamefont {Kapeliovich}},\ and\ \bibinfo {author} {\bibfnamefont {T.~L.}\ \bibnamefont {Perelman}},\ }\bibfield  {title} {\bibinfo {title} {{Electron emission from metal surfaces exposed to ultrashort laser pulses}},\ }\href {https://ui.adsabs.harvard.edu/abs/1974ZhETF..66..776A/abstract} {\bibfield  {journal} {\bibinfo  {journal} {ZhETF}\ }\textbf {\bibinfo {volume} {66}},\ \bibinfo {pages} {776} (\bibinfo {year} {1974})}\BibitemShut {NoStop}%
\bibitem [{\citenamefont {Koopmans}\ \emph {et~al.}(2009)\citenamefont {Koopmans}, \citenamefont {Malinowski}, \citenamefont {Dalla~Longa}, \citenamefont {Steiauf}, \citenamefont {F{\"{a}}hnle}, \citenamefont {Roth}, \citenamefont {Cinchetti},\ and\ \citenamefont {Aeschlimann}}]{Koopmans2009ExplainingDemagnetization}%
  \BibitemOpen
  \bibfield  {author} {\bibinfo {author} {\bibfnamefont {B.}~\bibnamefont {Koopmans}}, \bibinfo {author} {\bibfnamefont {G.}~\bibnamefont {Malinowski}}, \bibinfo {author} {\bibfnamefont {F.}~\bibnamefont {Dalla~Longa}}, \bibinfo {author} {\bibfnamefont {D.}~\bibnamefont {Steiauf}}, \bibinfo {author} {\bibfnamefont {M.}~\bibnamefont {F{\"{a}}hnle}}, \bibinfo {author} {\bibfnamefont {T.}~\bibnamefont {Roth}}, \bibinfo {author} {\bibfnamefont {M.}~\bibnamefont {Cinchetti}},\ and\ \bibinfo {author} {\bibfnamefont {M.}~\bibnamefont {Aeschlimann}},\ }\bibfield  {title} {\bibinfo {title} {{Explaining the paradoxical diversity of ultrafast laser-induced demagnetization}},\ }\href {https://doi.org/10.1038/nmat2593} {\bibfield  {journal} {\bibinfo  {journal} {Nature Materials 2010 9:3}\ }\textbf {\bibinfo {volume} {9}},\ \bibinfo {pages} {259} (\bibinfo {year} {2009})}\BibitemShut {NoStop}%
\bibitem [{\citenamefont {Chimata}\ \emph {et~al.}(2012)\citenamefont {Chimata}, \citenamefont {Bergman}, \citenamefont {Bergqvist}, \citenamefont {Sanyal},\ and\ \citenamefont {Eriksson}}]{Chimata2012MicroscopicDynamics}%
  \BibitemOpen
  \bibfield  {author} {\bibinfo {author} {\bibfnamefont {R.}~\bibnamefont {Chimata}}, \bibinfo {author} {\bibfnamefont {A.}~\bibnamefont {Bergman}}, \bibinfo {author} {\bibfnamefont {L.}~\bibnamefont {Bergqvist}}, \bibinfo {author} {\bibfnamefont {B.}~\bibnamefont {Sanyal}},\ and\ \bibinfo {author} {\bibfnamefont {O.}~\bibnamefont {Eriksson}},\ }\bibfield  {title} {\bibinfo {title} {{Microscopic model for ultrafast remagnetization dynamics}},\ }\href {https://doi.org/10.1103/PHYSREVLETT.109.157201/FIGURES/3/MEDIUM} {\bibfield  {journal} {\bibinfo  {journal} {Physical Review Letters}\ }\textbf {\bibinfo {volume} {109}},\ \bibinfo {pages} {157201} (\bibinfo {year} {2012})}\BibitemShut {NoStop}%
\bibitem [{\citenamefont {Dzyaloshinsky}(1958)}]{Dzyaloshinsky1958AAntiferromagnetics}%
  \BibitemOpen
  \bibfield  {author} {\bibinfo {author} {\bibfnamefont {I.}~\bibnamefont {Dzyaloshinsky}},\ }\bibfield  {title} {\bibinfo {title} {{A thermodynamic theory of “weak” ferromagnetism of antiferromagnetics}},\ }\href {https://doi.org/10.1016/0022-3697(58)90076-3} {\bibfield  {journal} {\bibinfo  {journal} {Journal of Physics and Chemistry of Solids}\ }\textbf {\bibinfo {volume} {4}},\ \bibinfo {pages} {241} (\bibinfo {year} {1958})}\BibitemShut {NoStop}%
\bibitem [{\citenamefont {Moriya}(1960)}]{Moriya1960AnisotropicFerromagnetism}%
  \BibitemOpen
  \bibfield  {author} {\bibinfo {author} {\bibfnamefont {T.}~\bibnamefont {Moriya}},\ }\bibfield  {title} {\bibinfo {title} {{Anisotropic Superexchange Interaction and Weak Ferromagnetism}},\ }\href {https://doi.org/10.1103/PhysRev.120.91} {\bibfield  {journal} {\bibinfo  {journal} {Physical Review}\ }\textbf {\bibinfo {volume} {120}},\ \bibinfo {pages} {91} (\bibinfo {year} {1960})}\BibitemShut {NoStop}%
\bibitem [{\citenamefont {Brand{\~{a}}o}\ \emph {et~al.}(2017)\citenamefont {Brand{\~{a}}o}, \citenamefont {Azzawi}, \citenamefont {Hindmarch},\ and\ \citenamefont {Atkinson}}]{Brandao2017UnderstandingNanowires}%
  \BibitemOpen
  \bibfield  {author} {\bibinfo {author} {\bibfnamefont {J.}~\bibnamefont {Brand{\~{a}}o}}, \bibinfo {author} {\bibfnamefont {S.}~\bibnamefont {Azzawi}}, \bibinfo {author} {\bibfnamefont {A.~T.}\ \bibnamefont {Hindmarch}},\ and\ \bibinfo {author} {\bibfnamefont {D.}~\bibnamefont {Atkinson}},\ }\bibfield  {title} {\bibinfo {title} {{Understanding the role of damping and Dzyaloshinskii-Moriya interaction on dynamic domain wall behaviour in platinum-ferromagnet nanowires}},\ }\href {https://doi.org/10.1038/s41598-017-04088-8} {\bibfield  {journal} {\bibinfo  {journal} {Scientific Reports 2017 7:1}\ }\textbf {\bibinfo {volume} {7}},\ \bibinfo {pages} {1} (\bibinfo {year} {2017})}\BibitemShut {NoStop}%
\bibitem [{\citenamefont {Thiaville}\ \emph {et~al.}(2012)\citenamefont {Thiaville}, \citenamefont {Rohart}, \citenamefont {Ju{\'{e}}}, \citenamefont {Cros},\ and\ \citenamefont {Fert}}]{Thiaville2012DynamicsFilms}%
  \BibitemOpen
  \bibfield  {author} {\bibinfo {author} {\bibfnamefont {A.}~\bibnamefont {Thiaville}}, \bibinfo {author} {\bibfnamefont {S.}~\bibnamefont {Rohart}}, \bibinfo {author} {\bibfnamefont {E.}~\bibnamefont {Ju{\'{e}}}}, \bibinfo {author} {\bibfnamefont {V.}~\bibnamefont {Cros}},\ and\ \bibinfo {author} {\bibfnamefont {A.}~\bibnamefont {Fert}},\ }\bibfield  {title} {\bibinfo {title} {{Dynamics of Dzyaloshinskii domain walls in ultrathin magnetic films}},\ }\href {https://doi.org/10.1209/0295-5075/100/57002} {\bibfield  {journal} {\bibinfo  {journal} {Europhysics Letters}\ }\textbf {\bibinfo {volume} {100}},\ \bibinfo {pages} {57002} (\bibinfo {year} {2012})}\BibitemShut {NoStop}%
\bibitem [{\citenamefont {Yoshimura}\ \emph {et~al.}(2015)\citenamefont {Yoshimura}, \citenamefont {Kim}, \citenamefont {Taniguchi}, \citenamefont {Tono}, \citenamefont {Ueda}, \citenamefont {Hiramatsu}, \citenamefont {Moriyama}, \citenamefont {Yamada}, \citenamefont {Nakatani},\ and\ \citenamefont {Ono}}]{Yoshimura2015Soliton-likeDzyaloshinskiiMoriyainteraction}%
  \BibitemOpen
  \bibfield  {author} {\bibinfo {author} {\bibfnamefont {Y.}~\bibnamefont {Yoshimura}}, \bibinfo {author} {\bibfnamefont {K.~J.}\ \bibnamefont {Kim}}, \bibinfo {author} {\bibfnamefont {T.}~\bibnamefont {Taniguchi}}, \bibinfo {author} {\bibfnamefont {T.}~\bibnamefont {Tono}}, \bibinfo {author} {\bibfnamefont {K.}~\bibnamefont {Ueda}}, \bibinfo {author} {\bibfnamefont {R.}~\bibnamefont {Hiramatsu}}, \bibinfo {author} {\bibfnamefont {T.}~\bibnamefont {Moriyama}}, \bibinfo {author} {\bibfnamefont {K.}~\bibnamefont {Yamada}}, \bibinfo {author} {\bibfnamefont {Y.}~\bibnamefont {Nakatani}},\ and\ \bibinfo {author} {\bibfnamefont {T.}~\bibnamefont {Ono}},\ }\bibfield  {title} {\bibinfo {title} {{Soliton-like magnetic domain wall motion induced by the interfacial Dzyaloshinskii–Moriya interaction}},\ }\href {https://doi.org/10.1038/nphys3535} {\bibfield  {journal} {\bibinfo  {journal} {Nature Physics 2015 12:2}\ }\textbf {\bibinfo {volume} {12}},\ \bibinfo {pages} {157} (\bibinfo {year} {2015})}\BibitemShut
  {NoStop}%
\bibitem [{\citenamefont {Wang}\ \emph {et~al.}(2015)\citenamefont {Wang}, \citenamefont {Albert}, \citenamefont {Beg}, \citenamefont {Bisotti}, \citenamefont {Chernyshenko}, \citenamefont {Cort{\'{e}}s-Ortu{\~{n}}o}, \citenamefont {Hawke},\ and\ \citenamefont {Fangohr}}]{Wang2015Magnon-drivenInteraction}%
  \BibitemOpen
  \bibfield  {author} {\bibinfo {author} {\bibfnamefont {W.}~\bibnamefont {Wang}}, \bibinfo {author} {\bibfnamefont {M.}~\bibnamefont {Albert}}, \bibinfo {author} {\bibfnamefont {M.}~\bibnamefont {Beg}}, \bibinfo {author} {\bibfnamefont {M.~A.}\ \bibnamefont {Bisotti}}, \bibinfo {author} {\bibfnamefont {D.}~\bibnamefont {Chernyshenko}}, \bibinfo {author} {\bibfnamefont {D.}~\bibnamefont {Cort{\'{e}}s-Ortu{\~{n}}o}}, \bibinfo {author} {\bibfnamefont {I.}~\bibnamefont {Hawke}},\ and\ \bibinfo {author} {\bibfnamefont {H.}~\bibnamefont {Fangohr}},\ }\bibfield  {title} {\bibinfo {title} {{Magnon-driven domain-wall motion with the Dzyaloshinskii-Moriya interaction}},\ }\href {https://doi.org/10.1103/PHYSREVLETT.114.087203/FIGURES/4/MEDIUM} {\bibfield  {journal} {\bibinfo  {journal} {Physical Review Letters}\ }\textbf {\bibinfo {volume} {114}},\ \bibinfo {pages} {087203} (\bibinfo {year} {2015})}\BibitemShut {NoStop}%
\bibitem [{\citenamefont {Cort{\'{e}}s-Ortu{\~{n}}o}\ \emph {et~al.}(2018)\citenamefont {Cort{\'{e}}s-Ortu{\~{n}}o}, \citenamefont {Beg}, \citenamefont {Nehruji}, \citenamefont {Breth}, \citenamefont {Pepper}, \citenamefont {Kluyver}, \citenamefont {Downing}, \citenamefont {Hesjedal}, \citenamefont {Hatton}, \citenamefont {Lancaster}, \citenamefont {Hertel}, \citenamefont {Hovorka},\ and\ \citenamefont {Fangohr}}]{Cortes-Ortuno2018ProposalInteraction}%
  \BibitemOpen
  \bibfield  {author} {\bibinfo {author} {\bibfnamefont {D.}~\bibnamefont {Cort{\'{e}}s-Ortu{\~{n}}o}}, \bibinfo {author} {\bibfnamefont {M.}~\bibnamefont {Beg}}, \bibinfo {author} {\bibfnamefont {V.}~\bibnamefont {Nehruji}}, \bibinfo {author} {\bibfnamefont {L.}~\bibnamefont {Breth}}, \bibinfo {author} {\bibfnamefont {R.}~\bibnamefont {Pepper}}, \bibinfo {author} {\bibfnamefont {T.}~\bibnamefont {Kluyver}}, \bibinfo {author} {\bibfnamefont {G.}~\bibnamefont {Downing}}, \bibinfo {author} {\bibfnamefont {T.}~\bibnamefont {Hesjedal}}, \bibinfo {author} {\bibfnamefont {P.}~\bibnamefont {Hatton}}, \bibinfo {author} {\bibfnamefont {T.}~\bibnamefont {Lancaster}}, \bibinfo {author} {\bibfnamefont {R.}~\bibnamefont {Hertel}}, \bibinfo {author} {\bibfnamefont {O.}~\bibnamefont {Hovorka}},\ and\ \bibinfo {author} {\bibfnamefont {H.}~\bibnamefont {Fangohr}},\ }\bibfield  {title} {\bibinfo {title} {{Proposal for a micromagnetic standard problem for materials with Dzyaloshinskii–Moriya interaction}},\ }\href
  {https://doi.org/10.1088/1367-2630/AAEA1C} {\bibfield  {journal} {\bibinfo  {journal} {New Journal of Physics}\ }\textbf {\bibinfo {volume} {20}},\ \bibinfo {pages} {113015} (\bibinfo {year} {2018})}\BibitemShut {NoStop}%
\bibitem [{\citenamefont {Lepadatu}(2020)}]{Lepadatu2020EmergenceDemagnetization}%
  \BibitemOpen
  \bibfield  {author} {\bibinfo {author} {\bibfnamefont {S.}~\bibnamefont {Lepadatu}},\ }\bibfield  {title} {\bibinfo {title} {{Emergence of transient domain wall skyrmions after ultrafast demagnetization}},\ }\href {https://doi.org/10.1103/PHYSREVB.102.094402/FIGURES/8/MEDIUM} {\bibfield  {journal} {\bibinfo  {journal} {Physical Review B}\ }\textbf {\bibinfo {volume} {102}},\ \bibinfo {pages} {094402} (\bibinfo {year} {2020})}\BibitemShut {NoStop}%
\bibitem [{\citenamefont {Yan}\ \emph {et~al.}(2011)\citenamefont {Yan}, \citenamefont {Wang},\ and\ \citenamefont {Wang}}]{Yan2011All-magnonicPropagation}%
  \BibitemOpen
  \bibfield  {author} {\bibinfo {author} {\bibfnamefont {P.}~\bibnamefont {Yan}}, \bibinfo {author} {\bibfnamefont {X.~S.}\ \bibnamefont {Wang}},\ and\ \bibinfo {author} {\bibfnamefont {X.~R.}\ \bibnamefont {Wang}},\ }\bibfield  {title} {\bibinfo {title} {{All-magnonic spin-transfer torque and domain wall propagation}},\ }\href {https://doi.org/10.1103/PHYSREVLETT.107.177207/FIGURES/4/MEDIUM} {\bibfield  {journal} {\bibinfo  {journal} {Physical Review Letters}\ }\textbf {\bibinfo {volume} {107}},\ \bibinfo {pages} {177207} (\bibinfo {year} {2011})}\BibitemShut {NoStop}%
\bibitem [{\citenamefont {Kovalev}\ and\ \citenamefont {Tserkovnyak}(2012)}]{Kovalev2012ThermomagnonicIOPscience}%
  \BibitemOpen
  \bibfield  {author} {\bibinfo {author} {\bibfnamefont {A.~A.}\ \bibnamefont {Kovalev}}\ and\ \bibinfo {author} {\bibfnamefont {Y.}~\bibnamefont {Tserkovnyak}},\ }\bibfield  {title} {\bibinfo {title} {{Thermomagnonic spin transfer and Peltier effects in insulating magnets - IOPscience}},\ }\href {https://iopscience.iop.org/article/10.1209/0295-5075/97/67002} {\bibfield  {journal} {\bibinfo  {journal} {Europhysics Letters}\ }\textbf {\bibinfo {volume} {97}} (\bibinfo {year} {2012})}\BibitemShut {NoStop}%
\bibitem [{\citenamefont {Yan}\ \emph {et~al.}(2013)\citenamefont {Yan}, \citenamefont {Kamra}, \citenamefont {Cao},\ and\ \citenamefont {Bauer}}]{Yan2013AngularFerromagnets}%
  \BibitemOpen
  \bibfield  {author} {\bibinfo {author} {\bibfnamefont {P.}~\bibnamefont {Yan}}, \bibinfo {author} {\bibfnamefont {A.}~\bibnamefont {Kamra}}, \bibinfo {author} {\bibfnamefont {Y.}~\bibnamefont {Cao}},\ and\ \bibinfo {author} {\bibfnamefont {G.~E.}\ \bibnamefont {Bauer}},\ }\bibfield  {title} {\bibinfo {title} {{Angular and linear momentum of excited ferromagnets}},\ }\href {https://doi.org/10.1103/PHYSREVB.88.144413/FIGURES/2/MEDIUM} {\bibfield  {journal} {\bibinfo  {journal} {Physical Review B - Condensed Matter and Materials Physics}\ }\textbf {\bibinfo {volume} {88}},\ \bibinfo {pages} {144413} (\bibinfo {year} {2013})}\BibitemShut {NoStop}%
\bibitem [{\citenamefont {Kim}\ and\ \citenamefont {Tserkovnyak}(2015)}]{Kim2015Landau-LifshitzTorque}%
  \BibitemOpen
  \bibfield  {author} {\bibinfo {author} {\bibfnamefont {S.~K.}\ \bibnamefont {Kim}}\ and\ \bibinfo {author} {\bibfnamefont {Y.}~\bibnamefont {Tserkovnyak}},\ }\bibfield  {title} {\bibinfo {title} {{Landau-Lifshitz theory of thermomagnonic torque}},\ }\href {https://doi.org/10.1103/PHYSREVB.92.020410/FIGURES/1/MEDIUM} {\bibfield  {journal} {\bibinfo  {journal} {Physical Review B - Condensed Matter and Materials Physics}\ }\textbf {\bibinfo {volume} {92}},\ \bibinfo {pages} {020410} (\bibinfo {year} {2015})}\BibitemShut {NoStop}%
\bibitem [{\citenamefont {Garcia-Sanchez}\ \emph {et~al.}(2014)\citenamefont {Garcia-Sanchez}, \citenamefont {Borys}, \citenamefont {Vansteenkiste}, \citenamefont {Kim},\ and\ \citenamefont {Stamps}}]{Garcia-Sanchez2014NonreciprocalInteractionb}%
  \BibitemOpen
  \bibfield  {author} {\bibinfo {author} {\bibfnamefont {F.}~\bibnamefont {Garcia-Sanchez}}, \bibinfo {author} {\bibfnamefont {P.}~\bibnamefont {Borys}}, \bibinfo {author} {\bibfnamefont {A.}~\bibnamefont {Vansteenkiste}}, \bibinfo {author} {\bibfnamefont {J.~V.}\ \bibnamefont {Kim}},\ and\ \bibinfo {author} {\bibfnamefont {R.~L.}\ \bibnamefont {Stamps}},\ }\bibfield  {title} {\bibinfo {title} {{Nonreciprocal spin-wave channeling along textures driven by the Dzyaloshinskii-Moriya interaction}},\ }\href {https://doi.org/10.1103/PHYSREVB.89.224408/FIGURES/4/MEDIUM} {\bibfield  {journal} {\bibinfo  {journal} {Physical Review B - Condensed Matter and Materials Physics}\ }\textbf {\bibinfo {volume} {89}},\ \bibinfo {pages} {224408} (\bibinfo {year} {2014})}\BibitemShut {NoStop}%
\bibitem [{\citenamefont {Evans}\ \emph {et~al.}(2014)\citenamefont {Evans}, \citenamefont {Fan}, \citenamefont {Chureemart}, \citenamefont {Ostler}, \citenamefont {Ellis},\ and\ \citenamefont {Chantrell}}]{Evans2014AtomisticNanomaterials}%
  \BibitemOpen
  \bibfield  {author} {\bibinfo {author} {\bibfnamefont {R.~F.~L.}\ \bibnamefont {Evans}}, \bibinfo {author} {\bibfnamefont {W.~J.}\ \bibnamefont {Fan}}, \bibinfo {author} {\bibfnamefont {P.}~\bibnamefont {Chureemart}}, \bibinfo {author} {\bibfnamefont {T.~A.}\ \bibnamefont {Ostler}}, \bibinfo {author} {\bibfnamefont {M.~O.~A.}\ \bibnamefont {Ellis}},\ and\ \bibinfo {author} {\bibfnamefont {R.~W.}\ \bibnamefont {Chantrell}},\ }\bibfield  {title} {\bibinfo {title} {{Atomistic spin model simulations of magnetic nanomaterials}},\ }\href {https://doi.org/10.1088/0953-8984/26/10/103202} {\bibfield  {journal} {\bibinfo  {journal} {Journal of Physics: Condensed Matter}\ }\textbf {\bibinfo {volume} {26}},\ \bibinfo {pages} {103202} (\bibinfo {year} {2014})}\BibitemShut {NoStop}%
\bibitem [{\citenamefont {Pershan}\ \emph {et~al.}(1966)\citenamefont {Pershan}, \citenamefont {Van Der~Ziel},\ and\ \citenamefont {Malmstrom}}]{Pershan1966TheoreticalPhenomena}%
  \BibitemOpen
  \bibfield  {author} {\bibinfo {author} {\bibfnamefont {P.~S.}\ \bibnamefont {Pershan}}, \bibinfo {author} {\bibfnamefont {J.~P.}\ \bibnamefont {Van Der~Ziel}},\ and\ \bibinfo {author} {\bibfnamefont {L.~D.}\ \bibnamefont {Malmstrom}},\ }\bibfield  {title} {\bibinfo {title} {{Theoretical Discussion of the Inverse Faraday Effect, Raman Scattering, and Related Phenomena}},\ }\href {https://doi.org/10.1103/PhysRev.143.574} {\bibfield  {journal} {\bibinfo  {journal} {Physical Review}\ }\textbf {\bibinfo {volume} {143}},\ \bibinfo {pages} {574} (\bibinfo {year} {1966})}\BibitemShut {NoStop}%
\bibitem [{\citenamefont {Belotelov}\ \emph {et~al.}(2010)\citenamefont {Belotelov}, \citenamefont {Bezus}, \citenamefont {Doskolovich}, \citenamefont {Kalish},\ and\ \citenamefont {Zvezdin}}]{Belotelov2010InverseHeterostructures}%
  \BibitemOpen
  \bibfield  {author} {\bibinfo {author} {\bibfnamefont {V.~I.}\ \bibnamefont {Belotelov}}, \bibinfo {author} {\bibfnamefont {E.~A.}\ \bibnamefont {Bezus}}, \bibinfo {author} {\bibfnamefont {L.~L.}\ \bibnamefont {Doskolovich}}, \bibinfo {author} {\bibfnamefont {A.~N.}\ \bibnamefont {Kalish}},\ and\ \bibinfo {author} {\bibfnamefont {A.~K.}\ \bibnamefont {Zvezdin}},\ }\bibfield  {title} {\bibinfo {title} {{Inverse Faraday effect in plasmonic heterostructures}},\ }\href {https://doi.org/10.1088/1742-6596/200/9/092003} {\bibfield  {journal} {\bibinfo  {journal} {Journal of Physics: Conference Series}\ }\textbf {\bibinfo {volume} {200}},\ \bibinfo {pages} {092003} (\bibinfo {year} {2010})}\BibitemShut {NoStop}%
\bibitem [{\citenamefont {Freimuth}\ \emph {et~al.}(2021)\citenamefont {Freimuth}, \citenamefont {Bl{\"{u}}gel},\ and\ \citenamefont {Mokrousov}}]{Freimuth2021Laser-inducedSpirals}%
  \BibitemOpen
  \bibfield  {author} {\bibinfo {author} {\bibfnamefont {F.}~\bibnamefont {Freimuth}}, \bibinfo {author} {\bibfnamefont {S.}~\bibnamefont {Bl{\"{u}}gel}},\ and\ \bibinfo {author} {\bibfnamefont {Y.}~\bibnamefont {Mokrousov}},\ }\bibfield  {title} {\bibinfo {title} {{Laser-induced torques in spin spirals}},\ }\href {https://doi.org/10.1103/PHYSREVB.103.054403/FIGURES/9/MEDIUM} {\bibfield  {journal} {\bibinfo  {journal} {Physical Review B}\ }\textbf {\bibinfo {volume} {103}},\ \bibinfo {pages} {054403} (\bibinfo {year} {2021})}\BibitemShut {NoStop}%
\end{thebibliography}%

\end{document}



\title{Supplemental Material: 
"Domain wall dynamics driven by a transient laser-induced magnetisation"}

\author{Paul-Iulian Gavriloaea}
\affiliation{Instituto de Ciencia de Materiales de Madrid, CSIC, Cantoblanco, 28049 Madrid, Spain}
	\author{Elías Saugar}
\affiliation{Instituto de Ciencia de Materiales de Madrid, CSIC, Cantoblanco, 28049 Madrid, Spain}
	\author{Rubén Otxoa}
\affiliation{Hitachi Cambridge Laboratory, CB3 OHE Cambridge, United Kingdom}
\affiliation{Donostia International Physics Center, 20018 Donostia San Sebastian, Spain}
	\author{Oksana Chubykalo-Fesenko}
\affiliation{Instituto de Ciencia de Materiales de Madrid, CSIC, Cantoblanco, 28049 Madrid, Spain}

\maketitle

\onecolumngrid
\newpage
\section{S1. Analysis of time dependent DW profile in a chain geometry. }
\begin{suppfigure}[t]
    \centering
    \includegraphics[width=1.0\linewidth]{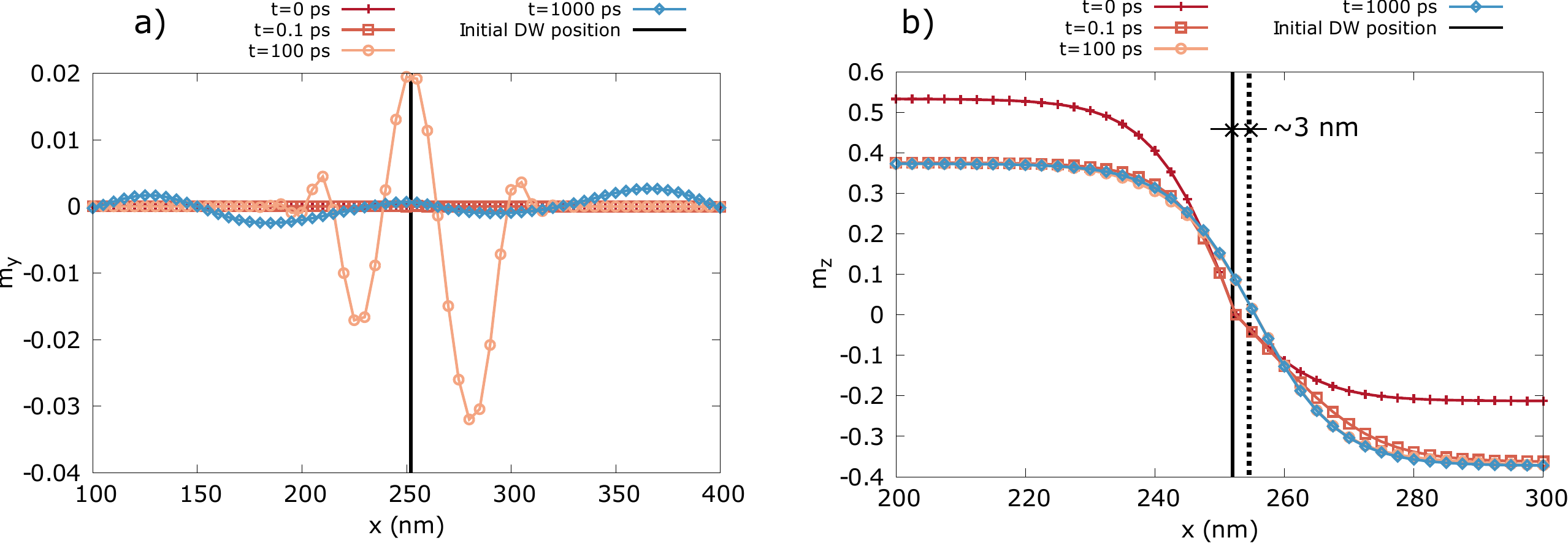}
    \caption{Time dependence of the N\'eel wall magnetisation profile along $Oy$ and $Oz$ during the dynamics induced by a transient magnetisation of amplitude $\delta m_z=0.16$. The x-axis range as well as the frequency of the point representation is adapted for each subplot for better visualisation. The continuous vertical line in subplots (b),(c) defines the initial DW position while the dotted vertical line in (c) marks an intermediate state ($t=1000~ps$) at which the wall finds itself approximately $3~nm$ away from its starting point according to Eq. (11) in the manuscript.} 
    \label{fig:Figure1_SI}
\end{suppfigure}
In Fig. S1 we analyse the time-dependent $Oy$ and $Oz$ magnetisation profiles of the N\'eel DW in the chain system under the influence of a transient magnetisation of amplitude $\delta m_z=0.16$. Although the initial DW structure is characterised by $m_y=0$ along its length as seen in subplot (a), the macro-spins will deviate away from the $Oxz$ plane due to the instability introduced in the system by the transient magnetisation. Thus, a non-zero  $m_y$ component will appear along the chain displaying an oscillatory behaviour which will continuously decrease in amplitude as a function of time. Though not directly shown, a similar behaviour is seen also in the $m_x$ and $m_z$ magnetisation profiles. The oscillations along $Ox$ will directly influence the DW position calculation, giving rise to the sinusoidal behaviour seen in Fig. 2(a) of the manuscript beyond the $100~ps$ time threshold. While slightly more difficult to identify visually, evidence of DW motion due to the induced transient magnetisation can also be extracted from the $m_z$ magnetisation profile in subplot (b) of Fig. S1. The symmetry point splitting the characteristic hyperbolic tangent profile in half clearly displaces in time as indicated by the vertical lines marking the initial state and the configuration we obtain at $t=1000~ps$. One should note that the range of the x-axes in the subplots of Fig. S1 as well as the frequency of the point representation have been individually adapted for better visualisation.
\bibliography{References.bib}